\documentclass[11pt,a4paper]{article}
\pdfoutput=1
\usepackage{jheppub}
\usepackage{slashed}

\makeatletter
\def\@fpheader{\relax}
\makeatother

\usepackage[czech,english]{babel}
\usepackage{graphicx}
\usepackage{amsmath,amsfonts,amssymb,mathcomp}
\DeclareMathOperator{\MyProd}{\scalebox{1.4}{$\mathrm{I\kern-0.2ex I}$}}

\title{Gravity Dual of Two-Dimensional $\mathcal{N} = (2,2)^*$ Supersymmetric Yang-Mills Theory and Integrable Models}

\author[a]{Jun Nian}

\emailAdd{nian@ihes.fr}

\affiliation[a]{Institut des Hautes \'Etudes Scientifiques\\
	Le Bois-Marie, 35 route de Chartres\\
         91440 Bures-sur-Yvette, France\\}

\abstract{The 2D $\mathcal{N}=(2,2)^*$ supersymmetric Yang-Mills theory can be obtained from the 2D $\mathcal{N}=(4,4)$ theory with a twisted mass deformation. In this paper we construct the gravity dual theory of the 2D $\mathcal{N}=(2,2)^*$ supersymmetric $U(N)$ Yang-Mills theory at the large $N$ and large 't Hooft coupling limit using the 5D gauged supergravity. In the UV regime, this construction also provides the gravity dual of the 2D $\mathcal{N}=(2,2)^*$ $U(N)$ topological Yang-Mills-Higgs theory. We propose a triality in the UV regime among integrable model, gauge theory and gravity, and we make some checks of this relation at classical level.}


\arxivnumber{}

\newcommand{\bea}{\begin{eqnarray}}
\newcommand{\eea}{\end{eqnarray}}

\newcommand{\be}{\begin{equation}}
\newcommand{\ee}{\end{equation}}

\begin{document}
\maketitle

\section{Introduction}\label{sec:introduction}

Many interesting and profound relations between integrable models and gauge theories have been revealed in recent years. A prototype of this relation is the celebrated AdS/CFT correspondence between the 10D type IIB superstring theory on $AdS_5 \times S^5$ and the 4D $\mathcal{N}=4$ supersymmetric Yang-Mills theory \cite{Maldacena}, where the 4D $\mathcal{N}=4$ supersymmetric Yang-Mills theory is believed to be a completely integrable model, and its integrability can be studied in the dual supergravity, which is the low-energy effective theory of the superstring theory (for a review see Ref.~\cite{AdSCFTinteg}).

More recently, some dualities between quantum integrable models and some 2D gauge theories have been established by Nekrasov and Shatashvili \cite{NS-1, NS-2, NS-3}. The integrable models are defined in (1+1)D, and they can be nonlinear partial differential equations or lattice spin models. The corresponding 2D gauge theories have $\mathcal{N}=(2,2)^*$ supersymmetry. In particular, the Bethe Ansatz equations of the quantum integrable models are equivalent to the vacuum equations of the gauge theories. The string dual of the Omega deformation and consequently the Nekrasov-Shatashvili duality has been constructed by Hellerman, Orlando and Reffert in Refs.~\cite{HOR-1, HOR-2, OR-2}, where they showed that various integrable models are dual to the NS5-D2-D4 systems in the fluxtrap background of the type IIA string theory. Besides the string dual, it would also be interesting to construct the gravity dual of the 2D gauge theories, which can provide us with a novel approach of studying the integrability on the gravity side.

We can start with the simplest example among the relations discovered by Nekrasov and Shatashvili, which is the one between the (1+1)D nonlinear Schr\"odinger equation and the 2D $\mathcal{N}=(2,2)^*$ $U(N)$ topological Yang-Mills-Higgs theory found by Gerasimov and Shatashvili \cite{GS-1, GS-2}. From the wave function of the 2D $\mathcal{N}=(2,2)^*$ $U(N)$ topological Yang-Mills-Higgs theory one can reproduce the wave function of the quantum nonlinear Schr\"odinger equation in the $N$-particle sector.

The 2D $\mathcal{N}=(2,2)^*$ $U(N)$ Yang-Mills-Higgs theory was constructed in Ref.~\cite{HiggsBundle}, and it is called topological when the coupling $g_{YM}$ is set to zero. This theory can be viewed as the dimensional reduction of the 4D topologically twisted $\mathcal{N}=2$ $U(N)$ super Yang-Mills theory with a deformation term, which provides the twisted mass and breaks $8$ supercharges into $4$ supercharges. Alternatively, it can also be viewed as the 2D $\mathcal{N}=(2,2)^*$ $U(N)$ super Yang-Mills theory deformed by some supersymmetry exact terms, which for supersymmetry closed observables (e.g. partition function, etc.) do not change the theory at quantum level. Hence, for supersymmetry closed observables the 2D $\mathcal{N}=(2,2)^*$ $U(N)$ Yang-Mills-Higgs theory is equivalent to the 2D $\mathcal{N}=(2,2)^*$ $U(N)$ super Yang-Mills theory, and we can study the latter one instead. Due to the asymptotic freedom of the 2D $\mathcal{N}=(2,2)^*$ $U(N)$ super Yang-Mills theory, the theory approaches the topological Yang-Mills-Higgs theory in the UV regime.

Based on the principle of gauge/gravity correspondence, we can construct the gravity dual of the 2D $\mathcal{N}=(2,2)^*$ $U(N)$ super Yang-Mills theory in the large $N$ and large 't Hooft coupling limit. In the UV regime, it also provides the gravity dual theory to the 2D topological Yang-Mills-Higgs theory. The basic idea is following. The gravity dual of the 2D $\mathcal{N}=(4,4)$ super Yang-Mills theory has been constructed in Ref.~\cite{GravDual-1}, and the solution can be embedded in the 10D type IIB supergravity uplifted from the 5D $\mathcal{N}=2$ gauged supergravity with the gauge group $U(1)^3$. Hence, we can first turn on an additional scalar field and a real parameter $\widetilde{c}$ corresponding to the twisted mass in the 5D gauged supergravity, and then uplift the gravity dual solution to the 10D type IIB supergravity. In this way, we obtain the mass-deformed supergravity solution, which is dual to the 2D $\mathcal{N}=(2,2)^*$ super Yang-Mills theory. The solution will be characterized by the 10D metric \eqref{eq:10DmetricGen} and the 5-form flux \eqref{eq:22flux}, both of which depend on the parameter $\widetilde{c}$. When $\widetilde{c} = 0$, the solution returns to the one constructed in Ref.~\cite{GravDual-1} that is dual to the 2D $\mathcal{N}=(4,4)$ super Yang-Mills theory, while for generic values of $\widetilde{c} \neq 0$ the gravity solution is dual to the 2D $\mathcal{N}=(2,2)^*$ super Yang-Mills theory. Various tests of the gravity dual solution can be made.

\begin{figure}[!htb]
      \begin{center}
        \includegraphics[width=0.52\textwidth]{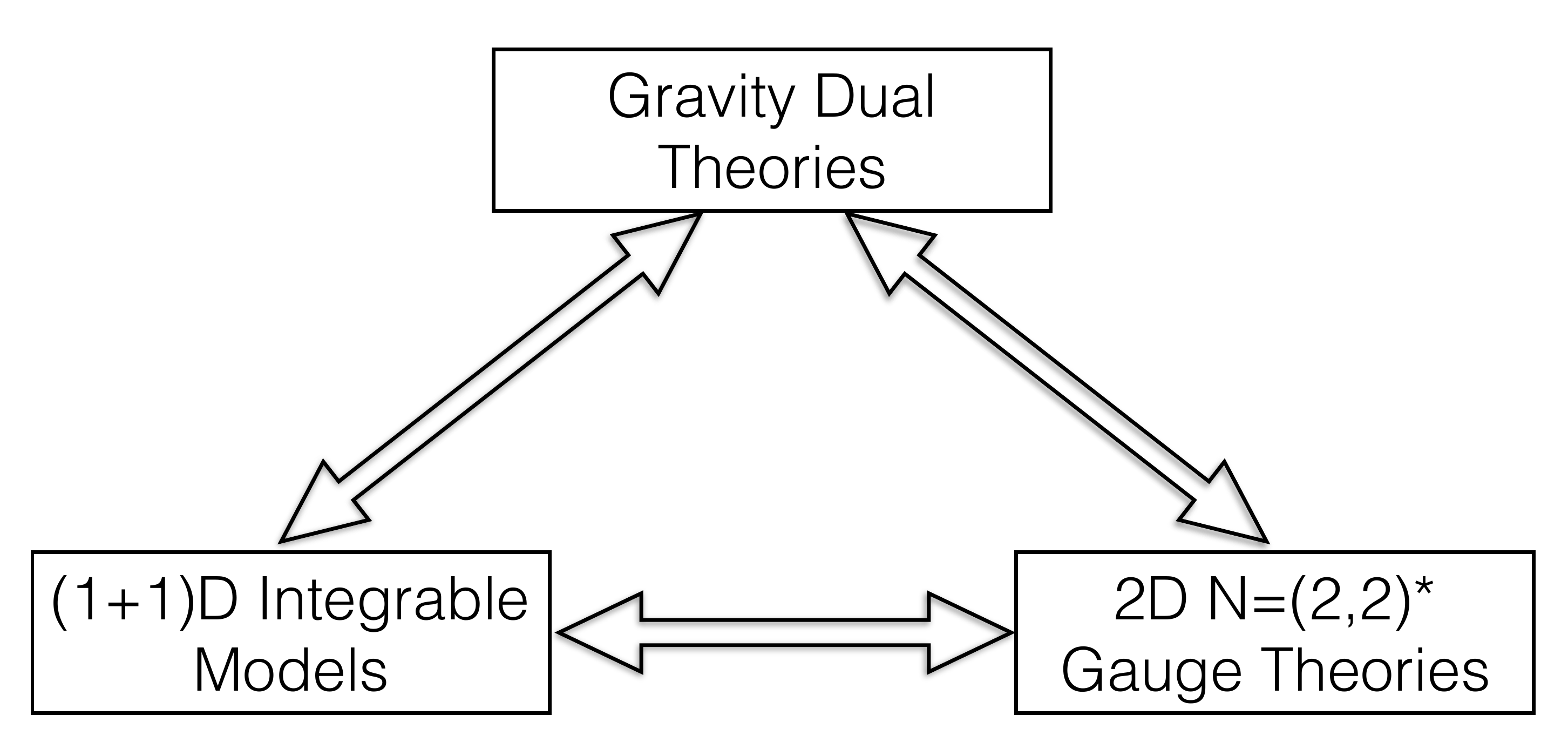}
      \caption{The triality among different theories}
      \label{fig:triangle}
      \end{center}
    \end{figure}

As discussed in Ref.~\cite{NS-1}, on top of the 2D $\mathcal{N}=(2,2)^*$ super Yang-Mills theory, if one turns on additional deformations (e.g. tree-level superpotential, matter multiplets in various representations, etc.), the resulting gauge theories correspond to a large class of integrable models. Based on our construction of the gravity dual of the 2D $\mathcal{N}=(2,2)^*$ super Yang-Mills theory with twisted mass, we propose a triality in the UV regime among gauge theories, integrable models and gravity theories (see Fig.~\ref{fig:triangle}).

By setting up this triality, in principle we can study the integrability on the gravity side, and at the same time study some properties of the gravity on the integrable model side. As the simplest example, we first consider the (1+1)D nonlinear Schr\"odinger equation, which corresponds to the 2D $\mathcal{N}=(2,2)^*$ super Yang-Mills theory with an additional tree-level twisted superpotential \cite{NS-1}. By taking the large $N$ limit in both the gravity dual and the nonlinear Schr\"odinger equation, we find the correspondence between classical solutions, i.e., $N$ D-branes in the supergravity and $N$ solitons in the classical nonlinear Schr\"odinger equation. The correspondence at quantum level will be investigated in the future work.

This paper is organized as follows. In Section~\ref{sec:reviewTYMH} the 2D $\mathcal{N}=(2,2)^*$ $U(N)$ topological Yang-Mills-Higgs theory and its relation with some other 2D gauge theories will be reviewed. In Section~\ref{sec:gravitydual}, we discuss the construction of the gravity dual of the 2D $\mathcal{N}=(2,2)^*$ $U(N)$ topological Yang-Mills-Higgs theory using the 5D gauged supergravity uplifted to 10D, and perform some checks of the gravity dual. In Section~\ref{sec:NLS}, we briefly review the nonlinear Schr\"odinger equation, in particular, how the $N$-particle solution to the quantum nonlinear Schr\"odinger equation reduces to the $N$-soliton solution in the large $N$ limit. The triality among different theories shown in Fig.~\ref{fig:triangle} will be established in Section~\ref{sec:correspondence}. Finally, in Section~\ref{sec:discussion} some prospects for the future research will be discussed. In Appendix~\ref{app:5DSUGRA} we review the consistently truncated 5D gauged supergravity, which has been applied the construction of the gravity dual in the main text. There are two 10D metrics appearing in the paper, one from the 5D gauged supergravity uplifted to 10D and the other from the brane construction. In Appendix~\ref{app:10Dmetric} we show that these two metrics can be identified. Moreover, the asymptotic forms of the metrics in the UV regime and the RR 5-form flux appearing in the 10D type IIB supergravity will be discussed in Appendix~\ref{app:UVmetric} and Appendix~\ref{app:5form} respectively. The preliminary results of this paper have also been reported in Ref.~\cite{IntegProceeding}, which appears in the proceeding of 24th International Conference on Integrable Systems and Quantum Symmetries.

\section{2D $\mathcal{N}=(2,2)^*$ Supersymmetric Yang-Mills Theory}\label{sec:reviewTYMH}

In this section we review the 2D $\mathcal{N}=(2,2)^*$ supersymmetric Yang-Mills theory and its relation with other 2D gauge theories. Also, we demonstrate that the 2D super Yang-Mills theory can be viewed as the dimensional reduction of the 4D topologically twisted $\mathcal{N}=2$ super Yang-Mills theory.

\subsection{Review of the 2D Gauge Theories}\label{sec:TYMH}

Following Ref.~\cite{Witten-2}, the 2D cohomological Yang-Mills theory for a compact group $G$ on a Riemann surface $\Sigma_h$ can be defined by the following path integral:
\be\label{eq:2DcohomYM}
  Z_{YM} (\Sigma) = \frac{1}{\textrm{Vol} (G)} \int D\varphi\, DA\, D\psi\, e^{S_{YM}}
\ee
with
\be\label{eq:2DcohomYMaction}
  S_{YM} = \frac{1}{2\pi} \int_\Sigma \textrm{Tr}\left[i \varphi F(A) + \frac{1}{2} \psi \wedge \psi - g_{YM}^2 \varphi^2 \, \textrm{vol}_{\Sigma_h} \right]\, ,
\ee
where $A$ is a connection on the principal $G$-bundle over $\Sigma_h$, while $\varphi$ and $\psi$ are a zero-form and a one-form on $\Sigma_h$ respectively taking values in the adjoint representation of the Lie algebra $\mathfrak{g} = \textrm{Lie} (G)$, and $\textrm{vol}_{\Sigma_h}$ is the volume form of $\Sigma_h$. The gauge coupling is $g_{YM}$. When $g_{YM}=0$, the theory is called topological Yang-Mills theory.

The theory \eqref{eq:2DcohomYM} is invariant under the following supersymmetry transformations:
\be
  Q A = i \psi\, ,\quad Q \psi = - \left(d\varphi + [A,\, \varphi] \right)\, ,\quad Q \varphi = 0\, ,
\ee
and the gauge transformations:
\be
  \mathcal{L} A = d \varphi + [A,\, \varphi]\, ,\quad \mathcal{L} \psi = -[\varphi,\, \psi]\, ,\quad \mathcal{L} \varphi = 0\, .
\ee

We will discuss in the next subsection, that the 2D cohomological Yang-Mills theory can be viewed as a consistent truncation of the dimensional reduction of the 4D topologically twisted $\mathcal{N}=2$ supersymmetric Yang-Mills theory, which preserves $\mathcal{N}=(4,4)$ supersymmetry in 2D. It was also demonstrated in Ref.~\cite{Witten-2}, that the 2D cohomological Yang-Mills theory is related to the $\mathcal{N}=0$ physical Yang-Mills theory
\be
  Z = \int DA\, D\varphi\, \textrm{exp} \left(\frac{1}{4 \pi^2} \int_{\Sigma_h} \textrm{Tr} \left[i \varphi F + \frac{\epsilon}{2} \varphi^2 d\mu \right] \right)\, ,
\ee
where $\epsilon > 0$ is a real number, and $d\mu$ is a measure defined by $\int_{\Sigma_h} d\mu = 1$. This theory has been studied a lot in the literature \cite{Migdal, Rusakov, Fine, Witten-1, Blau, Gross-1, Gross-2, Gross-3, Moore-1, Moore-2, Minahan, Gorsky, Douglas, Rudd, Vafa-1, Vafa-2}.

Using the technique of cohomological localization, one can evaluate the partition function of the 2D topological Yang-Mills theory, i.e. Eq.~\eqref{eq:2DcohomYM} with $g_{YM} = 0$, exactly, and the result for a Riemann surface $\Sigma_h$ of genus $h$ is \cite{Witten-2}
\be
  Z_{YM} (\Sigma_h) = \left(\frac{\textrm{Vol} (G)}{(2 \pi)^{\textrm{dim} (G)}} \right)^{2 h - 2} \sum_{\lambda} \left(\textrm{dim}\, R_\lambda \right)^{2 - 2 h}\, ,
\ee
where $\lambda$ is the highest weight of the irreducible representation $R_\lambda$ of the group $G$.

Related to the 2D cohomological Yang-Mills theory \eqref{eq:2DcohomYM}, the 2D $\mathcal{N}=(2,2)^*$ $U(N)$ Yang-Mills-Higgs theory was first constructed in Ref.~\cite{HiggsBundle}, and later also discussed in Refs.~\cite{GS-1, GS-2, PestunThesis}. It is defined by the path integral
\be\label{eq:2DTYMH}
  Z_{YMH} (\Sigma_h) = \frac{1}{\textrm{Vol} (\mathcal{G}_{\Sigma_h})} \int D \varphi_0\, D \varphi_\pm\, DA\, D\Phi\, D\psi_A\, D\psi_\Phi\, D\chi_\pm \, e^{S_{YMH}}\, ,
\ee
where
\be
  S_{YMH} = S_0 + S_1
\ee
with
\begin{align}
  S_0 & = \frac{1}{2\pi} \int_{\Sigma_h} \textrm{Tr} \bigg( i \varphi_0 (F(A) - \Phi \wedge \Phi) - c \, \Phi \wedge * \Phi - g_{YM}^2 \, \varphi_0^2\, \textrm{vol}_{\Sigma_h} \nonumber\\
  {} & \qquad\qquad\qquad\quad + \varphi_+ \nabla_A^{(1,0)} \Phi^{(0,1)}  + \varphi_- \nabla_A^{(0,1)} \Phi^{(1,0)} \bigg) \, ,\label{eq:TYMHS0}\\
  S_1 & = \frac{1}{2\pi} \int_{\Sigma_h} \textrm{Tr} \bigg(\frac{1}{2} \psi_A \wedge \psi_A + \frac{1}{2} \psi_\Phi \wedge \psi_\Phi + \chi_+ \left[\psi_A^{(1,0)},\, \Phi^{(0,1)} \right] \nonumber\\
  {} & \qquad \qquad \qquad + \chi_- \left[\psi_A^{(0,1)},\, \Phi^{(1,0)} \right] + \chi_+ \nabla_A^{(1,0)} \psi_\Phi^{(0,1)} + \chi_- \nabla_A^{(0,1)} \psi_\Phi^{(1,0)} \bigg)\, .\label{eq:TYMHS1}
\end{align}
Like in the 2D cohomological Yang-Mills theory \eqref{eq:2DcohomYM}, $A$ is a connection on the principal $G$-bundle over the Riemann surface $\Sigma_h$, while $\varphi_0$ and $\psi_A$ are a zero-form and a one-form respectively taking values in the adjoint representation of the Lie algebra $\mathfrak{g} = \textrm{Lie} (G)$. In addition, $\Phi$ and $\psi_\Phi$ are one-forms, while $\varphi_\pm$ and $\chi_\pm$ are zero-forms. From spin statistics, $\Phi$ and $\varphi_\pm$ are even fields, while $\psi_\Phi$ and $\chi_\pm$ are odd fields. Similar to the 2D cohomological Yang-Mills theory \eqref{eq:2DcohomYM}, the 2D Yang-Mills-Higgs theory \eqref{eq:2DTYMH} with $g_{YM}=0$ is called topological Yang-Mills-Higgs theory. When $g_{YM}=0$, it is obvious that the fields $\varphi_0$ and $\varphi_\pm$ in the action $S_0$ play the role of Lagrange multipliers and impose the Hitchin equations:
\be
  F(A) - \Phi \wedge \Phi = 0\, ,\quad \nabla_A^{(1,0)} \Phi^{(0,1)} = 0\, ,\quad \nabla_A^{(0,1)} \Phi^{(1,0)} = 0\, .
\ee

In Eq.~\eqref{eq:TYMHS0}, the term $\sim c\, \textrm{Tr} (\Phi \wedge * \Phi)$ in the action $S_0$ can be viewed as a mass deformation. As we will see in the next subsection, the 2D Yang-Mills-Higgs theory can be viewed as the dimensional reduction of the 4D topologically twisted $\mathcal{N}=2$ super-Yang-Mills theory. In terms of the ordinary 2D superfields, this mass deformation corresponds to giving mass to a 2D $\mathcal{N}=(2,2)$ chiral multiplet, as we will discuss in Subsection~\ref{sec:TwMass}. When $c=0$, the theory preserves $\mathcal{N}=(4,4)$ supersymmetry, while for generic values of $c \neq 0$ the theory preserves $\mathcal{N}=(2,2)$ supersymmetry. Hence, we would like to call the theory \eqref{eq:2DTYMH} with $g_{YM}=0$ and a generic value of $c$ the 2D $\mathcal{N}=(2,2)^*$ topological Yang-Mills-Higgs theory. Moreover, as discussed in Ref.~\cite{GS-1}, in the limit $c \to \infty$ the filed $\Phi$ and $\psi_\Phi$ drop out, and the 2D Yang-Mills-Higgs theory becomes the 2D Yang-Mills theory, while in another limit $c \to 0$ the 2D Yang-Mills-Higgs theory is equivalent to the partially gauge-fixed 2D Yang-Mills theory with the complex gauge group $G^c$ after integrating out some fermionic fields.

The topological Yang-Mills-Higgs theory \eqref{eq:2DTYMH} is invariant under the supersymmetry transformations given by
\be
  Q A = i \psi_A\, ,\quad Q \psi_A = - D \varphi_0\, ,\quad Q \varphi_0 = 0\, ,\label{eq:SUSY-1}
\ee
\be
  Q \Phi = i \psi_\Phi\, ,
\ee
\be
  Q \psi_\Phi^{(1,0)} = [\Phi^{(1,0)},\, \varphi_0] + c \Phi^{(1,0)}\, ,\quad Q \psi_\Phi^{(0,1)} = [\Phi^{(0,1)},\, \varphi_0] + c \Phi^{(0,1)}\, ,
\ee
\be
  Q \chi_\pm = i \varphi_\pm\, ,\quad Q \varphi_\pm = [\chi_\pm,\, \varphi_0] \pm c \chi_\pm \, .\label{eq:SUSY-4}
\ee
As we will see in the next subsection, this theory can also be understood as the dimensional reduction of the 4D topologically twisted $\mathcal{N}=2$ $U(N)$ super Yang-Mills theory with a deformation term.

From the supersymmetric transformations \eqref{eq:SUSY-1} $\sim$ \eqref{eq:SUSY-4}, one can show that the action of the 2D $\mathcal{N}=(2,2)^*$ Yang-Mills-Higgs theory \eqref{eq:2DTYMH} can be written as the 2D $\mathcal{N}=(2,2)^*$ cohomological Yang-Mills theory action \eqref{eq:2DcohomYMaction} with a supersymmetry exact deformation as follows:
\be
  S_{YMH} = S_{YM} + \left[Q,\, \int_{\Sigma_h} \textrm{Tr} \left(\frac{1}{2} \Phi\wedge \psi_\Phi + \chi_+ \nabla_A^{(1,0)} \Phi^{(0,1)} + \chi_- \nabla_A^{(0,1)} \Phi^{(1,0)} \right)\right]\, .
\ee
As discussed in Ref.~\cite{GS-1}, the second term in the equation above, which is a supersymmetry exact deformation term, can be replaced by other Lorentz- and gauge-invariant expressions without changing the theory for supersymmetry closed observables at quantum level. Also, we observe that the theory \eqref{eq:2DTYMH} does not contain kinetic terms explicitly, which can also be reintroduced by adding appropriate $Q$-exact terms, as shown in Ref.~\cite{Witten-2} (see also Refs.~\cite{S2-1, S2-2} for $\Sigma_h = S^2$).

Based on the discussions above, when we construct the gravity dual in the next section, on the field theory side we can consider the 2D $\mathcal{N}=(2,2)^*$ super Yang-Mills theory with the kinetic terms and the coupling $g_{YM} \neq 0$ instead of the topological Yang-Mills-Higgs theory, by deforming the original Yang-Mills theory with an appropriate $Q$-exact term. Due to the asympototic freedom of the 2D $\mathcal{N}=(2,2)^*$ super Yang-Mills theory, the 2D topological Yang-Mills theory with $g_{YM} = 0$ can be recovered in the UV regime.

As shown in Ref.~\cite{PestunThesis}, the 2D Yang-Mills-Higgs theory \eqref{eq:2DTYMH} can also be obtained from the so-called constrained Higgs-Yang-Mills theory, which is constructed using the symplectic structures $\omega_i$ and the moment maps $\mu_i$ on the field space $M = (A,\, \Phi)$, by introducing scalar auxiliary fields and their superpartners. Perturbatively, the constrained Higgs-Yang-Mills theory is equivalent to the 2D $\mathcal{N}=0$ physical Yang-Mills theory, which is also related to the 2D cohomological Yang-Mills theory \eqref{eq:2DcohomYM} as discussed in Ref.~\cite{Witten-2}.

Using the technique of cohomological localization, one can compute exactly the partition function of the 2D $\mathcal{N}=(2,2)^*$ $U(N)$ topological Yang-Mills-Higgs theory, and the result is
\be
  Z_{YMH} (\Sigma_h) = e^{(1-h)\, a(c)} \sum_{\lambda \in \mathcal{R}_N} D_\lambda^{2-2h}\, ,
\ee
where the factor $D_\lambda$ is given by
\be
  D_\lambda = \mu(\lambda)^{-1/2} \prod_{i<j} (\lambda_i - \lambda_j) \, \left(c^2 + (\lambda_i - \lambda_j)^2 \right)^{1/2} \, ,
\ee
and $\mathcal{R}_N$ denotes the set of $\lambda_i$'s satisfying the following equation:
\be\label{eq:BAE-1}
  e^{2 \pi i \lambda_j} \prod_{k\neq j} \frac{\lambda_k - \lambda_j + i c}{\lambda_k - \lambda_j - i c} = 1\, ,\quad k = 1,\, \cdots,\, N.
\ee
More precisely, in order to obtain the results above, one needs to consider the path integral in the presence of a nonlocal two-observable $\mathcal{O}^{(2)}$ to regularize it \cite{HiggsBundle, GS-1, GS-2}. We will encounter the same equation \eqref{eq:BAE-1} later in Section~\ref{sec:NLS}, which appears as the Bethe Ansatz equation of the (1+1)D nonlinear Schr\"odinger equation.

\subsection{Relation with 4D $\mathcal{N}=2$ super Yang-Mills Theory}

The topological twist of the 4D $\mathcal{N}=2$ super Yang-Mills theory was first studied by Witten in Ref.~\cite{TopQFT} (for a review see also Ref.~\cite{PestunReview}). Let us review it in the following. 

Before topological twist, the 4D $\mathcal{N}=2$ vector multiplet $(A_\mu, M, N, \psi_i, T_{ij})$ contains a gauge field $A_\mu$, two real scalars $M$ and $N$, the R-symmetry $SU(2)_I$-doublet of spinors $\psi_i$ $(i=1, 2)$ and the R-symmetry $SU(2)_I$-triplet of auxiliary fields $T_{ij}$, which is symmetric in $i$ and $j$. The gauge field and the scalars $(A_\mu, M, N)$ can also be viewed as the dimensional reduction of the 6D gauge field $A_m$. In the following, we adopt the notation used in Ref.~\cite{TopSYM}. Since the Lorentz group of the 4D Euclidean space is $SO(4) \cong SU(2)_L \times SU(2)_R$, the $SU(2)_L$ indices $\alpha$ and the $SU(2)_R$ indicies $\dot{\alpha}$ can be written explicitly, i.e., the spinors are $(\psi_{\alpha i}, \, \psi_{\dot{\alpha} i})$, while $A_{\alpha \dot{\alpha}} = A_\mu\, \sigma^\mu_{\alpha \dot{\alpha}}$. The supersymmetry transformations are given by
\begin{align}
  \delta A_{\alpha \dot{\alpha}} & = i \xi_\alpha\,^i \psi_{\dot{\alpha} i} - i \xi_{\dot{\alpha}}\,^i \psi_{\alpha i}\, ,\nonumber\\
  \delta \psi_{\alpha i} & = - \xi_\alpha\,^j T_{ij} + 2 \xi^\beta\,_i F_{\alpha\beta} + \frac{1}{4} \xi_{\alpha i} + \frac{1}{4} \xi_{\alpha i} [M,\, N] - \xi^{\dot{\beta}}\,_i D_{\alpha \dot{\beta}} N\, ,\nonumber\\
  \delta \psi_{\dot{\alpha} i} & = - \xi_{\dot{\alpha}}\,^j T_{ij} + 2 \xi^{\dot{\beta}}\,_i F_{\dot{\alpha} \dot{\beta}} + \frac{1}{4} \xi_{\dot{\alpha} i} + \frac{1}{4} \xi_{\dot{\alpha} i} [M,\, N] - \xi^{\beta}\,_i D_{\beta \dot{\alpha}} N\, ,\nonumber\\
  \delta M & = 2 i \xi^{\dot{\alpha} i} \psi_{\dot{\alpha} i}\, ,\nonumber\\
  \delta N & = 2 i \xi^{\alpha i} \psi_{\alpha i}\, ,\nonumber\\
  \delta T_{ij} & = i \xi^\alpha\,_i D_\alpha\,^{\dot{\beta}} \psi_{\dot{\beta} j} + i \xi^\alpha\,_j D_\alpha\,^{\dot{\beta}} \psi_{\dot{\beta} i} - \frac{i}{2} \xi^\alpha\,_i [\psi_{\alpha j},\, M] - \frac{i}{2} \xi^\alpha\,_j [\psi_{\alpha i},\, M] \nonumber\\
  {} & \quad - i \xi^{\dot{\alpha}}\,_i D^\beta\,_{\dot{\alpha}} \psi_{\beta j} - i \xi^{\dot{\alpha}}\,_j D^\beta\,_{\dot{\alpha}} \psi_{\beta i} - \frac{i}{2} \xi^{\dot{\alpha}}\,_i [\psi_{\dot{\alpha} j},\, N] - \frac{i}{2} \xi^{\dot{\alpha}}\,_j [\psi_{\dot{\alpha} i},\, N]\, .
\end{align}

Now let us consider the topological twist. In the presence of the R-symmetry group $SU(2)_I$ one can replace $SU(2)_R$ with the diagonal subgroup $SU(2)_D \subset SU(2)_R \times SU(2)_I$. Using the following notation
\begin{align}
  \xi & = \frac{1}{2} \epsilon^{\alpha i} \xi_{\alpha i}\, , \quad \hat{\xi}_{\alpha i} = \frac{1}{2} (\xi_{\alpha i} + \xi_{i \alpha})\, ,\nonumber\\
  \psi & = \epsilon^{\alpha i} \psi_{\alpha i}\, , \quad \chi_{\alpha i} = -\frac{1}{2} (\psi_{\alpha i} + \psi_{i \alpha})\, ,
\end{align}
one can express the supersymmetry transformations discussed above in terms of the new fields according to the representations of $SU(2)_L \times SU(2)_D$ and the transformation parameters $\xi$ and $\hat{\xi}_{\alpha i}$. It is observed in Ref.~\cite{TopQFT}, when the theory is minimally coupled to a gravitational background, the supersymmetry with the parameter $\xi$ can be defined for an arbitrary metric $g_{\mu\nu}$, i.e., the theory is topological. Hence, after the topological twist, the 4D $\mathcal{N}=2$ supersymmetry transformations can be formally expressed as
\be
  \delta X = \xi Q X\, ,
\ee
where $X$ stands for an arbitrary field in the 4D $\mathcal{N}=2$ vector multiplet after the topological twist. More explicitly,
\begin{align}
  Q A_\mu & = i \psi_\mu\, ,\nonumber\\
  Q \psi_\mu & = - D_\mu M\, ,\nonumber\\
  Q \psi & = \frac{1}{2} [M,\, N]\, ,\nonumber\\
  Q M & = 0\, ,\nonumber\\
  Q N & = 2 i \psi\, ,\nonumber\\
  Q \chi_{\mu\nu} & = T_{\mu\nu} + 2 F_{\mu\nu}^+\, ,\nonumber\\
  Q T_{\mu\nu} & = - 2 i (D_\mu \psi_\nu - D_\nu \psi_\mu)^+ - i [\chi_{\mu\nu},\, M]\, ,
\end{align}
where $\psi_\mu$ is a vector defined by $\psi_\mu = \sigma_\mu^{\alpha \dot{\alpha}} \psi_{\alpha \dot{\alpha}}$, $\psi$ is a scalar, and $\chi_{\mu\nu}$ is a self-dual rank-two anti-symmetric tensor satisfying $\chi_{\mu\nu} = (\sigma_{\mu\nu})^{\alpha\beta} \chi_{\alpha\beta}$, $\chi_{\mu\nu} = \frac{1}{2} \epsilon_{\mu\nu\rho\sigma} \chi^{\rho\sigma}$.

We can further reduce the 4D topological twisted $\mathcal{N}=2$ vector multiplet to 2D. In the reduction procedure, we also perform a consistent truncation on the components by setting $\psi = 0$ and $N = 0$, which conseqently leads to $Q \psi = 0$ and $Q N = 0$. After the dimensional reduction, the 4D gauge field $A_\mu$ becomes a 2D gauge field and a complex scalar, i.e. $(A,\, \Phi)$, where we suppress the 2D spacetime indices. Correspondingly, $\psi_\mu$ becomes a 2D vector field and a complex scalar denoted by $(\psi_A,\, \psi_\Phi)$ respectively. To treat $\chi_{\mu\nu}$ and $T_{\mu\nu}$, let us first define a new field $\widetilde{T}_{\mu\nu} \equiv T_{\mu\nu} + 2 F_{\mu\nu}^+$, then the supersymmetry transformations of $\chi_{\mu\nu}$ and $\widetilde{T}_{\mu\nu}$ become
\be
  Q \chi_{\mu\nu} = \widetilde{T}_{\mu\nu}\, ,\quad Q \widetilde{T}_{\mu\nu} = - i [\chi_{\mu\nu},\, M]\, .
\ee
Next, we can decompose $\chi_{\mu\nu}$ and $\widetilde{T}_{\mu\nu}$ into $(\chi_\pm,\, \chi_0)$ and $(\widetilde{T}_\pm,\, \widetilde{T}_0)$ respectively. We make a further consistent truncation by setting $\chi_0 = 0$ and $\widetilde{T}_0 = 0$. Moreover, let us rename the scalars $M$ and $\widetilde{T}_\pm$ to be $\varphi_0$ and $\varphi_\pm$ respectively, and assume that $\varphi_0$ depends only on the 2D coordinates. Finally, the 2D truncated $\mathcal{N}=(4,4)$ supersymmetry transformations obtained from the dimensional reduction are
\be\label{eq:red4DYM-1}
  Q A = i \psi_A\, ,\quad Q \psi_A = - D \varphi_0\, ,\quad Q \varphi_0 = 0\, ,
\ee
\be
  Q \Phi = i \psi_\Phi\, ,\quad Q \psi_\Phi = [\Phi,\, \varphi_0]\, ,
\ee
\be\label{eq:red4DYM-2}
  Q \chi_\pm = i \varphi_\pm\, ,\quad Q \varphi_\pm = [\chi_\pm,\, \varphi_0]\, .
\ee
As explained in Ref.~\cite{Witten-2}, these supersymmetry transformations can also be viewed as a 2D cohomological Yang-Mills theory of $(A,\, \psi,\, \phi)$ with two additional multiplets $(\lambda,\, \eta)$ and $(\chi,\, -i H)$, which satisfy
\be\label{eq:red4DYM-3}
  \delta A_i = i \epsilon \psi_i\, ,\quad \delta \psi_i = - \epsilon D_i \phi\, ,\quad \delta \phi = 0\, ,
\ee
\be
  \delta \lambda = i \epsilon \eta\, ,\quad \delta \eta = \epsilon [\phi,\, \lambda]\, ,
\ee
\be\label{eq:red4DYM-4}
  \delta \chi = \epsilon H\, ,\quad \delta H = i \epsilon [\phi,\, \chi]\, .
\ee
We see that the transformations Eqs.~\eqref{eq:red4DYM-1} $\sim$ \eqref{eq:red4DYM-2} or Eqs.~\eqref{eq:red4DYM-3} $\sim$ \eqref{eq:red4DYM-4} are the same as the ones for the 2D $\mathcal{N}=(2,2)^*$ Yang-Mills-Higgs theory given by Eqs.~\eqref{eq:SUSY-1} $\sim$ \eqref{eq:SUSY-4} with the mass deformation parameter $c$ turned off, i.e. $c=0$. Therefore, without the mass deformation the supersymmetry transformations of the 2D $\mathcal{N}=(2,2)^*$ Yang-Mills-Higgs theory coincide with the ones from the dimensional reduction of the 4D topologically twisted $\mathcal{N}=2$ supersymmetry transformations, which preserve 8 supercharges.

\section{Gravity Dual}\label{sec:gravitydual}

The gauge/gravity duality was initiated by the work of Maldacena \cite{Maldacena}, where it was conjectured that the 4D $\mathcal{N}=4$ supersymmetric $U(N)$ Yang-Mills theory is dual to the 10D type IIB supergravity on $AdS_5 \times S^5$ in the limit of large $N$ and large 't Hooft coupling $g_{YM}^2 N$. Afterwards, many more cases have been studied in the literature. In this section, we would like to construct the gravity dual of the 2D $\mathcal{N}=(2,2)^*$ $U(N)$ super Yang-Mills theory, which is equivalent to the 2D $\mathcal{N}=(2,2)^*$ $U(N)$ Yang-Mills-Higgs theory for supersymmetry closed observables.

In order to construct this gravity dual theory, we start from the gravity dual of 2D $\mathcal{N}=(4,4)$ super Yang-Mills theory, which was found in Ref.~\cite{GravDual-1}. By turning on an additional scalar field and choosing an appropriate scalar potential, the supersymmetry of the theory is broken to $\mathcal{N}=(2,2)$. The logic is similar to the case of the 4D super Yang-Mills theory. Starting from the gravity dual theory of the 4D $\mathcal{N}=4$ super Yang-Mills theory, one can turn on additional scalar fields and choose appropriate scalar potentials on the gravity side to preserve $\mathcal{N}=1$ \cite{PW-2} or $\mathcal{N}=2$ supersymmetry \cite{PW-1}. In particular, the latter one is known as the gravity dual theory of the 4D $\mathcal{N}=2^*$ super Yang-Mills theory.

\subsection{Gravity Dual of 2D $\mathcal{N} = (4,4)$ super Yang-Mills Theory}\label{44dual}

As explained in the beginning of this section, to construct the gravity dual of the 2D $\mathcal{N}=(2,2)^*$ $U(N)$ super Yang-Mills theory, we start with the known gravity dual of the 2D $\mathcal{N}=(4,4)$ $U(N)$ super Yang-Mills theory, which has been found in Ref.~\cite{GravDual-1}. Let us briefly review the construction in this subsection.

To realize the $\mathcal{N}=(4,4)$ supersymmetry, one considers $N$ D3-branes wrapped on the two-cycle of a CY 2-fold, which can be seen from the following table:
\begin{center}
\begin{tabular}{c|c|c|c|c|c|c|c|c|c|c}
  {} & \multicolumn{2}{c|}{$\mathbb{R}^{1,1}$} & \multicolumn{2}{c|}{$S^2$} & \multicolumn{2}{c|}{$N_2$} & \multicolumn{4}{c}{$\mathbb{R}^4$}\\
  \hline
  D3 & $\times$ & $\times$ & \textbigcircle & \textbigcircle & $\phantom{A}$ & $\phantom{A}$ & $\phantom{A}$ & $\phantom{A}$ & $\phantom{A}$ & $\phantom{A}$
\end{tabular}
\end{center}
Locally, this CY 2-fold is $S^2 \times N_2$. In a more general construction, $S^2$ can replaced by a Riemann surface $\Sigma$, which we will consider in the next subsection when we discuss the gravity dual of the 2D $\mathcal{N}=(2,2)^*$ super Yang-Mills theory. From the brane construction, one can propose an Ansatz of the metric in 10D type IIB supergravity:
\begin{align}
  ds^2 & = H(\rho, \sigma)^{-\frac{1}{2}} \left[dx_{1,1}^2 + \frac{z(\rho, \sigma)}{m^2} \left(d\theta^2 + \textrm{sin}^2 \theta\, (d\phi)^2 \right) \right] \nonumber\\
  {} & \quad + H(\rho, \sigma)^{\frac{1}{2}} \left[\frac{1}{z(\rho, \sigma)} d\sigma^2 + \frac{\sigma^2}{z(\rho, \sigma)} \left(d\psi + \textrm{cos} \theta\, d\phi \right)^2 + d\rho^2 + \rho^2 d\Omega_3^2 \right]\, ,\label{eq:44metric}
\end{align}
where
\be
  0 \leq \theta \leq \pi\, ,\quad 0 \leq \phi,\, \psi < 2 \pi\, ,\quad 0 \leq \rho,\, \sigma < \infty\, ,
\ee
while $z(\rho, \sigma)$ and $H(\rho, \sigma)$ are two factors that can be determined by solving the BPS equations, which will be discussed in the following. The constant $m$ has the dimension of mass, which will be fixed later by the quantization condition of the RR 5-form, and $m^{-1}$ can be viewed as a length scale in the metric \eqref{eq:44metric}. 
For a general Riemann surface $\Sigma$ instead of $S^2$ in the compactification, the metric \eqref{eq:44metric} always preserves an $U(1)\times SO(4)$ isometry, as expected from of the 2D $\mathcal{N}=(4,4)$ R-symmetry discussed in Ref.~\cite{MN-1}.

In addition to the metric \eqref{eq:44metric}, the RR 5-form in the 10D type IIB supergravity is given by
\be
  F_5 = \mathcal{F}_5 + * \mathcal{F}_5\, ,\label{eq:44flux}
\ee
where $\mathcal{F}_5 = d \mathcal{C}_4$ with
\be
  \mathcal{C}_4 = g(\rho,\, \sigma)\, \omega_3 \wedge (d\psi + \textrm{cos} \theta\, d\phi) \, ,
\ee
and $\omega_3$ is the volume element of the 3-sphere, i.e., for the metric of the 3-sphere given by
\be
  d\Omega_3^2 = d\beta_1^2 + \textrm{sin}^2 \beta_1 \left(d\beta_2^2 + \textrm{sin}^2 \beta_2\, d\beta_3^2 \right)
\ee
with
\be
  0 \leq \beta_1,\, \beta_2 \leq \pi\, ,\quad 0 \leq \beta_3 < 2 \pi\, ,
\ee
$\omega_3$ is defined as
\be
  \omega_3 = \textrm{sin}^2 \beta_1\, \textrm{sin} \beta_2\, d\beta_1 \wedge d\beta_2 \wedge d\beta_3\, .
\ee
The constant $m$ in the metric \eqref{eq:44metric} is fixed by the quantization condition of the RR 5-form $F_5$:
\be\label{eq:Fquantization}
  \frac{1}{2 \kappa_{10}^2} \int_{\mathcal{M}_5} F_5 = N\, T_3
\ee
with
\be
  2 \kappa_{10}^2 = (2 \pi)^7 \, g_s^2 \, (\alpha')^4\, ,\quad T_3 = \frac{1}{(2 \pi)^3 \, g_s \, (\alpha')^2}\, .
\ee
After some analyses shown in Appendix~\ref{app:5form}, one finds that the constant $m$ is fixed by
\be\label{eq:constm}
  \frac{1}{m^2} = \sqrt{4 \pi g_s\, N} \alpha'\, ,
\ee
where $g_s$ and $\alpha'$ are the string coupling constant and the Regge slope respectively.

From the metric \eqref{eq:44metric} and the flux \eqref{eq:44flux}, one can write down the BPS equations and try to solve them. It turns out that the BPS equations can be solved by using the results from the 5D $\mathcal{N}=2$ gauged supergravity discussed in Ref.~\cite{MN-1}. This is due to the fact that the metric \eqref{eq:44metric} can also be constructed from the 5D $\mathcal{N}=2$ gauged supergravity \cite{GravDual-1}. Briefly speaking, the coordinates $\rho$ and $\sigma$ in the metric \eqref{eq:44metric} can be recombined into two new variables $r$ and $\widetilde{\theta}$, and the radial coordinate $r$ together with the $\mathbb{R}^{1,1} \times S^2$ part of the metric \eqref{eq:44metric} becomes a warped $AdS_5$, while the remaining part of the metric becomes a warped $S^5$.

The 5D $\mathcal{N}=2$ gauged supergravity will be briefly reviewed in Appendix~\ref{app:5DSUGRA}. Let us recall some facts in the following. The bosonic part of the 5D $\mathcal{N}=2$ gauged supergravity with the gauge group $U(1)^3$ is given by \cite{Cvetic, Chamseddine, MN-1}:
\be\label{eq:5DGravL}
  \mathcal{L} = R - \frac{1}{2} (\partial_\mu \phi_1)^2 - \frac{1}{2} (\partial_\mu \phi_2)^2 + 4 \sum_{I=1}^3 e^{\alpha_I} - \frac{1}{4} \sum_{I=1}^3 e^{2 \alpha_I}\, F_{\mu\nu}^I F^{I, \mu\nu} + \frac{1}{4} \epsilon^{\mu\nu\alpha\beta\rho} F^1_{\mu\nu} F^2_{\alpha\beta} A^3_\rho\, ,
\ee
where
\be
  \alpha_1 = \frac{\phi_1}{\sqrt{6}} + \frac{\phi_2}{\sqrt{2}}\, ,\quad \alpha_2 = \frac{\phi_1}{\sqrt{6}} - \frac{\phi_2}{\sqrt{2}}\, ,\quad \alpha_3 = - \frac{2}{\sqrt{6}} \phi_1\, .
\ee
As shown by Maldacena and N\'u\~nez in Ref.~\cite{MN-1}, the theory can be compactified on a Riemann surface to provide the gravity duals of some 2D conformal field theories.

For the 5D gauged supergravity compactified on a Riemann surface of genus $g > 1$, there is the following condition to preserve at least 2D $\mathcal{N}=(0,2)$ supersymmetry:
\be
  a_1 + a_2 + a_3 = 1\, ,
\ee
where $a_I$ ($I = 1,\, 2,\, 3$) characterize the twist by picking up a special background
\be
  T = a_1 T_1 + a_2 T_2 + a_3 T_3
\ee
with $T_I$ ($I = 1,\, 2,\, 3$) denoting the generators of the $SO(2)$'s in the subgroup $SO(2) \times SO(2) \times SO(2)$ in the R-symmetry group $SO(6)$ of the 4D $\mathcal{N}=4$ super Yang-Mills theory. In this paper, we make the following choice of the parameters $a_I$'s:
\be\label{eq:aIchoice}
  a_1 = \widetilde{c}\, ,\quad a_2 = 0\, ,\quad a_3 = 1-\widetilde{c}
\ee
to describe the deformation of the 2D $\mathcal{N} = (4,4)$ gauge theory, where $\widetilde{c}$ is a real parameter. For $\widetilde{c}=0 \textrm{ or } 1$, both the 2D gauge theory and its gravity dual preserve $\mathcal{N} = (4,4)$ supersymmetry, while for generic values of $\widetilde{c} \neq 0,\, 1$ the supersymmetry is broken into $\mathcal{N} = (2,2)$ in both gauge theory and gravity.

By choosing $a_I = (0,\, 0,\, 1)$ and $H^2$ as the Riemann surface for compactification, it was constructed in Ref.~\cite{GravDual-1} the gravity solution with $\mathcal{N}=(4,4)$ supersymmetry in the 5D gauged supergravity:
\be
  ds_5^2 = e^{2 f(r)} \left(dx_{1,1}^2 +  dr^2 \right) + \frac{e^{2 g(r)}}{m^2} \left[d\theta^2 + \textrm{sinh}^2 \theta\, (d\phi)^2 \right]\, .
\ee
As explained in Ref.~\cite{MN-1}, for the compactification on the surface $S^2$, one can obtain the solution by replacing $\theta \to i \theta$:
\be\label{eq:44metricOnS2}
  ds_5^2 = e^{2 f(r)} \left(dx_{1,1}^2 +  dr^2 \right) + \frac{e^{2 g(r)}}{m^2} \left[d\theta^2 + \textrm{sin}^2 \theta\, (d\phi)^2 \right]\, .
\ee
Moreover, the three $U(1)$ gauge fields are chosen to be
\be\label{eq:44gaugefield}
  A^1 = 0\, ,\quad A^2 = 0\, ,\quad A^3 = \frac{1}{m} \, \textrm{cos} \theta\, d\phi\, .
\ee
Compared with the original Maldacena-N\'u\~nez solution (Ref.~\cite{MN-1}, see also Appendix~\ref{app:5DSUGRA}), we see that an additional parameter $m$ with dimension of mass has been introduced in both the metric \eqref{eq:44metricOnS2} and the gauge field \eqref{eq:44gaugefield}, and $m^{-1}$ plays the role of the length scale. As discussed above (see also Appendix~\ref{app:5form}), the value of $m$ is fixed by the quantization condition of the RR 5-form, and for the $\mathcal{N}=(4,4)$ case the expression of $m$ is given by Eq.~\eqref{eq:constm}.

The factors $f(r)$, $g(r)$ and the profiles of the scalar fields $\phi_1(r)$, $\phi_2(r)$ can be obtained by solving the BPS equations. As discussed in Ref.~\cite{MN-1}, if two of the three $a_I$'s are equal, e.g. $a_I = (0,\, 0,\, 1)$, the BPS equations can be simplified. We will discuss the BPS equations for generic $a_I$'s in the next subsection, while in this subsection we focus on the special case $a_I = (0,\, 0,\, 1)$. For this case one finds immediately that
\be
  \phi_2 = 0
\ee
is a solution, and we will argue in the next subsection that under the parametrization \eqref{eq:aIchoice} the special case $a_I = (0,\, 0,\, 1)$ has only the asymptotic solution $\phi_2 = 0$ near the boundary $r = 0$.

Defining $\varphi = \phi_1 / \sqrt{6}$, the BPS equations for $a_I = (0,\, 0,\, 1)$ become
\begin{align}
  f' & = -\frac{m}{3} e^f (2 e^{-\varphi} + e^{2 \varphi}) - \frac{m}{6} e^{f-2g} e^{-2 \varphi}\, ,\label{eq:BPSsp-1}\\
  g' & = -\frac{m}{3} e^f (2 e^{-\varphi} + e^{2 \varphi}) + \frac{m}{3} e^{f-2g} e^{-2 \varphi}\, ,\\
  \varphi' & = \frac{2 m}{3} e^f (- e^{-\varphi} + e^{2 \varphi}) + \frac{m}{3} e^{f-2g} e^{-2 \varphi}\, ,
\label{eq:BPSsp-3}
\end{align}
where the prime denotes the derivative with respect to $r$. This choice $(a_1,\, a_2,\, a_3) = (0,\, 0,\, 1)$ corresponds to the $\mathcal{N}=(4,4)$ case. For a different choice $(a_1,\, a_2,\, a_3) = (1/2,\, 0,\, 1/2)$, there are also solutions with $\phi_2 \neq 0$ to the BPS equations, and it corresponds to the $\mathcal{N}=(2,2)$ case, which we will discuss in more details in the next subsection.

We can study the asymptotic solutions to these BPS equations. By solving Eq.~\eqref{eq:BPSsp-1} $\sim$ Eq.~\eqref{eq:BPSsp-3} asymptotically near $r=0$ for $a_I = (0,\, 0,\, 1)$, one obtains \cite{MN-1}:
\begin{align}
  g(r) & = - \textrm{log} (r) + \frac{7}{36} r^2 + \cdots\, ,\\
  f(r) & = - \textrm{log} (m r) - \frac{1}{18} r^2 + \cdots\, ,\\
  \varphi (r) & = \frac{1}{3} r^2 \, \textrm{log}(r) + \cdots\, .
\end{align}
The asymptotic solution of $\varphi(r)$ implies the existence of a dual operator with dimension $\Delta=2$. However, compared to the generic case discussed in the next subsection, it also implies that the operator $\sim c\, \textrm{Tr} (\Phi \wedge * \Phi)$ appearing in the 2D Yang-Mils-Higgs theory \eqref{eq:2DTYMH} needs to be turned off, i.e. $c=0$ for this case.

Using the formulae in Ref.~\cite{Cvetic}, this gravity solution can be uplifted to 10D in the following way:
\be\label{eq:44metric-2}
  ds_{10}^2 = \sqrt{\Delta}\, ds_5^2 + \frac{3}{m^2 \sqrt{\Delta}} \sum_{I=1}^3 X_I \left[d\mu_I^2 + \mu_I^2 \left(d\phi^I + m A^I \right)^2 \right]\, ,
\ee
where $\phi^I$ ($I = 1, 2, 3$) are three angles with the range $[0, 2\pi)$. We emphasize that although similar in notation the angles $\phi^I$ are not related to the scalar fields $\phi_{1, 2}$ appearing in the action \eqref{eq:5DGravL}. Moreover,
\be\label{eq:defDelta}
  \Delta = \sum_{I=1}^3 X^I \mu_I^2\, ,\quad \textrm{with }  \sum_{I=1}^3 \mu_I^2 = 1\, .
\ee
One can parametrize $\mu_I$'s as follows:
\be
  \mu_1 = \textrm{cos} \widetilde{\theta}\, \textrm{sin} \widetilde{\psi}\, ,\quad \mu_2 = \textrm{cos} \widetilde{\theta}\, \textrm{cos} \widetilde{\psi}\, ,\quad \mu_3 = \textrm{sin} \widetilde{\theta}\, ,
\ee
where $0 \leq \widetilde{\theta} \leq \pi$ and $0 \leq \widetilde{\psi} < 2 \pi$. The quantities $X_I$ and $X^I$ are defined by
\be
  X_I = \frac{1}{3} \left(e^\varphi,\, e^\varphi,\, e^{-2 \varphi} \right)\, ,\quad X^I = \left(e^{-\varphi},\, e^{-\varphi},\, e^{2 \varphi} \right)\, .
\ee
It was shown in Ref.~\cite{GravDual-1} that indeed the metric \eqref{eq:44metric-2} can be rewritten into the expression of the metric \eqref{eq:44metric} discussed before by changing variables. We will also summarize some details in Appendix~\ref{app:10Dmetric}.

Finally, we would like to emphasize that the gravity dual solution becomes inconsistent in the IR regime. It can be seen from the following analysis. As shown in Ref.~\cite{GravDual-1}, by solving the BPS equations numerically one sees that the factor $z(\rho,\, \sigma)$ that controls the size of $S^2$ in the metric \eqref{eq:44metric} becomes negative for small values of $(\rho,\, \sigma)$, which corresponds to the IR regime. This fact indicates that the supergravity solution is inapplicable to this region.

\subsection{Gravity Dual of 2D $\mathcal{N} = (2,2)^*$ super Yang-Mills Theory}\label{22dual}

Now let us turn to the construction of the gravity dual of the 2D $\mathcal{N} = (2,2)^*$ $U(N)$ super Yang-Mills theory, which for supersymmetry closed observables is equivalent to the 2D $\mathcal{N} = (2,2)^*$ $U(N)$ Yang-Mills-Higgs theory at quantum level. As we discussed in the beginning of this section, we apply the same idea of constructing the gravity dual of 4D $\mathcal{N}=2^*$ super Yang-Mills theory \cite{PW-2, PW-1}, more specifically, we will turn on an additional scalar field and choose an appropriate scalar potential in the 5D gauged supergravity, and then uplift the solution to 10D type IIB supergravity.

\subsubsection{Solutions from 5D $\mathcal{N}=2$ Gauged Supergravity}

To construct the gravity dual of the 2D $\mathcal{N}=(2,2)^*$ Yang-Mills-Higgs theory, we start with the gravity dual of the $\mathcal{N}=(4,4)$ case discussed in the previous subsection and make use of the 5D $\mathcal{N}=2$ gauged supergravity (see Appendix~\ref{app:5DSUGRA} for a review).

\begin{figure}[!htb]
      \begin{center}
        \includegraphics[width=0.6\textwidth]{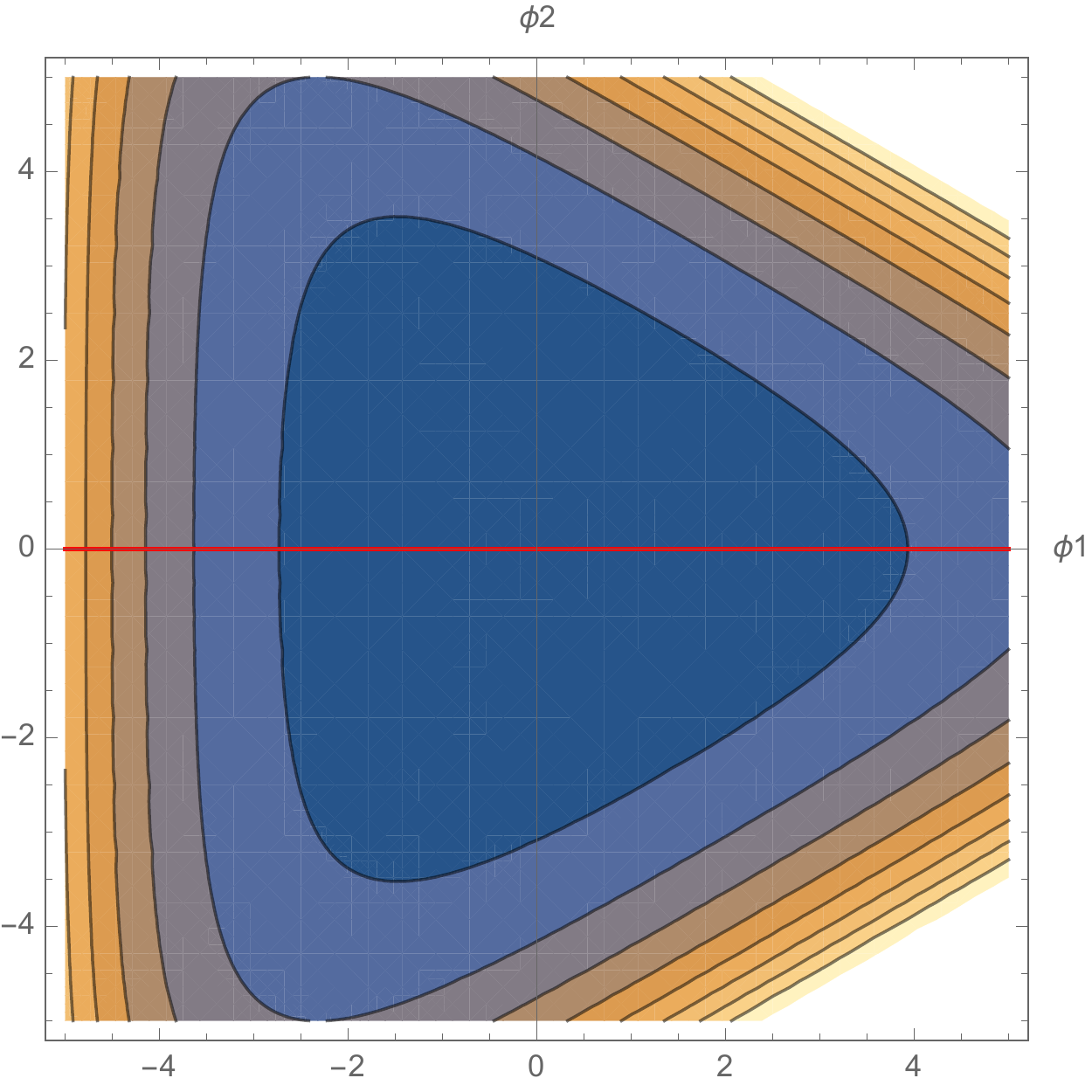}
      \caption{The scalar potential. The red line denotes $\phi_2 = 0$.}
      \label{fig:potential}
      \end{center}
    \end{figure}

Let us recall that the Lagrangian of the 5D $\mathcal{N}=2$ gauged supergravity is given by Eq.~\eqref{eq:5DGravL}. It has two scalars fields $\phi_1$ and $\phi_2$, and the scalar potential is shown in Fig.~\ref{fig:potential}.

As discussed in Subsection~\ref{44dual} and Appendix~\ref{app:5DSUGRA}, for the 5D gauged supergravity compactified on $H^2$ or more generally a Riemann surface $\Sigma$ of genus $g > 1$, to preserve at least 2D $\mathcal{N}=(0,2)$ supersymmetry the following condition should hold:
\be\label{eq:sumaI}
  a_1 + a_2 + a_3 = 1\, .
\ee
For 2D $\mathcal{N}=(2,2)$ supersymmetry, one of $a_I$'s should vanish. According to our choice \eqref{eq:aIchoice} made in this paper, when $\widetilde{c} \neq 0,\, 1$ the gravity solution corresponds to a 2D gauge theory with $\mathcal{N}=(2,2)$ supersymmetry.\footnote{In fact, the configuration with $\widetilde{c} = 1$ is equivalent to the one with $\widetilde{c} = 0$, which also corresponds to $\mathcal{N}=(4,4)$ and can be seen by interchanging $a_1$ and $a_3$.}

The $\mathcal{N}=(4,4)$ case discussed in the previous subsection can be viewed as a special case of the general 5D metric given by
\be\label{eq:5D22metric}
  ds^2 = e^{2 f(r)} (dx_{1,1}^2 + dr^2) + \frac{e^{2 g(r)}}{m^2 y^2} (dx^2 + dy^2)\, ,
\ee
where $f(r)$ and $g(r)$ are two factors determined by the BPS equations, and $m$ is a constant with the dimension of mass, which will be fixed later by the quantization condition of the RR 5-form in the 10D type IIB supergravity. Moreover, for the generic values of $a_I$'s, the three $U(1)$ gauge fields are given by
\be
  A^I = \frac{a_I}{m y}\, dx\, ,
\ee
where the parameters $a_I$ obey the condition \eqref{eq:sumaI}, and in addition they must be rational due to the quantization of the field strength on the compact Riemann surface $\Sigma$ of genus $g$ (see Appendix~\ref{app:5DSUGRA}), more precisely, for genus $g>1$:
\be
  2 a_I (g - 1) \in \mathbb{Z}\, .
\ee
Formally the metric \eqref{eq:5D22metric} looks the same as the one for the $\mathcal{N}=(4,4)$ case given by Eq.~\eqref{eq:44metricOnS2}, however, the factors $f(r)$ and $g(r)$ are determined by a set of BPS equations discussed in the following, which are different from the ones for the $\mathcal{N}=(4,4)$ case given by Eqs.~\eqref{eq:BPSsp-1} $\sim$ \eqref{eq:BPSsp-3}.

Besides the factors $f(r)$ and $g(f)$ appearing in the metric \eqref{eq:5D22metric}, one can also obtain the profiles of the scalar fields $\phi_1(r)$ and $\phi_2(r)$ by solving the BPS equations for generic values of $a_I$'s:
\begin{align}
  f' & = - m \left[\frac{e^f}{3} (X^1 + X^2 + X^3) + \frac{e^{f-2g}}{2} a_I X_I \right]\, ,\label{eq:BPSgen-1}\\
  g' & = - m \left[\frac{e^f}{3} (X^1 + X^2 + X^3) - e^{f-2g} a_I X_I \right]\, ,\\
  \phi_1' & = - m \left[\frac{\sqrt{6} e^f}{3} (X^1 + X^2 - 2 X^3) + \frac{\sqrt{6} e^{f-2g}}{2} (a_1 X_1 + a_2 X_2 - 2 a_3 X_3) \right]\, ,\\
  \phi_2' & = - m \left[\sqrt{2} e^f (X^1 - X^2) + \frac{3 \sqrt{2} e^{f-2g}}{2} (a_1 X_1 - a_2 X_2) \right]\, ,\label{eq:BPSgen-4}
\end{align}
where $X^I$ and $X_I$ are defined by
\be
  X^1 = e^{- \frac{\phi_1}{\sqrt{6}} - \frac{\phi_2}{\sqrt{2}}}\, ,\quad X^2 = e^{- \frac{\phi_1}{\sqrt{6}} + \frac{\phi_2}{\sqrt{2}}}\, ,\quad X^3 = e^{\frac{2}{\sqrt{6}} \phi_1}\, ,
\ee
\be
  X_1 = \frac{1}{3} e^{\frac{\phi_1}{\sqrt{6}} + \frac{\phi_2}{\sqrt{2}}}\, ,\quad X_2 = \frac{1}{3} e^{\frac{\phi_1}{\sqrt{6}} - \frac{\phi_2}{\sqrt{2}}}\, ,\quad X_3 = \frac{1}{3} e^{-\frac{2}{\sqrt{6}} \phi_1}\, .
\ee
For generic values of $a_I$'s the BPS equations do not have analytical solutions, but given boundary conditions one can solve the BPS equations \eqref{eq:BPSgen-1} $\sim$ \eqref{eq:BPSgen-4} numerically for arbitrary values of $r$.

By solving the equations \eqref{eq:BPSgen-1} $\sim$ \eqref{eq:BPSgen-4} near $r=0$, we obtain the asymptotic solutions:
\begin{align}
  g(r) & = - \textrm{log}(r) + \frac{7}{36} r^2 + \cdots\, ,\\
  f(r) & = - \textrm{log}(m r) - \frac{1}{18} r^2 + \cdots\, ,\\
  \phi_1(r) & = - \frac{1 - 3 a_3}{\sqrt{6}} r^2\, \textrm{log} (r) + \cdots\, ,\\
  \phi_2 (r) & = - \frac{a_1 - a_2}{\sqrt{2}} r^2 \, \textrm{log} (r) + \cdots\, .\label{eq:asymphi2}
\end{align}
The asymptotic solutions of $\phi_1$ and $\phi_2$ indicate that for generic values of $a_I$'s one can turn on two operators of dimension $\Delta=2$ dual to $\phi_1$ and $\phi_2$ respectively. Under our choice of $a_I$'s \eqref{eq:aIchoice} made in this paper, $a_1 - a_2 = \widetilde{c}$. Hence, $\phi_2$ vanishes asymptotically when $\widetilde{c} = 0$, or equivalently when $\mathcal{N}=(4,4)$, while a nonzero $\widetilde{c}$ will break the supersymmetry from $\mathcal{N}=(4,4)$ to $\mathcal{N}=(2,2)$ and at the same time allow a nonvanishing solution $\phi_2$ near the boundary $r=0$. Therefore, for the choice of $a_I$'s \eqref{eq:aIchoice} discussed in this paper, when the scalar field $\phi_2$ is turned off, i.e. $\phi_2 = 0$, which is denoted by the red line in Fig.~\ref{fig:potential}, each value of $\phi_1$ corresponds to a gravity solution with $\mathcal{N}=(4,4)$ supersymmetry. To break the supersymmetry from $\mathcal{N}=(4,4)$ to $\mathcal{N}=(2,2)$, we should turn on the scalar field $\phi_2$ in the scalar potential appearing in the Lagrangian \eqref{eq:5DGravL}.

From the discussions in Section~\ref{sec:reviewTYMH}, we know that to preserve $\mathcal{N} = (4,4)$ supersymmetry the operator $\sim c\, \textrm{Tr} (\Phi \wedge * \Phi)$ is turned off on the 2D gauge theory side, which corresponds to $(a_1,\, a_2,\, a_3) = (0,\, 0,\, 1)$ or a vanishing $\phi_2$ near $r=0$ on the gravity side. Deforming the 2D $\mathcal{N} = (4,4)$ gauge theory by turning on an additional operator $\sim c\, \textrm{Tr} (\Phi \wedge * \Phi)$ in the action \eqref{eq:TYMHS0} breaks the supersymmetry to $\mathcal{N} = (2,2)$, while correspondingly $\phi_2(r)$ has a nonvanishing asymptotic solution when the coefficient $a_1 - a_2 = \widetilde{c} \neq 0$. Hence, the parameters $c$ and $\widetilde{c}$ are correlated, and for small values of $\widetilde{c}$ there should be $\widetilde{c} \propto c$. The operator $\sim c\, \textrm{Tr} (\Phi \wedge * \Phi)$ in the gauge theory and the scalar field $\phi_2$ in the gravity are also correlated, although the dual operator of $\phi_2$ can be a linear combination of $c\, \textrm{Tr} (\Phi \wedge * \Phi)$ and some other dimension-two operators.

In summary, for the choice of the parameters $a_I$'s \eqref{eq:aIchoice}:
\begin{displaymath}
  a_I = (\widetilde{c},\, 0,\, 1 - \widetilde{c})\, .
\end{displaymath}
When $\widetilde{c}=0 \textrm{ or } 1$, it returns to the case analyzed in the previous subsection, which preserves $\mathcal{N}=(4,4)$ supersymmetry. When $\widetilde{c} \neq 0,\, 1$, the gravity solution preserves $\mathcal{N}=(2,2)$ supersymmetry. Consequently, the gauge fields now become
\be\label{eq:22gaugefield}
  A^1 = \frac{\widetilde{c}}{m y} \, dx\, ,\quad A^2 = 0\, ,\quad A^3 = \frac{1 - \widetilde{c}}{m y} \, dx\, .
\ee
As discussed in Appendix~\ref{app:5DSUGRA}, the parameter $\widetilde{c}$ should be rational due to the quantization of the field strength on the compact Riemann surface $\Sigma$ of the genus $g$, more precisely, for genus $g>1$:
\be
  2 \widetilde{c} (g - 1) \in \mathbb{Z}\, ,\quad 2 (1 - \widetilde{c}) (g - 1) \in \mathbb{Z}\, .
\ee
For a real deformation parameter $c$ in the 2D $\mathcal{N}=(2,2)^*$ super Yang-Mills theory, it can always be approached and approximated by the rational parameter $\widetilde{c}$ with increasing genus $g$ of the Riemann surface $\Sigma$ (see e.g. Refs.~\cite{Eoin-1, Eoin-2, Eoin-3}).

\subsubsection{Uplift 5D Solutions to 10D}

Like in the $\mathcal{N}=(4,4)$ case, after obtaining the factors $f(r)$, $g(r)$ and the scalar profiles $\phi_1(r)$, $\phi_2(r)$ by solving the BPS equations, we can use the formulae in Ref.~\cite{Cvetic} to uplift the solution for the $\mathcal{N}=(2,2)$ case in 5D gauged supergravity to a solution in 10D type IIB supergravity. The 10D metric is related to the 5D solution in the following way:
\begin{align}
  ds_{10}^2 & = \sqrt{\Delta}\, ds_5^2 + \frac{3}{m^2 \sqrt{\Delta}} \Bigg[\sum_{I=1}^3 X_I \, d\mu_I^2 + X_1 \mu_1^2 \left(d\phi^1 + \frac{\widetilde{c}}{y} \, dx \right)^2 \nonumber\\
  {} & \qquad\qquad\qquad\qquad\quad + X_2 \mu_2^2 \left(d\phi^2 \right)^2 + X_3 \mu_3^2 \left(d\phi^3 + \frac{1-\widetilde{c}}{y} \, dx \right)^2 \Bigg]\, ,\label{eq:10Dmetric}
\end{align}
where $ds_5^2$ is the 5D metric given by Eq.~\eqref{eq:5D22metric}, and $(\phi^1,\, \phi^2,\, \phi^3)$ are three angles with the range $[0, 2\pi)$, which are not related to the scalar fields $\phi_{1, 2}$ in the action \eqref{eq:5DGravL}. As defined before
\be
  \Delta = \sum_{I=1}^3 X^I \mu_I^2\, ,\quad \sum_{I=1}^3 \mu_I^2 = 1\, ,
\ee
with the parametrization
\be\label{eq:choosemu-1}
  \mu_1 = \textrm{cos} \widetilde{\theta}\, \textrm{sin} \widetilde{\psi}\, ,\quad \mu_2 = \textrm{cos} \widetilde{\theta}\, \textrm{cos} \widetilde{\psi}\, ,\quad \mu_3 = \textrm{sin} \widetilde{\theta}\, ,
\ee
where $0 \leq \widetilde{\theta} \leq \pi$ and $0 \leq \widetilde{\psi} < 2 \pi$. Hence, the 10D metric is
\begin{align}
  ds_{10}^2 & = \sqrt{\Delta} \left[e^{2f} (dx_{1,1}^2 + dr^2) + \frac{e^{2g}}{m^2 y^2} \left(dx^2 + dy^2 \right) \right] \nonumber\\
  {} & \quad + \frac{1}{m^2 \sqrt{\Delta}} \Bigg[e^{\varphi_1 + \varphi_2} d\mu_1^2 + e^{\varphi_1 - \varphi_2} d\mu_2^2 + e^{- 2 \varphi_1} d\mu_3^2 + e^{\varphi_1 + \varphi_2} \textrm{cos}^2 \widetilde{\theta}\, \textrm{sin}^2 \widetilde{\psi} \left(d\phi^1 + \frac{\widetilde{c}}{y} \, dx \right)^2 \nonumber\\
  {} & \qquad\qquad\qquad + e^{\varphi_1 - \varphi_2} \textrm{cos}^2 \widetilde{\theta}\, \textrm{cos}^2 \widetilde{\psi} (d\phi^2)^2 + e^{- 2 \varphi_1} \textrm{sin}^2 \widetilde{\theta} \left(d\phi^3 + \frac{1-\widetilde{c}}{y} \, dx \right)^2 \Bigg]\, ,\label{eq:10DmetricGen}
\end{align}
where $\varphi_1 \equiv \phi_1 / \sqrt{6}$ and $\varphi_2 \equiv \phi_2 / \sqrt{2}$ are the two scalar fields after rescaling, and the constant $m$ is fixed by the quantization condition of the RR 5-form given by Eq.~\eqref{eq:Fquantization}. For a generic value of $\widetilde{c}$ the metric above preserves an $SO(2) \times SO(2) \times SU(2)$ isometry, which will become manifest for the special value $\widetilde{c}=1/2$ discussed in the next subsection.

Moreover, the RR 5-form $F_5$ in 10D type IIB supergravity is given by
\be
  F_5 = \mathcal{F}_5 + * \mathcal{F}_5\, ,\label{eq:22flux}
\ee
where
\be
  \mathcal{F}_5 = \sum_{I=1}^3 \left[2 m X^I (X^I \mu_I^2 - \Delta) \epsilon_5 + \frac{1}{2 m^2 (X^I)^2} d(\mu_I^2) \left((d \phi^I + A^I) \wedge *_5 F^I + m X^I *_5 dX^I \right) \right]\, ,
\ee
and $\epsilon_5$ and $*_5$ are the volume form and the Hodge dual of the 5D space respectively, while $F^I = dA^I$ are the field strengths of the gauge fields given by Eq.~\eqref{eq:22gaugefield}. $\phi^I$ ($I = 1, 2, 3$) are three angles with the range $[0, 2\pi)$, which should be distinguished from the scalar fields $\phi_{1, 2}$ appearing in the supergravity action \eqref{eq:5DGravL}. Similar to the $\mathcal{N}=(4,4)$ case, the quantization condition of the RR 5-form $F_5$ \eqref{eq:Fquantization} fixes the constant $m$, as shown in Appendix~\ref{app:5form}.

\subsubsection{Solutions from Brane Construction}\label{sec:SolFromBrane}

The 10D supergravity solutions with $\mathcal{N}=(2,2)^*$ supersymmetry have been constructed in the previous subsections. We would like to rewrite the 10D metrics into the form similar to the ones given in Refs.~\cite{GravDual-1, GravDual-3}, from which the brane constructions and consequently the supersymmetry are more transparent.

First, the value $\widetilde{c}=0$ corresponds to the original undeformed theory discussed in Subsection~\ref{44dual}, which is the gravity dual of the 2D $\mathcal{N} = (4,4)$ super Yang-Mills theory. It was shown in Ref.~\cite{GravDual-1} that by changing variables the 10D metric \eqref{eq:10DmetricGen} with $\widetilde{c}=0$ and $\varphi_2 = 0$ can be identified with the one from the brane construction \eqref{eq:44metric}, as discussed in Appendix~\ref{app:10Dmetric}. We have also discussed in Subsection~\ref{44dual} that this configuration can be viewed as $N$ D3-branes wrapped on a two-cycle of a CY 2-fold, and it preserves 8 supercharges, i.e. $\mathcal{N}=(4,4)$ supersymmetry.

For a generic value of $\widetilde{c} \neq 0,\, 1$, the metric \eqref{eq:10DmetricGen} can also be rewritten into the form from the construction of branes wrapped on Calabi-Yau spaces. Since the explicit form of the metric is very complicated, which makes the relevant physics less transparent, we will skip the generic case. Instead we consider the special case $\widetilde{c}=1/2$ in the following to demonstrate the procedure.

Consider the special case $\widetilde{c} = 1/2$. For simplicity we take the Riemann surface $\Sigma$ to be $S^2$, then the 10D metric \eqref{eq:10DmetricGen} can be written as
\begin{align}
  ds_{10}^2 & = \sqrt{\Delta} \left[e^{2f} (dx_{1,1}^2 + dr^2) + \frac{e^{2g}}{m^2} \left(d\theta^2 + \textrm{sin}^2 \theta\, d\phi^2 \right) \right] \nonumber\\
  {} & \quad + \frac{1}{m^2 \sqrt{\Delta}} \Bigg[e^{\varphi_1 + \varphi_2} d\mu_1^2 + e^{\varphi_1 - \varphi_2} d\mu_2^2 + e^{- 2 \varphi_1} d\mu_3^2 + e^{\varphi_1 + \varphi_2} \textrm{cos}^2 \widetilde{\theta}\, \textrm{sin}^2 \widetilde{\psi} \left(d\phi^1 + \frac{1}{2}\, \textrm{cos} \theta \, d\phi \right)^2 \nonumber\\
  {} & \qquad\qquad\qquad + e^{\varphi_1 - \varphi_2} \textrm{cos}^2 \widetilde{\theta}\, \textrm{cos}^2 \widetilde{\psi} (d\phi^2)^2 + e^{- 2 \varphi_1} \textrm{sin}^2 \widetilde{\theta} \left(d\phi^3 + \frac{1}{2} \, \textrm{cos} \theta\, d\phi \right)^2 \Bigg]\, ,\label{eq:10DmetricSp}
\end{align}
where again $\varphi_1 \equiv \phi_1 / \sqrt{6}$ and $\varphi_2 \equiv \phi_2 / \sqrt{2}$ are the two scalar fields after rescaling. As discussed in Ref.~\cite{MN-1}, the metric \eqref{eq:10DmetricSp} has an $SO(2) \times SO(2) \times SU(2)$ isometry from two angles $\phi$ and $\phi^2 (\equiv \psi)$ as well as the $S^3$ parametrized by $(\widetilde{\psi},\, \phi^1,\, \phi^3)$ as a Hopf fibration on $S^2$, which corresponds to the remaining R-symmetry after the twist by picking up appropriate $U(1)$'s from the original R-symmetry group $SO(6)$ of the 4D $\mathcal{N}=4$ super Yang-Mills theory. Consequently, the dual field theory has an $\mathcal{N}=(2,2)$ supersymmetry.

After some changes of variables, the metric above can be further brought into the following form, from which the brane construction is clearer. The details of the derivation will be summarized in Appendix~\ref{app:10Dmetric}.
\begin{align}
  ds^2 & = H(\rho, \sigma)^{-\frac{1}{2}} \left[dx_{1,1}^2 + \frac{z(\rho, \sigma)}{m^2} \left(d\theta^2 + \textrm{sin}^2 \theta\, (d\phi)^2 \right) \right] \nonumber\\
  {} & \quad + H(\rho, \sigma)^{\frac{1}{2}} \Bigg[\frac{\sigma^2}{\sqrt{z(\rho, \sigma)}} \left(d\widetilde{\psi}^2 + \textrm{sin}^2 \widetilde{\psi} \left(d\phi^3 + \frac{1}{2} \textrm{cos} \theta\, d\phi \right)^2 + \textrm{cos}^2 \widetilde{\psi} \left(d\phi^1 + \frac{1}{2} \textrm{cos} \theta\, d\phi \right)^2 \right) \nonumber\\
  {} & \qquad\qquad\qquad + \frac{1}{\sqrt{z(\rho, \sigma)}} d\sigma^2 + d\rho^2 + \rho^2 d\psi^2 \Bigg]\, ,\label{eq:22metric}
\end{align}
where $H(\rho,\, \sigma)$ and $z(\rho,\, \sigma)$ are two factors that can be determined by solving the BPS equations, as discussed in Section~\ref{sec:UVmetric} and Appendix~\ref{app:UVmetric}. This metric can be interpreted as $N$ D3-branes wrapped on a two-cycle of a CY 3-fold. Hence, the theory manifestly preserves $\mathcal{N}=(2,2)$ supersymmetry, and the D3-branes can be viewed as solitons in the 10D type IIB supergravity. The configuration can be schematically presented in the following table.
\begin{center}
\begin{tabular}{c|c|c|c|c|c|c|c|c|c|c}
  {} & \multicolumn{2}{c|}{$\mathbb{R}^{1,1}$} & \multicolumn{2}{c|}{$S^2$} & \multicolumn{4}{c|}{$N_4$} & \multicolumn{2}{c}{$\mathbb{R}^2$}\\
  \hline
  D3 & $\times$ & $\times$ & \textbigcircle & \textbigcircle & $\phantom{A}$ & $\phantom{A}$ & $\phantom{A}$ & $\phantom{A}$ & $\phantom{A}$ & $\phantom{A}$
\end{tabular}
\end{center}
Locally, the D3-branes are $\mathbb{R}^{1,1} \times S^2$, and the CY 3-fold is $S^2 \times N_4$. The coordinates for $S^2$, $N_4$ and $\mathbb{R}^2$ are $(\theta,\, \phi)$, $(\sigma,\, \widetilde{\psi},\, \phi^1,\, \phi^3)$ and $(\rho,\, \psi)$ respectively.

From the analyses above, we have seen that turning on the mass deformation on the field theory side corresponds to a change of topology on the gravity side, i.e., from a CY 2-fold at $\widetilde{c}=0$ becomes a CY 3-fold at $\widetilde{c} \neq 0$, and consequently the supersymmetry is broken from $\mathcal{N}=(4,4)$ to $\mathcal{N}=(2,2)$. This story is quite well-known in the literature, for instance, the 4D $\mathcal{N}=1$ super Yang-Mills theory as deformations of the $\mathcal{N}=2$ super Yang-Mills theory (see Ref.~\cite{Bertolini} for a review). On the gravity side, one can start with $N$ D5-branes wrapped on a two-cycle of a CY 2-fold, e.g. a K3 surface, and the 10D spacetime is $\mathbb{R}^{1,3} \times \textrm{K3} \times \mathbb{C}$. By adding a scalar potential $W(\phi)$, one can break the supersymmetry from $\mathcal{N}=2$ to $\mathcal{N}=1$. On the gravity side, it corresponds to making the direct product $\textrm{K3} \times \mathbb{C}$ into a nontrivial fibration, i.e., a CY 3-fold with $SU(3)$ holonomy.

Similar to the $\mathcal{N}=(4,4)$ case, for generic $\mathcal{N}=(2,2)$ gravity dual solutions the factor $z(\rho,\, \sigma)$ in the metric from the brane construction also becomes negative at very small values of $(\rho,\, \sigma)$, which corresponds to the IR regime. Because $z(\rho,\, \sigma)$ controls the size of $S^2$, $z(\rho,\, \sigma)$ becomes negative at very small values of $(\rho,\, \sigma)$ implies that the gravity dual solution is inapplicable to the IR region. However, as an exception, the special case $\widetilde{c}=1/2$ flows to a good $AdS_3$ vacuum, which is dual to an $\mathcal{N}=(2,2)$ conformal field theory \cite{BB}.

\subsubsection{Twisted Mass}\label{sec:TwMass}

In this subsection we would like to highlight the twisted mass in the 2D super Yang-Mills theory and its gravity dual.

A typical 2D $\mathcal{N}=(2,2)$ supersymmetric gauge theory can be constructed in the superspace using the the chiral multiplet $\bf{X}$, the vector multiplet $\bf{V}$ and the twisted chiral multiplet $\bf{\Sigma} = \mathcal{D}_+ \overline{\mathcal{D}}_- \bf{V}$. For the most general 2D $\mathcal{N}=(2,2)$ supersymmetric gauge theory, one should also consider the semi-chiral multiplets $\mathbb{X}_L$ and $\mathbb{X}_R$ as well as the semi-chiral vector multiplets ($\mathbb{V},\, \widetilde{\mathbb{V}}$), which have been studied in the literature (see e.g. Refs.~\cite{Martin-1, Martin-2, Martin-3, Martin-4, Martin-5, T2, S2}). In this subsection, we restrict our discussion to the typical case with $\bf{X}$, $\bf{V}$ and $\bf{\Sigma}$. Using these multiplets, the supersymmetric actions can be expressed in terms of the $D$-term, the $F$-term and the twisted $F$-term as follows:
\begin{align}
  D\textrm{-term}: & \int d^2 x\, d^4 \theta\, \textrm{Tr} \left[\bf{K} \left(e^{\bf{V}/2} \bf{X},\, \overline{\bf{X}} e^{\bf{V}/2} \right) + \bf{\Sigma} \overline{\bf{\Sigma}} \right]\, ,\\
  F\textrm{-term}: & \int d^2 x\, d\theta^+ d\theta^- \, W(\bf{X}) + \textrm{c.c.}\, ,\\
  F^{\textrm{tw}}\textrm{-term}: & \int d^2 x\, d\theta^+ d\bar{\theta}^- \, \widetilde{W} (\bf{\Sigma}) + \textrm{c.c.}\, .
\end{align}

Following Ref.~\cite{NS-1}, to introduce the twisted mass we first consider the chiral multiplet $\bf{X}$ transforming in a linear representation $\mathcal{R}$ of the gauge group $G$, and $\mathcal{R}$ can be decomposed as
\be
  \mathcal{R} = \bigoplus_i \bf{M}_i \otimes R_i\, ,
\ee
denoting the irreducible representation $R_i$ with the multiplicity space $\bf{M}_i$. The global symmetry group $H$ is a subgroup of of $H^{\textrm{max}}$ defined by
\be
  H^{\textrm{max}} \equiv \bigotimes_i U(\bf{M}_i)\, .
\ee
The twisted masses are the deformation parameters:
\be
  \widetilde{m} = (\widetilde{m}_i)\, ,\quad \textrm{with } \widetilde{m}_i \in \textrm{End} (\bf{M_i}) \cap \textrm{$H$}\, .
\ee
The twisted mass term in the superspace is
\be
  \int d^2 x\, d^4 \theta\, \textrm{Tr}_{\mathcal{R}} \bf{X}^\dagger \left(\sum_i e^{\widetilde{V}_i + \textrm{h.c.}} \otimes \mathbb{I}_{R_i} \right) \bf{X}
\ee
with
\be
  \widetilde{V}_i = \widetilde{m}_i \theta_+ \bar{\theta}_- \, .
\ee
The twisted mass is a real parameter, which should be distinguished from the complex masses defined by a $F$-term with the superpotential
\be
  W = \sum_{a, b} m_a^b \widetilde{Q}_b Q^a\, ,
\ee
where $Q^a$ and $\widetilde{Q}_b$ denote $n_{\bf{f}}$ chiral multiplets in the fundamental representation of $G$ and $n_{\bar{\bf{f}}}$ chiral multiplets in the anti-fundamental representation of $G$ respectively.

The string dual of the twisted mass was discussed in Ref.~\cite{HananyHori}, and more recently in Refs.~\cite{HOR-1, Reffert, OR-1, HOR-2}, where the string dual of the Omega deformation was constructed using an NS5-D2-D4 system in a fluxtrap background of the type IIA string theory. For the 2D $\mathcal{N}=(2,2)^*$ case, the brane configuration can be shown in the following table:
\begin{center}
\begin{tabular}{c|cccccccccc}
  {} & 0 & 1 & 2 & 3 & 4 & 5 & 6 & 7 & 8 & 9\\
  \hline
  fluxbrane & $\times$ & $\times$ & $\times$ & $\times$ & & & & & & $\times$\\
  NS5 & $\times$ & $\times$ & & & & & $\times$ & $\times$ & $\times$ & $\times$\\
  D2 & $\times$ & $\times$ & $\times$ & & & & & & &\\
  D4 & $\times$ & $\times$ & & $\times$ & $\times$ & $\times$ & & & &
\end{tabular}
\end{center}
Schematically, there are $r+1$ parallel NS5-branes placed perpendicular to the $x_2$-direction, depending on the rank $r$ of the symmetry group of the spin chain. Between each pair of nearby NS5-branes, there can be a stack of $N_a$ D2-branes suspended between them, where $N_a$ ($a \in \{1,\, \cdots,\, r \}$) becomes the number of particles for the $a$-th color in the spin chain. There can also be a stack of $L_a$ D4-branes hanging on each NS5-brane, and $L_a$ ($a \in \{1,\, \cdots,\, r \}$) denotes the effective length of the spin chain for the $a$-th color. On the gauge theory side, the configuration corresponds to a quiver gauge theory with the gauge group $U(N_1) \times U(N_2) \times \cdots \times U(N_r)$, and attached to each node of the quiver there is a flavor group $U(L_a)$ ($a \in \{1,\, \cdots,\, r \}$). In this brane configuration, the separation of the D4-branes in the $x_6$-, $x_7$-directions can be interpreted as the twisted masses.

For the 2D Yang-Mills-Higgs theory \eqref{eq:2DTYMH} written in terms of the fields after topological twist, the twisted mass term is $\sim c\, \textrm{Tr} (\Phi \wedge * \Phi)$ in the action \eqref{eq:TYMHS0}, and the parameter $c$ can be viewed as the twisted mass, which also appears in the Bethe Ansatz equation of the nonlinear Schr\"odinger equation discussed in Section~\ref{sec:NLS}. Based on the discussions in this section, we have found the counter-part of the twisted mass in the type IIB gravity dual, which is proportional to the parameter $\widetilde{c}$. We can also justify this statement by analyzing the R-symmetry of the gravity dual solution. For a generic value of $\widetilde{c}$, the metric of the $\mathcal{N}=(2,2)^*$ gravity dual solution \eqref{eq:10DmetricGen} preserves the isometry $SO(2)\times SO(2)\times SU(2)$, which becomes manifest for the special case \eqref{eq:10DmetricSp} with $\widetilde{c}=1/2$. The parameter $\widetilde{c}$ is invariant under the two $SO(2)$'s in the isometry, which is supported by the analysis in Ref.~\cite{HananyHori} that the twisted masses are neutral under the $U(1)_V$ R-symmetry, while the complex masses are charged. Hence, the interpretation of $\widetilde{c}$ as the counter-part of the twisted mass in the gravity dual is consistent with the previous results. It would be nice to connect the IIB gravity dual considered in this paper with the IIA string theory dual discussed in Refs.~\cite{HOR-1, Reffert, OR-1, HOR-2}, which we would like to explore in the future research.

\subsection{Some Checks}

After constructing the gravity dual of the 2D $\mathcal{N}=(2,2)^*$ $U(N)$ super Yang-Mills theory in the previous subsection, in this subsection we perform some checks of the gravity dual solution by calculating some quantities, for instance, the running coupling and the entanglement entropy.

\subsubsection{UV Metric}\label{sec:UVmetric}
To compute the quantities of interest, we need to first analyze the metric of the gravity dual in the UV regime.

The $\mathcal{N}=(4,4)$ case with $\widetilde{c} = 0$ was analyzed in Ref.~\cite{GravDual-1}, and we will summarize the steps in Appendix~\ref{app:UVmetric}. In the UV regime, $z$ approaches a constant $z_*$ defined in Appendix~\ref{app:UVmetric}, and the values of $\rho$ and $\sigma$ are large. The final result of the UV metric for the $\mathcal{N}=(4,4)$ case is
\begin{align}
  ds_{UV}^2 & \approx \frac{m^2}{z_*} \left[dx_{1,1}^2 + \frac{z_*}{m^2} \left(d\theta^2 + \textrm{sin}^2 \theta\, d\phi^2 \right) \right] + \frac{1}{m^2} \frac{du^2}{u^2} \nonumber\\
  {} & \quad + \frac{1}{m^2} \left[d \hat{\alpha}^2 + \textrm{sin}^2 \hat{\alpha} \left(d\psi + \textrm{cos} \theta\, d\phi \right)^2 + \textrm{cos}^2 \hat{\alpha} \, d\Omega_3^2 \right]\, ,\label{eq:44UVmetric}
\end{align}
where $u$ and $\hat{\alpha}$ are two new variables related to the variables $\rho$ and $\sigma$ in the following way:
\be
  u = \sqrt{\sigma^2 + z_* \rho^2}\, ,\quad \textrm{tan} \hat{\alpha} = \frac{\sigma}{\sqrt{z_*} \rho}\, ,\quad 0 \leq \hat{\alpha} \leq \frac{\pi}{2}\, .
\ee
One can solve for the factors $z(\rho,\, \sigma)$ and $H(\rho,\, \sigma)$ in the metric \eqref{eq:44metric} near the asymptotic value $z_*$, and they have the expressions:
\be\label{eq:44UVfactor}
  z(\rho,\, \sigma) \approx z_* - \frac{z_*}{2 m^2 (\sigma^2 + z_* \rho^2)}\, ,\quad H(\rho,\, \sigma) \approx \frac{z_*^2}{m^4 \left(\sigma^2 + z_* \rho^2 \right)^2}\, .
\ee

We would like to apply the same approach to analyze the $\mathcal{N}=(2,2)^*$ case with $\widetilde{c} = 1/2$ discussed in Subsection~\ref{sec:SolFromBrane}, i.e. $a_I = (1/2,\, 0,\, 1/2)$. The final result for the UV metric in this case is
\begin{align}
  ds^2 & = \frac{m^2 u^2}{\sqrt{z_*}} \left[dx_{1,1}^2 + \frac{z_*}{m^2} \left(d\theta^2 + \textrm{sin}^2 \theta\, (d\phi)^2 \right) \right] + \frac{1}{m^2} \frac{du^2}{u^2} \nonumber\\
  {} & \quad + \frac{1}{m^2} \Bigg[d\hat{\alpha}^2 + \textrm{sin}^2 \hat{\alpha} \left(d\widetilde{\psi}^2 + \textrm{sin}^2 \widetilde{\psi} \left(d\phi^3 + \frac{1}{2} \textrm{cos} \theta\, d\phi \right)^2 + \textrm{cos}^2 \widetilde{\psi} \left(d\phi^1 + \frac{1}{2} \textrm{cos} \theta\, d\phi \right)^2 \right) \nonumber\\
  {} & \qquad\qquad + \textrm{cos}^2 \hat{\alpha}\, d\psi^2 \Bigg]\, .\label{eq:22UVmetric}
\end{align}
For this case, the relations between the new variables $u$, $\hat{\alpha}$ and the old variables $\rho$, $\sigma$ are slightly different from the ones for the $\mathcal{N}=(4,4)$ case:
\be
  u = \sqrt{\sigma^2 + \sqrt{z_*} \rho^2}\, ,\quad \textrm{tan} \hat{\alpha} = \frac{\sigma}{(z_*)^{1/4} \rho}\, ,\quad 0 \leq \hat{\alpha} \leq \frac{\pi}{2}\, .
\ee
Again, in the UV region $u$ is large, and $z$ approaches $z_*$. One can solve for the factors $z(\rho,\, \sigma)$ and $H(\rho,\, \sigma)$ in the metric \eqref{eq:22metric} near the asymptotic value $z_*$. For the $\mathcal{N}=(2,2)^*$ case with $\widetilde{c} = 1/2$ they become
\be\label{eq:22UVfactor}
  z(\rho,\, \sigma) \approx z_* - \frac{\sqrt{z_*}}{2 m^2 (\sigma^2 + \sqrt{z_*} \rho^2)}\, ,\quad H(\rho,\, \sigma) \approx \frac{z_*}{m^4 \left(\sigma^2 + \sqrt{z_*} \rho^2 \right)^2}\, .
\ee
More details of the derivations are shown in Appendix~\ref{app:UVmetric}.

\subsubsection{Running Coupling}\label{sec:RunningCoupling}

As discussed in Ref.~\cite{GravDual-1}, to compute the running coupling of the 2D $\mathcal{N}=(4,4)$ super Yang-Mills theory in the gravity dual, one can study the dynamics of a D3-brane probe moving in the background of the metric and the RR form. The action is given by the DBI and the WZ terms:
\be\label{eq:probe}
  S = - T_3 \int d^4 \xi\, e^{-\Phi} \sqrt{- \textrm{det} \left(\hat{G}_4 + 2 \pi \alpha' F \right)} + T_3 \int \hat{C}_4\, ,
\ee
where $\xi^a = (x^0,\, x^1,\, \theta,\, \phi)$ denote the coordinates on the world volume of the D3-brane, and $F$ is the field strength of the world volume gauge field, while $\hat{G}_4$ and $\hat{C}_4$ denote the induced metric on the D3-brane world volume and the pullback of the RR 4-form potential respectively, which are given by
\be
  \hat{G}_{ab} d\xi^a d\xi^b = H^{-\frac{1}{2}} dx_{1,1}^2 + \frac{z H^{-\frac{1}{2}}}{m^2} \left[(d\theta)^2 + \textrm{sin}^2 \theta \left(1 + \sigma^2 \frac{m^2 H}{z^2} \, \textrm{cot}^2 \theta \right) d\phi^2 \right]\, ,
\ee
\be
  \hat{C}_4 = \frac{z\, \textrm{sin} \theta}{m^2 H} \, dx^0 \wedge dx^1 \wedge d\theta \wedge d\phi\, .
\ee
Plugging these terms into the effective action \eqref{eq:probe}, in the absense of the gauge field we obtain
\be
  S = - T_3 \int d^2 x\, d\theta\, d\phi\, \frac{z}{m^2 H} \, \textrm{sin} \theta \left(\sqrt{1 + \sigma^2 \frac{m^2 H}{z^2}\, \textrm{cot}^2 \theta} - 1 \right)\, .
\ee
This potential vanishes at $\sigma = 0$, which can be interpreted as the supersymmetric locus of the brane inside the CY space.

Next, at $\sigma = 0$ we switch on the world volume gauge field, and assume that the only nonvanishing components of the gauge field are those along the unwrapped directions $x^\mu = (x^0,\, x^1)$. Also, we consider the flat directions $Z^i$ in the transverse directions of the metric \eqref{eq:44metric}:
\be
  d\rho^2 + \rho^2 d\Omega_3^2 = (dZ^i)^2
\ee
with $i = 1, \cdots, 4$, and relate them with the scalar fields $n^i$ of the gauge theory living on the brane:
\be
  Z^i = 2 \pi \alpha' n^i\, .
\ee
At $\sigma = 0$ the Lagrangian of the DBI term for the probe brane action becomes
\be
  \mathcal{L}_{DBI} = - T_3 \frac{z}{m^2 H} \, \textrm{sin} \theta \left[1 + \frac{(2 \pi \alpha')^2}{2} H F_{\mu\nu} F^{\mu\nu} + H (\partial_\mu Z^i)^2 \right]^{\frac{1}{2}}\, .
\ee
Generalizing this Lagrangian to the non-Abelian case and integrating it over $(\theta,\, \phi)$, we obtain at quadratic order:
\be
  \int d\theta\, d\phi\, \mathcal{L}_{DBI} \bigg|_{\textrm{quadratic}} = - \frac{(2\pi)^3 (\alpha')^2 T_3}{m^2} z \, \textrm{Tr} \left(\frac{1}{2} F_{\mu\nu} F^{\mu\nu} + \partial_\mu n^i \partial^\mu n^i \right)\, .
\ee
Therefore, we obtain
\be\label{eq:coupling}
  \frac{1}{g_{YM}^2} = \frac{(2\pi)^3 (\alpha')^2 T_3}{m^2}\, z(\rho,\, \sigma=0) = \frac{z(\rho,\, \sigma=0)}{m^2 g_s}\, ,
\ee
where in the last step we used $(2 \pi)^3 (\alpha')^2 T_3 = 1 / g_s$.

To compare the result above with the one from field theory, we relate the energy scale $\mu$ to the holographic coordinate $\rho$ in the following way:
\be\label{eq:scalerel}
  \rho = 2 \pi \alpha' \mu\, .
\ee
Moreover, we use the expression of the factor $z(\rho,\, \sigma)$ obtained from the analysis of the UV metric for the $\mathcal{N}=(4,4)$ case given by Eq.~\eqref{eq:44UVfactor} at $\sigma=0$:
\be\label{eq:Zrel}
  z(\rho,\, \sigma = 0) \approx z_* - \frac{1}{2 m^2 \rho^2}\, .
\ee
Taking into the account the relation \eqref{eq:constm}, finally we obtain for the 2D $\mathcal{N}=(4,4)$ pure super Yang-Mills theory:
\be\label{eq:44couplingGrav}
  \frac{z(\rho,\, \sigma=0)}{m^2 g_s} = \frac{z_*}{m^2 g_s} - \frac{N}{2 \pi \mu^2}\, ,
\ee
or equivalently,
\be\label{eq:44couplingField}
  \frac{1}{g_{YM}^2 (\mu)} = \frac{1}{g_{YM}^2} \left(1 - \frac{g_{YM}^2}{2 \pi \mu^2} N \right)\, ,
\ee
where the UV coupling constant is defined as
\be
  g_{YM}^2 \equiv m^2 g_s / z_*\, .
\ee
This expression of the running coupling implies the negative beta-function and consequently the asymptotic freedom, i.e., when $\mu \to \infty$, $g_{YM}^2 \to 0$, and it matches the field theory result \cite{2DYMcoupling, GravDual-1}:
\be
  \frac{1}{g_{YM}^2 (\mu)} = \frac{1}{g_{YM}^2} \left(1 + \frac{g_{YM}^2}{4 \pi \mu^2} b \right)\, ,
\ee
where for the vector multiplet with gauge group $SU(N)$:
\be
  b = \left(\frac{1}{6} n_s - 4\, n_v + \frac{2}{3} n_f \right) N\, ,
\ee
with $(n_v,\, n_f,\, n_s)$ denoting the number of vector fields, Dirac fermions and real scalar fields respectively, which is $(n_v,\, n_f,\, n_s) = (1,\, 2,\, 4)$ for the 2D $\mathcal{N} = (4, 4)$ super Yang-Mills theory and $(n_v,\, n_f,\, n_s) = (1,\, 1,\, 2)$ for the 2D $\mathcal{N} = (2, 2)$ super Yang-Mills theory.

For the special $\mathcal{N}=(2,2)^*$ case with $\widetilde{c}=1/2$ considered in Subsection~\ref{sec:SolFromBrane}, we can repeat the same steps. The results are similar but slightly different, for instance, the effective action \eqref{eq:probe} now becomes
\be
  S = - T_3 \int d^2 x\, d\theta\, d\phi\, \frac{z}{m^2 H} \, \textrm{sin} \theta \left(\sqrt{1 + \sigma^2 \frac{m^2 H}{z^{3/2}} \, \textrm{cot}^2 \theta} - 1 \right)\, .
\ee
Also, in the metric \eqref{eq:22metric} the flat directions in the transverse direction are:
\be
  d\rho^2 + \rho^2 d\psi^2 = dZ^2\, .
\ee
Similar analyses lead to the same result as Eq.~\eqref{eq:coupling}, and based on the expression \eqref{eq:22UVfactor} of the factor $z(\rho,\, \sigma)$ for the $\mathcal{N}=(2,2)^*$ case with $\widetilde{c} = 1/2$ we obtain
\be\label{eq:22couplingGrav}
  \frac{1}{g_{YM}^2} = \frac{z_*}{m^2 g_s} - \frac{2 \pi (\alpha')^2}{\rho^2} N\, ,
\ee
which is essentially the same as the result \eqref{eq:44couplingGrav} for the $\mathcal{N}=(4,4)$ case obtained from the gravity side.

To compare this result with the one from field theory, we would like to first recall the relation between the 4D $\mathcal{N}=2^*$ super-Yang-Mills theory and the 4D $\mathcal{N}=4$, $\mathcal{N}=2$ super-Yang-Mills theories. As discussed in Refs.~\cite{RussoZarembo-1, RussoZarembo-2}, the 4D $\mathcal{N}=2^*$ super-Yang-Mills theory can be obtained by giving equal masses to the two hypermultiplets in the $\mathcal{N}=4$ super-Yang-Mills theory, which can be integrated out in the IR, leaving a pure $\mathcal{N}=2$ super-Yang-Mills theory. Hence, the 4D $\mathcal{N}=2^*$ super-Yang-Mills theory can be viewed as a flow from the $\mathcal{N}=4$ super-Yang-Mills in the UV to the $\mathcal{N}=2$ super-Yang-Mills in the IR, and the difference between the $\mathcal{N}=2^*$ and the $\mathcal{N}=4$ super-Yang-Mills theories disappears in the UV.

Similar to the 4D story briefly mentioned above, the 2D $\mathcal{N}=(2,2)^*$ super-Yang-Mills theory can be viewed as a flow from the $\mathcal{N}=(4,4)$ super-Yang-Mills in the UV to the $\mathcal{N}=(2,2)$ super-Yang-Mills in the IR, and the difference between the $\mathcal{N}=(2,2)^*$ and the $\mathcal{N}=(4,4)$ super-Yang-Mills theories vanishes in the UV. Hence, the running coupling \eqref{eq:22couplingGrav} of the 2D $\mathcal{N}=(2,2)^*$ super-Yang-Mills theory should have the same expression as the $\mathcal{N}=(4,4)$ theory in the UV given by Eq.~\eqref{eq:44couplingField}, which is true as long as the relation \eqref{eq:scalerel} holds.\footnote{The author would like to thank Saebyeok Jeong for discussions on this point.}

When the mass deformation $c \to \infty$, one obtains the pure $\mathcal{N}=(2,2)$ super-Yang-Mills theory. On the gravity side, since the parameter $\widetilde{c}$ does not show up in the result \eqref{eq:22couplingGrav}, we expect that Eq.~\eqref{eq:22couplingGrav} from gravity side still holds in the pure $\mathcal{N}=(2,2)$ case. However, as explained in Ref.~\cite{GravDual-3}, to match the field theory result for the $\mathcal{N}=(2,2)$ case, one cannot adopt the same relation \eqref{eq:scalerel} beween the energy scale and the holographic coordinate as the $\mathcal{N}=(4,4)$ case. Instead, for the $\mathcal{N}=(2,2)$ case we require
\be
  \rho^2 = \frac{8}{3} \pi^2 (\alpha')^2 \mu^2\, ,
\ee
then the result matches the field theory expectation for the 2D $\mathcal{N}=(2,2)$ super Yang-Mills theory that we have discussed above:
\be
  \frac{1}{g_{YM}^2 (\mu)} = \frac{1}{g_{YM}^2} \left(1 - \frac{3 g_{YM}^2}{4 \pi \mu^2}\, N \right)\, ,
\ee
where again $g_{YM}^2 \equiv m^2 g_s / z_*$.

\subsubsection{Entanglement Entropy}

Another quantity one can compute in the gravity dual is the entanglement entropy. Let us first summarize the results for the $\mathcal{N}=(4,4)$ case considered in Ref.~\cite{GravDual-1}, and then discuss the $\mathcal{N}=(2,2)^*$ case.

Consider two complementary regions $A$ and $B$ in the Hilbert space of a quantum field theory. For simplicity, one can consider two spatially complementary regions. The reduced density matrix $\rho_A$ is defined as the density matrix traced over the degrees of freedom in $B$:
\be
  \rho_A \equiv \textrm{tr}_B\, \rho\, .
\ee
The entanglement entropy is then defined as the von Neumann entropy of $\rho_A$:
\be
  S_E \equiv - \textrm{tr} \rho_A \textrm{log} \rho_A\, .
\ee
For a quantum field theory with gravity dual, a holographic way of computing the entanglement in a $(d+1)$-dimensional conformal field theory was proposed by Ryu and Takayanagi in Ref.~\cite{RT}:
\be\label{eq:RT}
  S_E = \frac{\textrm{area of } \gamma_A}{4 G_{d+2}}\, ,
\ee
where $\gamma_A$ is the minimal surface spanned by the spatial region $A$ in the $(d+2)$-dimensional AdS space, and $G_{d+2}$ is the $(d+2)$-dimensional Newton's constant.

Applying a generalized version of the formula \eqref{eq:RT} to the 2D case, one obtains:
\be\label{eq:10DRT}
  S_E = \frac{1}{4 G_{10}} \int_\Omega d^8 \xi\, e^{-2 \phi} \sqrt{\textrm{det} \, \hat{G}_8}\, ,
\ee
where the spatial region $A$ is taken to be $- \ell / 2 \leq x^1 \leq \ell / 2$, and $\Omega$ is the 8-dimensional minimal surface with $A$ as its boundary. $G_{10} = 8 \pi^6 \alpha'^4 g_s^2$ is the 10-dimensional Newton's constant, and $\hat{G}_8$ is the induced metric on $\Omega$.

Let us define the spatial coordinate to be $x \equiv x^1$, and the region $A$ is taken to be an interval $- \ell / 2 \leq x \leq \ell / 2$. For the $\mathcal{N}=(4,4)$ case, using the UV metric \eqref{eq:44UVmetric}, we assume that the 8D surface is described by
\be
  u = u(x)
\ee
in the 8D space parametrized by the coordinates
\be
  \xi^a = (x,\, \theta,\, \phi,\, \hat{\alpha},\, \psi,\, \beta^i)\, .
\ee
Plugging the UV metric \eqref{eq:44UVmetric} into the formula of the entanglement entropy \eqref{eq:10DRT}, we obtain
\be\label{eq:EEtemp}
  S_E = \frac{\pi^4}{m^6 G_{10}} \int_{-\ell / 2}^{\ell / 2} dx\, u \left(u'^2 + \frac{m^4 u^4}{z_*} \right)^{\frac{1}{2}}\, .
\ee
For this theory, the first integral is conserved, which leads to
\be
  \frac{u^5}{\left(u'^2 + \frac{m^4 u^4}{z_*} \right)^{\frac{1}{2}}} = \frac{\sqrt{z_*}}{m^2} \, u_0^3\, ,
\ee
where $u_0$ is a constant corresponding to the maximal value of $u$ on the surface. The equation above can be written as
\be\label{eq:uprime}
  u' = \pm \frac{m^2}{\sqrt{z_*}} u^2 \sqrt{\left(\frac{u}{u_0} \right)^6 - 1}\, .
\ee
Consequently, one can express the length $\ell$ of the interval $- \ell / 2 \leq x \leq \ell / 2$ as
\be
  \ell = 2 \int_{u_0}^\infty \frac{du}{|u'(x)|} = 2 \int_{u_0}^\infty \frac{du}{\frac{m^2}{\sqrt{z_*}} u^2 \sqrt{\left(\frac{u}{u_0} \right)^6 - 1}}\, .
\ee
This integral can be evaluated analytically, and the final result is
\be\label{eq:ell}
  \ell = \frac{2 \sqrt{\pi z_*}}{m^2 u_0}\, \frac{\Gamma \left(\frac{2}{3} \right)}{\Gamma \left(\frac{1}{6} \right)}\, .
\ee
Next, one can try to express the entanglement entroy $S_E$ \eqref{eq:EEtemp} also in terms of $u_0$ by plugging \eqref{eq:uprime} back into Eq.~\eqref{eq:EEtemp}:
\be
  S_E = \frac{\pi^4}{m^6 G_{10}} \int_{-\ell / 2}^{\ell / 2} dx\, \frac{m^2 u^6}{\sqrt{z_*} u_0^3} = \frac{2 \pi^4 u_0^2}{m^6 G_{10}} \int_1^\infty d\xi \, \frac{\xi^4}{\sqrt{\xi^6 - 1}}\, .
\ee
where $\xi \equiv u / u_0$. This integral is divergent. To regularize it, one can introduce a cutoff $u_\infty$ and integrate $\xi$ over $[1,\, u_\infty / u_0]$. The result of the regularized integral is
\be
  S_E = \frac{2 \pi^4 u_0^2}{m^6 G_{10}} \left[\frac{1}{2} \left(\frac{u_\infty}{u_0} \right)^2 \phantom{|}_2 F_1 \left(-\frac{1}{3},\, \frac{1}{2},\, \frac{2}{3},\, \left(\frac{u_0}{u_\infty} \right)^6 \right) - \frac{\sqrt{\pi}\, \Gamma \left(\frac{2}{3} \right)}{2\, \Gamma \left(\frac{1}{6} \right)} \right]\, .
\ee
One can expand the result in powers of $u_0 / u_\infty$, and at the leading order the result is
\be
  S_E = \frac{\pi^4 u_\infty^2}{m^6 G_{10}} - \frac{\pi^4 \sqrt{\pi}}{m^6 G_{10}} \frac{\Gamma \left(\frac{2}{3} \right)}{\Gamma \left(\frac{1}{6} \right)} u_0^2\, .
\ee
Neglecting the divergent first term, we obtain the finite contribution to the entanglement entropy at leading order in terms of $u_0$:
\be\label{eq:EEtemp-2}
  S_E^{\textrm{finite}} = - \frac{\pi^4 \sqrt{\pi}}{m^6 G_{10}} \frac{\Gamma \left(\frac{2}{3} \right)}{\Gamma \left(\frac{1}{6} \right)} u_0^2\, .
\ee
Combining Eq.~\eqref{eq:EEtemp-2} with Eq.~\eqref{eq:ell}, we obtain a result for $S_E^{\textrm{finite}}$ in terms of $\ell$:
\be\label{eq:EEfinal}
  S_E^{\textrm{finite}} = - \frac{8 \pi \sqrt{\pi} z_*}{m^2} \left(\frac{\Gamma \left(\frac{2}{3} \right)}{\Gamma \left(\frac{1}{6} \right)} \right)^2 \frac{N_c^2}{\ell^2}\, ,
\ee
where the relation \eqref{eq:constm} is used. As discussed in Ref.~\cite{GravDual-1}, in the UV regime this result matches the one from (3+1)D gauge theory compactified on a sphere \cite{Klebanov}, and is consistent with the gravity dual construction of D3-branes wrapped on a two-cycle of CY 2-fold.

For the $\mathcal{N}=(2,2)^*$ case with $\widetilde{c}=1/2$ considered in Subsection~\ref{sec:SolFromBrane}, we can apply the same steps to the UV metric \eqref{eq:22UVmetric}. We find that for the $\mathcal{N}=(2,2)^*$ case Eq.~\eqref{eq:EEtemp} now becomes
\be
  S_E = \frac{\pi^4 \sqrt{z_*}}{m^6 G_{10}} \int_{-\ell / 2}^{\ell / 2} dx\, u \left(u'^2 + \frac{m^4 u^4}{\sqrt{z_*}} \right)^{\frac{1}{2}}\, .
\ee
Taking care of the different powers of $z_*$, in the end we find the same result as the $\mathcal{N}=(4,4)$ case given by Eq.~\eqref{eq:EEfinal}, and the deformation of the theory does not affect the finite part of the entanglement entropy. This result is also consistent with the gravity dual construction of D3-branes wrapped on a two-cycle of CY 3-fold for the $\mathcal{N}=(2,2)^*$ case, as expected from the gauge theory side in the UV regime \cite{Klebanov}.

Because later in this paper we will relate the gravity dual solution to the nonlinear Schr\"odinger equation, 
we would like to recall the entanglement entropy for the nonlinear Schr\"odinger equation, which has been computed for the ground state in Ref.~\cite{Korepin}, and the result at zero temperature is
\be
  S_E (\ell) = \frac{c}{3}\, \textrm{log} (\ell)\, ,\quad \ell \to \infty\, .
\ee
This expression can be understood as the IR result, and the difference between this result and the one in the UV regime obtained earliear in this subsection suggests a phase transition, which is consistent with the dicussions in Ref.~\cite{Klebanov}.

\section{Nonlinear Schr\"odinger Equation}\label{sec:NLS}

In this section we briefly review the (1+1)D quantum nonlinear Schr\"odinger equation and its relation with the 2D $\mathcal{N}=(2,2)^*$ $U(N)$ topological Yang-Mills-Higgs theory, following Ref.~\cite{GS-1, GS-2}.

\subsection{Review of the Theory}
After choosing an appropriate system of units, the (1+1)D nonlinear Schr\"odinger equation is
\be\label{eq:NLS}
  i \partial_t \phi = -\frac{1}{2} \partial_x^2 \phi + 2 c (\phi^* \phi) \phi\, .
\ee
The Hamiltonian of the theory is given by
\be
  \mathcal{H} = \int dx \, \left[\frac{1}{2} \frac{\partial \phi^*}{\partial x} \frac{\partial \phi}{\partial x} + c \left(\phi^* \phi \right)^2 \right] \, ,
\ee
where the field $\phi$ has the Poisson structure
\be
  \{\phi^* (x),\, \phi(x') \} = \delta (x-x') \, .
\ee
In (1+1)D, this theory is integrable both at the classical level and at the quantum level.

For the (1+1)D quantum nonlinear Schr\"odinger equation, if we consider the $N$-particle sector in the domain $x_1 \leq x_2 \leq \cdots \leq x_N$, the $N$-particle wave function satisfies the equation
\be
  \left(-\frac{1}{2} \sum_{i=1}^N \frac{\partial^2}{\partial x_i^2} \right) \Phi_\lambda (x) = 2 \pi^2 \left(\sum_{i=1}^N \lambda_i^2 \right) \Phi_\lambda (x)\, ,
\ee
and the normalized wave function is given by
\be
  \Phi_\lambda (x) = \sum_{\omega \in W} (-1)^{l(w)} \prod_{i<j} \left(\frac{\lambda_{\omega(i)} - \lambda_{\omega(j)} + i c\, \textrm{sgn} (x_i - x_j) }{\lambda_{\omega(i)} - \lambda_{\omega(j)} - i c\, \textrm{sgn} (x_i - x_j) } \right)^{\frac{1}{2}}\, \textrm{exp} \left(2 \pi i \sum_i \lambda_{\omega(k)}\, x_k \right) \, ,
\ee
where $\lambda_i$ denotes the momentum of the $i$-th particle, satisfying the Bethe Ansatz equation:
\be\label{eq:BAE-2}
  e^{2 \pi i \lambda_j} \prod_{k \neq j} \frac{\lambda_k - \lambda_j - i c}{\lambda_k - \lambda_j + i c} = 1\, ,\quad j = 1,\, \cdots,\, N\, ,
\ee
which is the same as the equation \eqref{eq:BAE-1} for the configurations contributing to the partition function of the 2D topological Yang-Mills-Higgs theory that we discussed in Subsection~\ref{sec:TYMH}.

From this analysis, we see the equivalence between the wave function of the 2D $\mathcal{N}=(2,2)^*$ $U(N)$ topological Yang-Mills-Higgs theory and the wave function of the (1+1)D quantum nonlinear Schr\"odinger equation in the $N$-particle sector. Hence, the duality between these two theories at quantum level is implied.

More generally, as discussed in Ref.~\cite{NS-1}, one can find dualities between a large class of integrable models and certain deformations of the 2D $\mathcal{N}=(2,2)^*$ super-Yang-Mills theory (by twisted mass, tree-level superpotential, matter fields in various representations, etc.). In particular, the topological Yang-Mills-Higgs theory discussed in Refs.~\cite{HiggsBundle, GS-1, GS-2}, which is dual to the nonlinear Schr\"odinger equation as mentioned above, corresponds to the $\mathcal{N}=(2,2)^*$ super-Yang-Mills theory with the following tree-level twisted superpotential:
\be
  \widetilde{W} (\sigma) = \frac{\lambda}{2} \textrm{Tr} \sigma^2\, ,
\ee
where $\sigma$ denotes the complex scalar in the 2D $\mathcal{N}=(2,2)$ vector multiplet, and in Refs.~\cite{GS-1, GS-2} the parameter $\lambda$ has been chosen to be $\lambda = 1$. This tree-level twisted superpotential corresponds to the insertion of a nonlocal two-observable $O^{(2)}$ in the path integral of the 2D $\mathcal{N}=(2,2)^*$ super-Yang-Mills theory, which also regularizes the path integral \cite{HiggsBundle, GS-1, GS-2}. Without this insertion, the Bethe Ansatz equation \eqref{eq:BAE-1} or \eqref{eq:BAE-2} will not have the phase factor on the left-hand side of the equation \cite{NS-1}.\footnote{The author would like to thank Xinyu Zhang for discussions on this point.}

\subsection{Soliton Solutions to Nonlinear Schr\"odinger Equation}

There are some well-known soliton solutions to the (1+1)D nonlinear Schr\"odinger equation \eqref{eq:NLS}. For the attractive interaction, i.e. $c<0$, the nonlinear Schr\"odinger equation has the so-called bright soliton solution, while for the repulsive interaction, i.e. $c>0$, it has the so-called dark soliton solution. We focus on the bright soliton solution in the following, since it has been known in the literature that for the attractive interaction, the quantum $N$ particles become $N$ solitons when $N$ is large \cite{NLS-1, NLS-2, NLS-thesis}, which makes it convenient to compare with the gravity dual.

A bright soliton solution to the nonlinear Schr\"odinger equation \eqref{eq:NLS} is given by
\be
  \phi = \sqrt{\frac{|c|}{2}} \, \textrm{sech} (|c| (x - x_0)) \, \textrm{exp} \left(\frac{i}{2 c^2} t \right) \, .
\ee
One can also generalize this solution to the $N$ coincident solitons, which is
\be
  \phi = N \sqrt{\frac{|c|}{2}} \, \textrm{sech} (|c| N (x - x_0)) \, \textrm{exp} \left(\frac{i N^2}{2 c^2} t \right) \, .
\ee
Let us choose the unit such that $|c| N = 1/2$ and set $x_0 = 0$, then the $N$ coincident soliton solution becomes
\be
  \phi = \frac{\sqrt{N}}{2} \, \textrm{sech} \left(\frac{x}{2} \right) \, \textrm{exp} \left(\frac{i}{8 c^4} t \right) \, .
\ee
Consequently, the soliton density is
\be
  \rho^{\textrm{sol}} (x) =  |\phi|^2 = \frac{N}{4} \textrm{sech}^2 \left(\frac{x}{2} \right)\, .
\ee
Using the following identity
\be
  \frac{1}{2} \, \textrm{sech} \left(\frac{x}{2} \right) = \frac{e^{-|x/2|}}{1 + e^{-|x|}} = e^{-|x/2|} \sum_{k=0}^\infty (-1)^k\, e^{-k |x|}\, ,
\ee
one can show that
\be\label{eq:Nsolitondensity}
  \rho^{\textrm{sol}} (x) = N \sum_{k=0}^\infty (-1)^k (k+1) \, e^{-(k+1) |x|}\, .
\ee

For the quantum nonlinear Schr\"odinger equation, it has been shown that in the soliton units ($\hbar = m = 2 |c| N = 1$) the density of $N$ particles is \cite{NLS-1, NLS-2, NLS-thesis}:
\be
  \rho (x) = N \sum_{k=0}^{N-2} \left[\prod_{j=0}^k \frac{N - j - 1}{N + j} \right] (-1)^k (k+1)\, e^{-(k+1) |x|}\, .
\ee
Comparing this expression with the $N$ coincident soliton density \eqref{eq:Nsolitondensity}, we see that
\begin{align}
  \frac{\rho^{\textrm{sol}} (x) - \rho (x)}{N} & = \sum_{k=0}^\infty \left[1 - \prod_{j=0}^k \frac{N - j - 1}{N + j} \right] (-1)^k (k+1)\, e^{-(k+1) |x|} \nonumber\\
  {} & \sim \sum_{k=0}^\infty \left[\frac{1}{N} + \mathcal{O} \left(\frac{k}{N} \right) \right] (-1)^k (k+1)\, e^{-(k+1) |x|}\, .
\end{align}
Hence, the $N$-particle density in the quantum nonlinear Schr\"odinger equation approaches the $N$-soliton density when $N\to \infty$, which implies that the $N$-particle solution to the quantum nonlinear Schr\"odinger equation becomes the $N$-soliton solution for the attractive interaction.

\section{Correspondence at Large $N$}\label{sec:correspondence}

In Section~\ref{sec:gravitydual} we have constructed the gravity dual of the 2D $\mathcal{N}=(2,2)^*$ topological Yang-Mills-Higgs theory. Together with the duality between the 2D $\mathcal{N}=(2,2)^*$ topological Yang-Mills-Higgs theory and the (1+1)D nonlinear Schr\"odinger equation, we would like to propose a more general triality mentioned in the introduction (see Fig.~\ref{fig:triangle}) among gauge theories, integrable models and gravity theories.

In the 2D topological Yang-Mills-Higgs theory the coupling constant is set to zero, i.e. $g_{YM}^2 = 0$. Based on our construction, we should require more precisely that $g_{YM}^2 \to 0$ while keeping the size of $S^2$ wrapped by the D3 branes fixed. On the other hand, the gauge/gravity duality is valid in the limit of large 't Hooft coupling $\lambda = g_{YM}^2 N$. Therefore, the proper limit for the triality to hold is
\begin{align}
  g_{YM}^2 \to 0 & \quad\textrm{keeping the size of $S^2$ wrapped by the D3 branes fixed,} \nonumber\\
  \quad N \to \infty\, , & \quad\lambda = g_{YM}^2 N \to \infty\, .\label{eq:limits}
\end{align}
As we analyzed in Subsection~\ref{sec:RunningCoupling}, the 2D $\mathcal{N}=(2,2)^*$ super Yang-Mills theory has asymptotic freedom, hence in the UV regime $g_{YM}^2 \to 0$. Consequently, this triality should hold in the UV regime.

\begin{figure}[!htb]
      \begin{center}
        \includegraphics[width=0.5\textwidth]{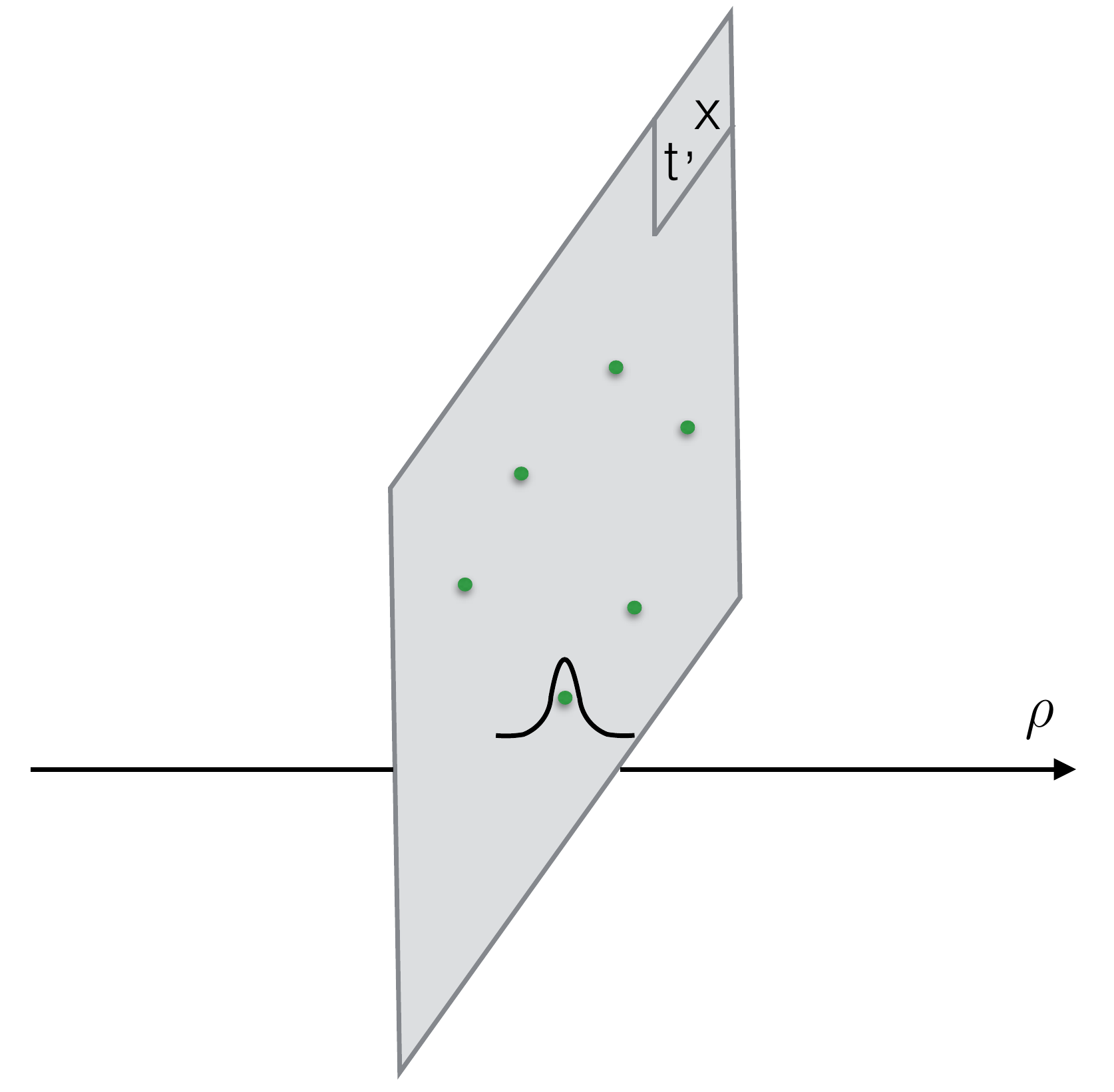}
      \caption{Schematic plot of the solitons and the overlapping D3-branes}
      \label{fig:schematic}
      \end{center}
    \end{figure}

We also expect that in the limit \eqref{eq:limits} the triality provides us with dualities between each two corners in Fig.~\ref{fig:triangle} at quantum level. It requires more detailed work to check this proposal. As a first step, let us consider the classical solutions of these theories in the large $N$ limit. We have seen in the previous section that, when $N$ is large, the $N$-particle solution to the quantum nonlinear Schr\"odinger equation becomes the $N$-soliton solution for the attractive interaction. In the gravity, this solution corresponds to $N$ overlapping D3-branes, and the $N$ solitons live in the world volume of the D3-branes. On the gauge theory side, $N$ becomes the rank of the gauge group, and the insertion of the nonlocal two-observable $\mathcal{O}^{(2)}$ in the gauge theory path integral corresponds to adding some probes on the gravity side. Schematically, the solitons to the nonlinear Schr\"odinger equation and the D3-branes are shown in Fig.~\ref{fig:schematic}, where $N$ solitons are lying in the two extended directions $(t, x)$ of $N$ overlapping D3-branes, which are placed perpendicular to the $\rho$-direction. The profiles of the D3-branes can be read off from the factor $H(\rho,\, \sigma)$ appearing in the 10D metric (see e.g. Eqs.~\eqref{eq:44metric} \eqref{eq:22metric}).

\section{Discussion}\label{sec:discussion}

In this paper, we construct the gravity dual of the 2D $\mathcal{N}=(2,2)^*$ supersymmetric $U(N)$ Yang-Mills theory using the 5D gauged supergravity uplifted to 10D. In the UV regime, it also provides the gravity dual theory to the 2D $\mathcal{N}=(2,2)^*$ topological Yang-Mills-Higgs theory. In some special cases, we have shown that these gravity solutions can also be obtained from $N$ D3-branes wrapped on a two-cycle of some CY-manifolds, in the same spirit of Refs.~\cite{GravDual-1, GravDual-2, GravDual-3, CYAdS-1, CYAdS-2}. In this way, we propose a triality among gauge theories, integrable models and gravity theories. This may provide some new perspectives and hopefully a new way of studying the connections of these different theories.

To explore the triality (see Fig.~\ref{fig:triangle}) proposed in this paper, some further studies are definitely needed. An immediate generalization is to add matters in the fundamental representation of the gauge group, i.e., add flavors in the gravity dual. This will allow us to generalize the duality between 2D topological Yang-Mills-Higgs theory and nonlinear Schr\"odinger equation to the larger class of dualities found by Nekrasov and Shatashvili \cite{NS-1, NS-2}, and consequently to incorporate 4D $\mathcal{N}=2$ gauge theories into the story and study its integrability \cite{NS-3}. Another related question is to study the integrability on the gravity side both at the classical level and at the quantum level, especially to reproduce the Bethe Ansatz equation on the gravity side, which requires the analysis of the chiral ring structure on the gravity side. It would also be interesting to understand the relation between the gravity dual theory that we construct and the Yang-Baxter deformations studied in the literature.

It was suggested in Ref.~\cite{GS-1} that the origin of the duality between the nonlinear Schr\"odinger equation and the topological Yang-Mills-Higgs theory can be understood using the Nahm transformation. We would like to make this argument more precisely at quantitative level, and relate it to the Hitchin systems as dimensional reduction of the 4D self-dual Yang-Mills theory to lower dimensions \cite{Hitchin, Mason}.

Another unexpected relation between integrable models comes from the boson/vortex duality discussed in Refs.~\cite{Zee, Gubser} and recently revisited in Refs.~\cite{BEC, BECstring, KdV}. Using this duality, one can show that in (3+1)D nonlinear Schr\"odinger equation can be mapped into an effective string theory. This duality can also be applied to the (1+1)D nonlinear Schr\"odinger equation, which is an integrable model, and the dual theory in a certain limit was found to be another integrable model \cite{KdV}, the KdV equation. This novel approach unveils a lot of interesting features, and many apparently different theories are related in a larger duality web.

Finally, our construction of the gravity theory dual to the 2D $\mathcal{N}=(2,2)^*$ theory brings some new ingredients to the study of the 2D gauge theories, for which there have been already a huge amount of literature and plenty of results. Therefore, besides integrable models we also anticipate some interesting results relating gravity theories discussed in this paper with knot theory, topological string theory, etc., which hopefully can shed light to some problems (e.g. the OSV conjecture \cite{OSV}) in these fields.

\section*{Acknowledgements}

The author would like to thank Daniel Are\'an, Andr\'e Coimbra, Heng-Yu Chen, Ilmar Gahramanov, Song He, Simeon Hellerman, Albrecht Klemm, Hai Lin, Saebyeok Jeong, Vladimir Korepin, Peter Koroteev, Nuno Rom\~ao, Martin Ro\v cek, Vatche Sahakian, Olof Ohlsson Sax, Stefan Vandoren, Linus Wulff, Takehiko Yasuda, Xinyu Zhang and Peng Zhao for many useful discussions, and also thank Susanne Reffert and Domenico Orlando for communications. In particular, the author is very grateful to Vasily Pestun for carefully proofreading the preliminary version of the manuscript and providing very constructive suggestions, and the author also would like to express special thanks to Sungjay Lee and Masahito Yamazaki for very enlightening discussions in resolving some key issues in the paper.

\appendix

\section{Review of the 5D Gauged Supergravity}\label{app:5DSUGRA}

In this appendix, we briefly review the 5D gauged supergravity that is used in this paper to construct the gravity dual. Originally, the 5D maximal gauged supergravity was studied in Refs.~\cite{5DSUGRA-1, 5DSUGRA-2, 5DSUGRA-3}. Later, it was found that this theory can be consistently truncated to an $\mathcal{N}=2$ subsection, which contains three $U(1)$ gauge fields $A_\mu^I$ and two real scalars $\phi_{1,2}$ \cite{Cvetic, Chamseddine} (see also Ref.~\cite{MN-1, BB}). This consistently truncated model is also sometimes called the STU model.

The bosonic part of the 5D $\mathcal{N}=2$ gauged supergravity is given by:
\be
  \mathcal{L} = R - \frac{1}{2} (\partial_\mu \phi_1)^2 - \frac{1}{2} (\partial_\mu \phi_2)^2 + 4 \sum_{I=1}^3 e^{\alpha_I} - \frac{1}{4} \sum_{I=1}^3 e^{2 \alpha_I}\, F_{\mu\nu}^I F^{I, \mu\nu} + \frac{1}{4} \epsilon^{\mu\nu\alpha\beta\rho} F^1_{\mu\nu} F^2_{\alpha\beta} A^3_\rho\, ,
\ee
where
\be
  \alpha_1 = \frac{\phi_1}{\sqrt{6}} + \frac{\phi_2}{\sqrt{2}}\, ,\quad \alpha_2 = \frac{\phi_1}{\sqrt{6}} - \frac{\phi_2}{\sqrt{2}}\, ,\quad \alpha_3 = - \frac{2}{\sqrt{6}} \phi_1\, .
\ee
The supersymmetry transformations of the fermionic fields are following \cite{5DSUGRA-4, MN-1}:
\begin{align}
  \delta \psi_\mu & = \left[\partial_\mu + \frac{1}{4} \omega_\mu^{ab} \gamma_{ab} + \frac{i}{8} X_I \left(\gamma_\mu\,^{\nu\rho} - 4 \delta_\mu^\nu \gamma^\rho \right) F_{\nu\rho}^I + \frac{1}{2} X^I V_I \gamma_\mu - \frac{3i}{2} V_I A_\mu^I \right] \epsilon\, ,\label{eq:FermionTrafo-1}\\
  \delta \chi_{(j)} & = \left[\frac{3}{8} (\partial_{\phi_j} X_I) F^I_{\mu\nu} \gamma^{\mu\nu} + \frac{3i}{2} V_I \partial_{\phi_j} X^I - \frac{i}{4} \delta_{jk} \partial_\mu \phi_k \gamma^\mu \right] \epsilon\, ,\quad (j = 1,\, 2)\, ,\label{eq:FermionTrafo-2}
\end{align}
where
\be
  X^I = e^{-\alpha_I}\, ,\quad V_I = \frac{1}{3}\, ,\quad X_I = \frac{1}{3} (X^I)^{-1}
\ee
for $I = 1,\, 2,\, 3$. Hence, $X^I$ satisfy
\be
  X^1 X^2 X^3 = 1\, .
\ee
One should impose some constraints on the Killing spinor $\epsilon$ to obtain an $\mathcal{N}=2$ truncation of the maximal supersymmetry. A possible choice of the constraints is following:
\be
  \gamma_{\hat{r}} \epsilon = \epsilon\, ,\quad \gamma_{\hat{x} \hat{y}} \epsilon = i \epsilon\, ,\quad \partial_t \epsilon = \partial_z \epsilon = \partial_x \epsilon = \partial_y \epsilon = 0\, ,
\ee
where the hat denotes the flat indices.

Using the consistently truncated 5D gauged supergravity discussed above, Maldacena and N\'u\~nez have studied the supergravity solution dual to the 4D superconformal field theory on $\mathbb{R}^2 \times \Sigma$ \cite{MN-1}, which can be uplifted to the 10D type IIB supergravity. The basic idea is to consider D3-branes wrapped on $\mathbb{R}^2 \times \Sigma$ with a specific normal bundle, and the gauge connection on the normal bundle will twist the theory and cancel the spin connection of $\Sigma$, such that some supersymmetries can still be preserved on the curve background. Depending on different ways of twisting, there can be $\mathcal{N}=(4,4)$, $(2,2)$, $(0,2)$ supersymmetries perserved in the construction. Starting from the 4D $\mathcal{N}=4$ super Yang-Mills theory, whose R-symmetry group is $SO(6)$, one can characterize the twist by picking up a special background
\be
  T = a_1 T_1 + a_2 T_2 + a_3 T_3
\ee
with $T_I$ ($I = 1,\, 2,\, 3$) denoting the generators of the Cartan subgroup $SO(2) \times SO(2) \times SO(2)$ of the R-symmetry group $SO(6)$. To preserve at least 2D $\mathcal{N}=(0,2)$ supersymmetry, the parameters $a_I$'s should satisfy
\be
  a_1 + a_2 + a_3 = - \kappa\, ,
\ee
where
\be
  \kappa = \Bigg\{ \begin{array}{cc}
    1\, , & \textrm{ for } g=0\, ; \\
    0\, , & \textrm{ for } g=1\, ; \\
    -1\, , & \textrm{ for } g>1\, .
  \end{array}
\ee
Hence, different choices of $a_I$'s lead to different twists of the theory, in order to cancel the spin connections from the curved background. In general, turning on more $a_I$'s correspond to picking up a subset from the original 16 supercharges, which will reduce the number of supersymmetries. When one of $a_I$'s equals zero, the gravity preserves $\mathcal{N}=(2,2)$ supersymmetry. When two of $a_I$'s equal zero, the gravity preserves $\mathcal{N}=(4,4)$ supersymmetry. When all of $a_I$'s equal zero, the gravity preserves $\mathcal{N}=(8,8)$ supersymmetry. In particular, the $\mathcal{N}=(4,4)$ case is dual to a 2D supersymmetric nonlinear sigma model on the Hitchin moduli space of the Riemann surface $\Sigma$ studied in Ref.~\cite{BershadskyVafa}.

According to the uniformization theorem, we can express the metrics for the three types of Riemann surfaces ($g=0$, $g=1$, $g>1$) in the following form:
\be
  ds_\Sigma^2 = e^{2 h(x,y)} (dx^2 + dy^2)\, ,
\ee
where
\be
  h(x,y) = \Bigg\{ \begin{array}{ll}
    - \textrm{log} \frac{1+x^2+y^2}{2} \, , & \textrm{ for } g=0\, ; \\
    \frac{1}{2} \textrm{log}\, 2\pi \, , & \textrm{ for } g=1\, ; \\
    - \textrm{log}\, y \, , & \textrm{ for } g>1\, .
  \end{array}
\ee
Considering the D3-branes wrapped on $\mathbb{R}^2 \times \Sigma$, we can take the following Ans\"atze for the 5D metric and the field strengths from the normal bundle as twists:
\begin{align}
  ds_5^2 & = e^{2 f(r)} \left(-dt^2 + dz^2 + dr^2 \right) + e^{2 g(r)}\, ds_\Sigma^2\, ,\\
  F^I & = -a_I\, e^{2 h(x,y)} dx \wedge dy\, ,\label{eq:FieldStrength}
\end{align}
where $I = 1,\, 2,\, 3$, and $ds_\Sigma^2$ is the metric of the Riemann surface discussed above. Moreover, we assume that the two scalars in the model are functions of the coordinate $r$, i.e. $\phi_{1,2} (r)$. Pay attention to that in Subsection~\eqref{22dual} we use slightly different expressions of the metric $ds_5^2$ and the field strengths $F^I$'s by explicitly introducing a length scale $m^{-1}$, which can be fixed by the quantization condition of the RR 5-form flux in the 10D type IIB supergravity, as discussed in Appendix~\ref{app:5form}.

Using the Ans\"atze above and setting the supersymmetry transformations of the fermionic fields \eqref{eq:FermionTrafo-1} \eqref{eq:FermionTrafo-2} to zero, we obtain the following BPS equations:
\begin{align}
  f' & = - e^f (X^1 + X^2 + X^3) / 3 - e^{f-2g} a_I X_I / 2\, ,\\
  g' & = - e^f (X^1 + X^2 + X^3) / 3 + e^{f-2g} a_I X_I\, ,\\
  \phi_1' & = -\sqrt{6} e^f (X^1 + X^2 - 2 X^3) / 3 - \sqrt{6} e^{f-2g} (a_1 X_1 + a_2 X_2 - 2 a_3 X_3) / 2\, ,\\
  \phi_2' & = -\sqrt{2} e^f (X^1 - X^2) - 3 \sqrt{2} e^{f-2g} (a_1 X_1 - a_2 X_2) / 2\, .
\end{align}
In general, given boundary conditions these equations can be solved numerically for fixed $a_I$'s. For some special choices of $a_I$'s these equations also take simpler forms, for instance Eqs.~\eqref{eq:BPSsp-1} $\sim$ \eqref{eq:BPSsp-3} for $a_I = (0,\, 0,\, 1)$. Moreover, in Subsection~\ref{22dual} we use slightly different expressions of the BPS equations compared to the ones above by introducing a length scale $m^{-1}$, which will be fixed in Appendix~\ref{app:5form}.

After obtaining the factors $f(r)$, $g(r)$ and the profiles of the fields $\phi_{1,2} (r)$ by solving the BPS equations, we can use the formulae in Ref.~\cite{Cvetic} to uplift the solution in 5D $\mathcal{N}=2$ gauged supergravity to a solution in 10D type IIB supergravity. The uplifted 10D metric is given by
\be
  ds_{10}^2 = \Delta^{1/2} ds_5^2 + \Delta^{-1/2} \sum_{I=1}^3 \frac{1}{X^I} \left(d\mu_I^2 + \mu_I^2 (d\phi^I + A^I)^2 \right)\, ,
\ee
where $A^I$ are the three $U(1)$ gauge fields corresponding to the field strengths $F^I$ discussed above, and
\be
  \Delta = \sum_{I=1}^3 X^I \mu_I^2\, ,
\ee
with $\mu_I$ ($I=1,\, 2,\, 3$) satisfying
\be
  \sum_{I=1}^3 \mu_I^2 = 1\, .
\ee
One can parametrize $\mu_I$'s as follows:
\be
  \mu_1 = \textrm{cos} \widetilde{\theta}\, \textrm{sin} \widetilde{\psi}\, ,\quad \mu_2 = \textrm{cos} \widetilde{\theta}\, \textrm{cos} \widetilde{\psi}\, ,\quad \mu_3 = \textrm{sin} \widetilde{\theta}\, ,
\ee
where $0 \leq \widetilde{\theta} \leq \pi$ and $0 \leq \widetilde{\psi} < 2 \pi$. The self-dual 5-form flux in the uplifted 10D solution is given by
\be
  F_5 = \mathcal{F}_5 + * \mathcal{F}_5\, ,
\ee
where
\be
  \mathcal{F}_5 = \sum_{I=1}^3 \left[2 X^I (X^I \mu_I^2 - \Delta) \epsilon_5 + \frac{1}{2 (X^I)^2} d(\mu_I^2) \left((d \phi^I + A^I) \wedge *_5 F^I + X^I *_5 dX^I \right) \right]\, ,
\ee
and $\epsilon_5$ and $*_5$ are the volume form and the Hodge dual of the 5D space respectively, while $F^I = dA^I$ are the field strengths of the gauge fields given by Eq.~\eqref{eq:22gaugefield}. $\phi^I$ ($I = 1, 2, 3$) are three angles with the range $[0, 2\pi)$, which should be distinguished from the scalar fields $\phi_{1, 2}$ discussed above. In the main text, we also slightly modify the uplifted 10D solution by explicitly introducing a length scale $m^{-1}$.

There is an important constraint that the parameters $a_I$'s should satisfy. Due to the compactness of the Riemann surface $\Sigma$, the field strengths $F^I$ should obey the quantization condition
\be
  \frac{1}{2\pi} \int_\Sigma F^I \in \mathbb{Z}\, .
\ee
Taking into account the expression of the field strength \eqref{eq:FieldStrength}, we obtain the following constraint on $a_I$'s for the Riemann surface $\Sigma$ of genus $g$:
\begin{align}
  \textrm{For } g \neq 1: & \quad 2 a_I |g-1| \in \mathbb{Z}\, ,\label{eq:Constraint}\\
  \textrm{For } g = 1: & \quad a_I \in \mathbb{Z}\, .
\end{align}
For the genus $g>1$, the constraint \eqref{eq:Constraint} essentially means that $a_I$ should be rational numbers, because it can be satisfied by appropriately choosing the genus $g$. Although $a_I$'s are not real numbers as we expected from the deformation of the 2D super Yang-Mills theory, they can approach any real number by increasing the genus $g$ (see e.g. Refs.~\cite{BB, Eoin-1, Eoin-2, Eoin-3}).

As discussed in Refs.~\cite{MN-1}, the 5D supergravity solutions constructed in this way flow from $AdS_5$ in the UV to $AdS_3$ in the IR. However, in order that the $AdS_3$ solutions are well-defined, they should satisfy
\be
  X^1 > 0\, ,\quad X^2 > 0\, ,\quad e^{2g} > 0\, ,\quad r e^f > 0\, ,
\ee
and consequently only certain ranges of $a_I$'s can provide good $AdS_3$ vacua satisfying the conditions above (see Refs.~\cite{BB, Eoin-1, Eoin-2, Eoin-3}). In this paper we are interested in the gravity duals of the 2D non-conformal super Yang-Mills theory, hence we do not need to consider the $AdS_3$ vacua, which correspond to conformal field theories, and we refer to Refs.~\cite{BB, Eoin-1, Eoin-2, Eoin-3} for the discussions on the relation between $a_I$'s and good $AdS_3$ vacua.

\section{Identify the 10D Metrics}\label{app:10Dmetric}

In this appendix we show that by changing variables the metric obtained from the 5D $\mathcal{N}=2$ gauged supergravity uplifted to 10D can be brought into the form of $N$ D3-branes wrapped on a two-cycle of a CY 2-fold or 3-fold, depending on the number of preserved supercharges.

For generic values of $\widetilde{c}$, the 10D metric is given by Eq.~\eqref{eq:10Dmetric}. With an explicit choice of $\mu_I$'s given by Eq.~\eqref{eq:choosemu-1}, the metric \eqref{eq:10Dmetric} becomes Eq.~\eqref{eq:10DmetricGen}. In the following we consider two special cases $\widetilde{c}=0$ and $\widetilde{c}=1/2$ with $S^2$ as the Riemann surface for compactification, and we demonstrate how the metric \eqref{eq:10DmetricGen} can be identified with the ones from the brane construction.

For $\widetilde{c}=0$ and $S^2$ as the Riemann surface for compactification, the metric \eqref{eq:10DmetricGen} corresponds to the gravity dual of the 2D $\mathcal{N}=(4,4)$ super Yang-Mills theory. As shown in Appendix~B of Ref.~\cite{GravDual-1}, for this case the metric \eqref{eq:10DmetricGen} can be simplified to
\begin{align}
  ds_{10}^2 & = \sqrt{\Delta} \left[e^{2f} (dx_{1,1}^2 + dr^2) + \frac{e^{2g}}{m^2} \left(d\theta^2 + \textrm{sin}^2 \theta\, (d\phi)^2 \right) \right] \nonumber\\
  {} & \quad + \frac{1}{m^2 \sqrt{\Delta}} \left[e^{-\varphi} \Delta\, d\widetilde{\theta}^2 + e^\varphi \, \textrm{cos}^2 \widetilde{\theta} \, d\Omega_3^2 + \textrm{sin}^2 \widetilde{\theta}\, e^{-2 \varphi} \left(d\phi^3 + \textrm{cos}\theta d\phi \right)^2 \right]\, ,\label{eq:app10D44metric}
\end{align}
where one scalar field $\varphi_1 \equiv \varphi$, and the other one $\varphi_2$ is set to zero, while
\be
  \Delta = \sum_{I=1}^3 X^I \mu_I^2 = e^{-\varphi}\, \textrm{cos}^2 \widetilde{\theta} + e^{2 \varphi} \, \textrm{sin}^2 \widetilde{\theta}\, .
\ee
To identify this metric with the one from the brane construction \eqref{eq:44metric}:
\begin{align}
  ds^2 & = H^{-\frac{1}{2}} \left[dx_{1,1}^2 + \frac{z}{m^2} \left(d\theta^2 + \textrm{sin}^2 \theta\, (d\phi)^2 \right) \right] \nonumber\\
  {} & \quad + H^{\frac{1}{2}} \left[\frac{1}{z} d\sigma^2 + \frac{\sigma^2}{z} \left(d\psi + \textrm{cos} \theta\, d\phi \right)^2 + d\rho^2 + \rho^2 d\Omega_3^2 \right]\, ,\nonumber
\end{align}
we can first compare the coefficients in front of $dx_{1,1}^2$ and $d\theta^2 + \textrm{sin}^2 \theta\, (d\phi)^2$, which lead to
\be\label{eq:HSol}
  e^{2 f} \sqrt{\Delta} = H^{-\frac{1}{2}}\, ,\quad \frac{e^{2g}}{m^2} \sqrt{\Delta} = \frac{H^{-\frac{1}{2}} z}{m^2}\, .
\ee
Combining these two relations, we obtain
\be\label{eq:zSol}
  z = e^{2 (g - f)}\, .
\ee
Also, we observe that $\phi^3$ can be identified with $\psi$. By comparing the coefficients of $\left(d\phi^3 + \textrm{cos}\theta d\phi \right)^2$ and $\left(d\psi + \textrm{cos}\theta d\phi \right)^2$ as well as the coefficients of $d\Omega_3^2$, we obtain
\be\label{eq:44rhosigmaSol}
  \frac{\textrm{sin}^2 \widetilde{\theta}\, e^{-2 \varphi}}{m^2 \sqrt{\Delta}} = \frac{H^{\frac{1}{2}} \sigma^2}{z}\, ,\quad \frac{\textrm{cos}^2 \widetilde{\theta}\, e^\varphi}{m^2 \sqrt{\Delta}} = H^{\frac{1}{2}} \rho^2\, .
\ee
Together with Eq.~\eqref{eq:HSol} and Eq.~\eqref{eq:zSol}, the equations above lead to
\be\label{eq:44rhosigmaSol-2}
  \rho = \frac{\textrm{cos} \widetilde{\theta}\, e^{f + \frac{\varphi}{2}}}{m}\, ,\quad \sigma = \frac{\textrm{sin} \widetilde{\theta}\, e^{g - \varphi}}{m}\, .
\ee
The differentials $d\rho$ and $d\sigma$ are then
\begin{align}
  d\rho & = \frac{e^{f + \frac{\varphi}{2}}}{m} \left[\left(f' + \frac{\varphi'}{2} \right) \textrm{cos} \widetilde{\theta}\, dr - \textrm{sin} \widetilde{\theta} \, d\widetilde{\theta} \right]\, ,\nonumber\\
  d\sigma & = \frac{e^{g - \varphi}}{m} \left[\left(g' - \varphi' \right) \textrm{sin} \widetilde{\theta}\, dr + \textrm{cos} \widetilde{\theta}\, d\widetilde{\theta} \right]\, .
\end{align}
Using the BPS equations \eqref{eq:BPSsp-1} $\sim$ \eqref{eq:BPSsp-3}, one can express the terms with derivatives in the equations above as
\be
  f' + \frac{\varphi'}{2} = m e^{f - \varphi}\, ,\quad g' - \varphi' = m e^{f + 2 \varphi}\, .
\ee
Hence,
\begin{align}
  d\rho & = e^{2 f - \frac{\varphi}{2}} \, \textrm{cos} \widetilde{\theta}\, dr - \frac{e^{f + \frac{\varphi}{2}}}{m} \, \textrm{sin} \widetilde{\theta}\, d\widetilde{\theta}\, ,\nonumber\\
  d\sigma & = e^{f + g + \varphi}\, \textrm{sin} \widetilde{\theta}\, dr + \frac{e^{g - \varphi}}{m}\, \textrm{cos} \widetilde{\theta}\, d\widetilde{\theta}\, .
\end{align}
Finally, one can prove that
\be
  H^{\frac{1}{2}}\, d\rho^2 + \frac{H^{\frac{1}{2}}}{z} d\sigma^2 = \frac{\sqrt{\Delta}}{m^2}\, e^{-\varphi} \, d\widetilde{\theta}^2 + \sqrt{\Delta} \, e^{2 f}\, dr^2\, .
\ee
Therefore, all the terms in the metric \eqref{eq:app10D44metric} and \eqref{eq:44metric} are identified, i.e., they are indeed the same metric by some changes of variables.

For $\widetilde{c} = 1/2$ and $S^2$ as the Riemann surface for compactification, we can perform the similar analysis. To simplify the final expression, we first make a permutation of the $\mu_I$'s chosen in Eq.~\eqref{eq:choosemu-1}, and we call the new ones $\widetilde{\mu}_I$'s:
\be\label{eq:choosemu-2}
  \widetilde{\mu}_1 = \mu_2 = \textrm{cos} \widetilde{\theta}\, \textrm{cos} \widetilde{\psi}\, ,\quad \widetilde{\mu}_2 = \mu_3 = \textrm{sin} \widetilde{\theta}\, ,\quad \widetilde{\mu}_3 = \mu_1 = \textrm{cos} \widetilde{\theta}\, \textrm{sin} \widetilde{\psi}\, ,
\ee
The metric \eqref{eq:10Dmetric} for $\widetilde{c} = 1/2$ now becomes
\begin{align}
  ds_{10}^2 & = \sqrt{\Delta} \left[e^{2f} (dx_{1,1}^2 + dr^2) + \frac{e^{2g}}{m^2} \left(d\theta^2 + \textrm{sin}^2 \theta\, (d\phi)^2 \right) \right] \nonumber\\
  {} & \quad + \frac{1}{m^2 \sqrt{\Delta}} \Bigg[e^{\varphi_1 + \varphi_2} d\mu_2^2 + e^{\varphi_1 - \varphi_2} d\mu_3^2 + e^{- 2 \varphi_1} d\mu_1^2 + e^{\varphi_1 + \varphi_2} \textrm{cos}^2 \widetilde{\theta}\, \textrm{cos}^2 \widetilde{\psi} \left(d\phi^1 + \frac{1}{2}\, \textrm{cos} \theta \, d\phi \right)^2 \nonumber\\
  {} & \qquad\qquad\qquad + e^{\varphi_1 - \varphi_2} \textrm{sin}^2 \widetilde{\theta} (d\phi^2)^2 + e^{- 2 \varphi_1} \textrm{cos}^2 \widetilde{\theta}\, \textrm{sin}^2 \widetilde{\psi} \left(d\phi^3 + \frac{1}{2} \, \textrm{cos} \theta\, d\phi \right)^2 \Bigg]\, .\label{eq:app10D22metric}
\end{align}
Furthermore, we define two new scalar fields $\hat{\varphi}_1$ and $\hat{\varphi}_2$, which are related to the scalar fields $\varphi_1$ and $\varphi_2$ in the following way:
\be
  \hat{\varphi}_1 - \hat{\varphi}_2 = \varphi_1 + \varphi_2\, ,\quad - 2 \hat{\varphi}_1 = \varphi_1 - \varphi_2\, ,\quad
\hat{\varphi}_1 + \hat{\varphi}_2 = - 2 \varphi_1\, ,
\ee
i.e.,
\be
  \hat{\varphi}_1 = - \frac{\varphi_1 - \varphi_2}{2}\, ,\quad \hat{\varphi}_2 = - \frac{3 \varphi_1 + \varphi_2}{2}\, .
\ee
One can check that the $\hat{\varphi}_2 = 0$ is a solution to the BPS equations \eqref{eq:BPSgen-1} $\sim$ \eqref{eq:BPSgen-4}, hence $\hat{\varphi}_2$ can be consistently turned off, which we will assume in the following. When we set $\hat{\varphi}_2 = 0$, the BPS equations \eqref{eq:BPSgen-1} $\sim$ \eqref{eq:BPSgen-4} reduce to the ones \eqref{eq:BPSsp-1} $\sim$ \eqref{eq:BPSsp-3} with $a_I = (1/2,\, 1/2,\, 0)$ and $\varphi$ replaced by $\hat{\varphi} \equiv \hat{\varphi}_1$. Consequently, the metric \eqref{eq:app10D22metric} can be simplified to be
\begin{align}
  ds_{10}^2 & = \sqrt{\Delta} \left[e^{2f} (dx_{1,1}^2 + dr^2) + \frac{e^{2g}}{m^2} \left(d\theta^2 + \textrm{sin}^2 \theta\, (d\phi)^2 \right) \right] \nonumber\\
  {} & \quad + \frac{1}{m^2 \sqrt{\Delta}} \Bigg[e^{-\hat{\varphi}} \Delta\, d\widetilde{\theta}^2 + e^{\hat{\varphi}} \textrm{cos}^2 \widetilde{\theta} d\widetilde{\psi}^2  + e^{\hat{\varphi}} \textrm{cos}^2 \widetilde{\theta}\, \textrm{cos}^2 \widetilde{\psi} \left(d\phi^1 + \frac{1}{2}\, \textrm{cos} \theta \, d\phi \right)^2 \nonumber\\
  {} & \qquad\qquad\qquad + e^{- 2 \hat{\varphi}} \textrm{sin}^2 \widetilde{\theta} (d\phi^2)^2 + e^{\hat{\varphi}} \textrm{cos}^2 \widetilde{\theta}\, \textrm{sin}^2 \widetilde{\psi} \left(d\phi^3 + \frac{1}{2} \, \textrm{cos} \theta\, d\phi \right)^2 \Bigg]\, ,\label{eq:app10D22metric-2}
\end{align}
where
\be
  \Delta = e^{-\hat{\varphi}}\, \textrm{cos}^2 \widetilde{\theta} + e^{2 \hat{\varphi}} \, \textrm{sin}^2 \widetilde{\theta}\, .
\ee

Using the new scalar field $\hat{\varphi}$, we can propose the final expression of the metric for $\widetilde{c}=1/2$ similar to the one discussed in Ref.~\cite{GravDual-3}:
\begin{align}
  ds^2 & = H^{-\frac{1}{2}} \left[dx_{1,1}^2 + \frac{z}{m^2} \left(d\theta^2 + \textrm{sin}^2 \theta\, (d\phi)^2 \right) \right] \nonumber\\
  {} & \quad + H^{\frac{1}{2}} \Bigg[\frac{1}{\sqrt{z}} d\sigma^2 + \frac{\sigma^2}{\sqrt{z}} \left(d\widetilde{\psi}^2 + \textrm{sin}^2 \widetilde{\psi} \left(d\phi^3 + \frac{1}{2} \textrm{cos} \theta\, d\phi \right)^2 + \textrm{cos}^2 \widetilde{\psi} \left(d\phi^1 + \frac{1}{2} \textrm{cos} \theta\, d\phi \right)^2 \right) \nonumber\\
  {} & \qquad\qquad + d\rho^2 + \rho^2 d\psi^2 \Bigg]\, .\label{eq:app22metric}
\end{align}
By comparing this metric with Eq.~\eqref{eq:app10D22metric-2}, we see that the relations \eqref{eq:HSol} and \eqref{eq:zSol} remain the same for this case. We can identify $\phi^2$ with $\psi$, then the relation \eqref{eq:44rhosigmaSol} becomes different for this case:
\be\label{eq:22rhosigmaSol}
  \frac{\textrm{sin}^2 \widetilde{\theta}\, e^{-2 \hat{\varphi}}}{m^2 \sqrt{\Delta}} = H^{\frac{1}{2}} \rho^2\, ,\quad \frac{\textrm{cos}^2 \widetilde{\theta}\, e^{\hat{\varphi}}}{m^2 \sqrt{\Delta}} = \frac{H^{\frac{1}{2}} \sigma^2}{\sqrt{z}}\, .
\ee
Together with the relations above, one can solve for $\rho$ and $\sigma$ in this case:
\be\label{eq:22rhosigmaSol-2}
  \rho = \frac{e^{f - \hat{\varphi}} \, \textrm{sin} \widetilde{\theta}}{m}\, ,\quad \sigma = \frac{e^{\frac{1}{2} (f + g + \hat{\varphi})} \, \textrm{cos} \widetilde{\theta}}{m}\, .
\ee
Using the BPS equations, we obtain
\begin{align}
  d\rho & = \frac{e^{f - \hat{\varphi}}}{m} \left(- e^{f + 2 \hat{\varphi}} m \, \textrm{sin} \widetilde{\theta}\, dr + \textrm{cos} \widetilde{\theta}\, d\widetilde{\theta} \right)\, ,\\
  d\sigma & = \frac{e^{\frac{1}{2} (f + g + \hat{\varphi})}}{m} \left(- e^{f - \hat{\varphi}} m \, \textrm{cos} \widetilde{\theta}\, dr - \textrm{sin} \widetilde{\theta}\, d\widetilde{\theta} \right)\, .
\end{align}
Hence, one can prove
\be
  H^{\frac{1}{2}} \, d\rho^2 + \frac{H^{\frac{1}{2}}}{\sqrt{z}} \, d\sigma^2 = \sqrt{\Delta}\, e^{2f}\, dr^2 + \frac{\sqrt{\Delta}}{m^2} e^{-\hat{\varphi}}\, d\widetilde{\theta}^2\, .
\ee
Therefore, the metrics \eqref{eq:app10D22metric-2} and \eqref{eq:app22metric} can indeed be identified.

\section{UV Metrics}\label{app:UVmetric}

In this appendix, let us discuss how to obtain the approximate metric in the UV regime. We have found the metrics from the brane construction in Section~\ref{sec:gravitydual}. The $\mathcal{N}=(4,4)$ and the $\mathcal{N}=(2,2)^*$ case both contain the factors $z(\rho,\, \sigma)$ and $H(\rho,\, \sigma)$. Hence, to find the UV metric is equivalent to determine these factors in the UV regime.

Let us start with the $\mathcal{N}=(4,4)$ case, which is discussed in Ref.~\cite{GravDual-1}. First, from the BPS equations \eqref{eq:BPSsp-1} $\sim$ \eqref{eq:BPSsp-3} with $a_I = (0,\, 0,\, 1)$, one can derive that
\be
  2 g' + \varphi' = m\, e^{f - \varphi}\, \left(e^{-2 g - \varphi} + 2 \right)\, ,\quad 2 f' + \varphi' = 2 m \, e^{f - \varphi}\, .
\ee
If we define
\be\label{eq:def44tau}
  \frac{d}{dr} \equiv m\, e^{f - \varphi} \frac{d}{d\tau}\, ,
\ee
and
\be
  \Lambda_1 \equiv 2 g + \varphi\, ,\quad \Lambda_2 \equiv 2 f + \varphi\, ,
\ee
the equations above become
\be
  \frac{d\Lambda_1}{d\tau} - e^{-\Lambda_1} = 2\, ,\quad \frac{d\Lambda_2}{d\tau} = 2\, ,
\ee
which can easily be integrated. Using the solutions, one can rewrite the BPS equations \eqref{eq:BPSsp-1} $\sim$ \eqref{eq:BPSsp-3} with $a_I = (0,\, 0,\, 1)$:
\begin{align}
  e^{2 g + \varphi} & = \alpha\, e^{2 \tau} - \frac{1}{2}\, ,\label{eq:appBPSsp-1}\\
  e^{2 f + \varphi} & = \beta\, e^{2 \tau}\, ,\\
  e^{-3 \varphi} & = \frac{\alpha\, e^{2 \tau} - \tau - \gamma}{\alpha\, e^{2 \tau} - \frac{1}{2}}\, ,\label{eq:appBPSsp-3}
\end{align}
where $\alpha$, $\beta$ and $\gamma$ are integration constants. From the first two equations above, one can obtain an expression for $z$ given by Eq.~\eqref{eq:zSol}:
\be
  z = e^{2 (g - f)} = \frac{e^{2 g + \varphi}}{e^{2 f + \varphi}} = \frac{\alpha\, e^{2 \tau} - \frac{1}{2}}{\beta\, e^{2 \tau}}\, .
\ee
Consequently, one can solve for $e^{2 \tau}$:
\be
  e^{2 \tau} = \frac{1}{2 \beta (z_* - z)}\, ,
\ee
where $z_* \equiv \alpha / \beta$. Using this expression, one can further bring the BPS equations \eqref{eq:appBPSsp-1} $\sim$ \eqref{eq:appBPSsp-3} into the following expressions:
\begin{align}
  e^{2 g + \varphi} & = \frac{z}{2 (z_* - z)}\, ,\\
  e^{2 f + \varphi} & = \frac{1}{2 (z_* - z)}\, ,\\
  e^{-3 \varphi} & = \frac{\Gamma(z)}{z}\, ,
\end{align}
where $\Gamma(z) \equiv z_* + (z_* - z) \left[\textrm{log} (z_* - z) + \kappa \right]$ with $\kappa \equiv \textrm{log} (2 \beta) - 2 \gamma$. Moreover, combining Eq.~\eqref{eq:44rhosigmaSol-2} with the new BPS equations, one can derive
\be
  \left[\rho^2 + \frac{\sigma^2}{\Gamma(z)} \right] (z_* - z) = \frac{1}{2 m^2}\, .
\ee
which implicitly determines the factor $z$. Also, using the new BPS equations, one can solve for the factor $H$ given by Eq.~\eqref{eq:HSol}:
\be
  H = \frac{e^{-4 f}}{\Delta} = \frac{2 z (z_* - z)}{m^2\, \Gamma(z) \left[\rho^2 + \frac{z}{\Gamma^2 (z)} \sigma^2 \right]}\, .
\ee

In the UV regime, $z$ approaches $z_*$, and $\Gamma(z_*) \approx z_*$. Correspondingly, the values of $\rho$ and $\sigma$ are large. At the leading order, one has
\be
  z (\rho,\, \sigma) \approx z_* - \frac{z_*}{2 m^2 (\sigma^2 + z_* \rho^2)}\, ,\quad H(\rho,\, \sigma) \approx \frac{z_*^2}{m^4 \left(\sigma^2 + z_* \rho^2 \right)^2}\, .
\ee
Moreover, one can define new variables
\be
  u = \sqrt{\sigma^2 + z_* \rho^2}\, ,\qquad \textrm{tan} \hat{\alpha} = \frac{\sigma}{\sqrt{z_*}\, \rho}\quad (0 \leq \hat{\alpha} \leq \frac{\pi}{2})\, .
\ee
At large $u$,
\be
  z \to z_*\, ,\quad H \to \frac{z_*^2}{m^4 u^4}\, .
\ee
Plugging these expressions into the metric \eqref{eq:44metric}, one obtains the approximate metric for the $\mathcal{N}=(4,4)$ case in the UV regime:
\begin{align}
  ds_{UV}^2 & \approx \frac{m^2}{z_*} \left[dx_{1,1}^2 + \frac{z_*}{m^2} \left(d\theta^2 + \textrm{sin}^2 \theta\, d\phi^2 \right) \right] + \frac{1}{m^2} \frac{du^2}{u^2} \nonumber\\
  {} & \quad + \frac{1}{m^2} \left[d \hat{\alpha}^2 + \textrm{sin}^2 \hat{\alpha} \left(d\psi + \textrm{cos} \theta\, d\phi \right)^2 + \textrm{cos}^2 \hat{\alpha} \, d\Omega_3^2 \right]\, .
\end{align}

For the $\mathcal{N}=(2,2)^*$ case with $\widetilde{c}=1/2$ there is a little obstacle, because the BPS equations in this case cannot be completely integrated analytically like in the $\mathcal{N}=(4,4)$ case. Nevertheless, we can start with the relation \eqref{eq:22rhosigmaSol-2}, from which we can derive the following relation:
\be\label{eq:app22rel}
  \frac{\rho^2}{e^{2 \tau}} - \frac{\sigma^2}{\sqrt{z} \, \frac{d\tau}{dz}} = \frac{1}{m^2}\, ,
\ee
where for this case
\be\label{eq:def22tau}
  \tau \equiv f - \hat{\varphi}\, ,
\ee
which satisfies
\begin{align}
  \frac{d\tau}{dz} & = - e^{2 f + \hat{\varphi}}\, ,\\
  \frac{d^2 \tau}{dz^2} & = - \left(2 e^{2 \tau} + \frac{1}{2 z} \right) \frac{d \tau}{d z}\, .\label{eq:app22tau2prime}
\end{align}
Also, we can derive that
\be\label{eq:app22HSol}
  H = \frac{\sqrt{z}}{m^2 \left[\rho^2 \sqrt{z}\, e^{-2 \tau} \left(\frac{d \tau}{d z} \right)^2 + \sigma^2\, e^{2 \tau} \right]}\, .
\ee

In the UV regime, $\tau$ is large, hence Eq.~\eqref{eq:app22tau2prime} can be approximated as
\be
  \frac{d^2 \tau}{d z^2} = - 2\, e^{2 \tau}\, \frac{d \tau}{d z} = - \frac{d}{dz} \left(e^{2 \tau} \right)\, ,
\ee
which leads to
\be
  e^{2 \tau} = \frac{C}{1 - e^{2 C (z - z_*)}}\, ,
\ee
where $C$ and $z_*$ are two integration constants. Plugging it into Eq.~\eqref{eq:app22rel}, we can solve for the factor $z$ in the UV regime:
\be
  z = z_* - \frac{\sqrt{z_*}}{2 m^2 \left(\sigma^2 + \sqrt{z_*}\, \rho^2 \right)}\, .
\ee
Also, in the UV regime the factor $H$ \eqref{eq:app22HSol} has the following approximate expression:
\be
  H = \frac{z_*}{m^4 \left(\rho^2 \sqrt{z_*} + \sigma^2 \right)^2}\, .
\ee
Like for the $\mathcal{N}=(4,4)$ case, we define two new variables for the $\mathcal{N}=(2,2)^*$ case:
\be
  u = \sqrt{\sigma^2 + \sqrt{z_*} \rho^2}\, ,\quad \textrm{tan} \hat{\alpha} = \frac{\sigma}{(z_*)^{1/4} \rho}\, ,\quad 0 \leq \hat{\alpha} \leq \frac{\pi}{2}\, .
\ee
At large $u$,
\be
  z \to z_*\, ,\quad H \to \frac{z_*}{m^4 u^4}\, .
\ee
Applying these expressions in the metric \eqref{eq:22metric}, we obtain the approximate metric for the $\mathcal{N}=(2,2)^*$ case with $\widetilde{c}=1/2$ in the UV regime:
\begin{align}
  ds^2 & = \frac{m^2 u^2}{\sqrt{z_*}} \left[dx_{1,1}^2 + \frac{z_*}{m^2} \left(d\theta^2 + \textrm{sin}^2 \theta\, (d\phi)^2 \right) \right] + \frac{1}{m^2} \frac{du^2}{u^2} \nonumber\\
  {} & \quad + \frac{1}{m^2} \Bigg[d\hat{\alpha}^2 + \textrm{sin}^2 \hat{\alpha} \left(d\widetilde{\psi}^2 + \textrm{sin}^2 \widetilde{\psi} \left(d\phi^3 + \frac{1}{2} \textrm{cos} \theta\, d\phi \right)^2 + \textrm{cos}^2 \widetilde{\psi} \left(d\phi^1 + \frac{1}{2} \textrm{cos} \theta\, d\phi \right)^2 \right) \nonumber\\
  {} & \qquad\qquad + \textrm{cos}^2 \hat{\alpha}\, d\psi^2 \Bigg]\, .
\end{align}

\section{RR 5-Form Flux}\label{app:5form}

The gravity solution in 10D type IIB supergravity includes an RR 5-form flux. In this appendix, we discuss this RR flux and its quantization condition.

Let us first review the $\mathcal{N}=(4,4)$ case, which was discussed in Ref.~\cite{GravDual-1}. One starts with the following Ansatz:
\be
  F_5 = \mathcal{F}_5 + * \mathcal{F}_5
\ee
with $\mathcal{F}_5 = d\mathcal{C}_4$ and
\be
  \mathcal{C}_4 = g(\rho,\, \sigma)\, \omega_3 \wedge (d\psi + \textrm{cos} \theta\, d\phi)\, ,
\ee
where $\omega_3$ is the volume form of the 3-sphere defined by the metric
\be
  d\Omega_3^2 = d\beta_1^2 + \textrm{sin}^2 \beta_1 \left(d\beta_2^2 + \textrm{sin}^2 \beta_2\, d\beta_3^2 \right)
\ee
with
\be
  0 \leq \beta_1,\, \beta_2 \leq \pi\, ,\quad 0 \leq \beta_3 < 2 \pi\, ,
\ee
which is given by
\be
  \omega_3 = \textrm{sin}^2 \beta_1 \, \textrm{sin} \beta_2\, d\beta_1 \wedge d\beta_2 \wedge d\beta_3\, .
\ee
Moreover, the BPS equations imply the Bianchi identity $dF_5 = 0$, which consequently leads to
\be
  F_5 = dC_4
\ee
with
\be
  C_4 = g \omega_3 \wedge (d\psi + \textrm{cos}\theta \, d\phi) + dx^0 \wedge dx^1 \wedge \left[\frac{z}{m^2 H} \omega_2 - \frac{\sigma}{z}\, d\sigma \wedge (d\psi + \textrm{cos} \theta\, d\phi) \right]\, ,
\ee
where
\be
  \omega_2 = \textrm{sin}\theta\, d\theta\wedge d\phi\, .
\ee

The results above can also be obtained from the solution in 5D gauged supergravity uplifed to 10D. In this way, the factor $g$ has the expression:
\be
  g = \frac{e^{-\varphi} \, \textrm{cos}^4 \widetilde{\theta}}{m^4 \Delta}\, ,
\ee
which leads to
\begin{align}
  \mathcal{F}_5 & = - \frac{2}{m^4} \frac{e^\varphi + e^{-\varphi} \Delta}{\Delta^2} \, \textrm{sin}\widetilde{\theta}\, \textrm{cos}^3 \widetilde{\theta}\, d\widetilde{\theta} \wedge \omega_3 \wedge (d\psi + \textrm{cos} \theta\, d\phi) \nonumber\\
  {} & \quad - \frac{3}{m^4} \frac{e^\varphi\, \textrm{sin}^2 \widetilde{\theta} \, \textrm{cos}^4 \widetilde{\theta}}{\Delta^2} \, \varphi' \, dr \wedge \omega_3 \wedge (d\psi + \textrm{cos} \theta \, d\phi) + \frac{e^{-\varphi}}{m^4} \frac{\textrm{cos}^4 \widetilde{\theta}}{\Delta}\, \omega_3 \wedge \omega_2\, .\label{eq:appF5}
\end{align}
$F_5$ satisfies the quantization condition:
\be
  \frac{1}{2 \kappa_{10}^2} \int_{\mathcal{M}_5} F_5 = N\, T_3
\ee
with
\be
  2 \kappa_{10}^2 = (2 \pi)^7 \, g_s^2 \, (\alpha')^4\, ,\quad T_3 = \frac{1}{(2 \pi)^3 \, g_s \, (\alpha')^2}\, .
\ee
We should perform the integration at $\tau \to \infty$ along the transverse 5-sphere parametrized by $(\widetilde{\theta},\, \psi,\, \beta_i)$, where $\tau$ is defined in Eq.~\eqref{eq:def44tau}, hence only the first term in Eq.~\eqref{eq:appF5} contributes. When $\tau \to \infty$,
\be
  \varphi \to 0\, ,\quad \Delta \to 1\, ,
\ee
then
\be
  \int_{S^5} \mathcal{F}_5 \Big|_{S^5} = \frac{4 \pi^3}{m^4}\, .
\ee
Therefore, the quantization condition of the flux fixes the constant $m$ to be
\be
  \frac{1}{m^2} = \sqrt{4 \pi g_s\, N} \alpha'\, ,
\ee
where $g_s$ and $\alpha'$ are the string coupling constant and the Regge slope respectively.

Now let us turn to the $\mathcal{N}=(2,2)^*$ case. As we discussed in the text, the way of constructing the gravity dual of the 2D $\mathcal{N}=(2,2)^*$ super Yang-Mills theory is to first find the solution in the 5D $\mathcal{N}=2$ gauged supergravity and then uplift it to 10D using the formulae in Ref.~\cite{Cvetic}. The result is \eqref{eq:22flux}:
\be
  F_5 = \mathcal{F}_5 + * \mathcal{F}_5\, ,
\ee
where
\be
  \mathcal{F}_5 = \sum_{I=1}^3 \left[2 m X^I (X^I \mu_I^2 - \Delta) \epsilon_5 + \frac{1}{2 m^2 (X^I)^2} d(\mu_I^2) \left((d \phi^I + A^I) \wedge *_5 F^I + m X^I *_5 dX^I \right) \right]\, ,
\ee
and $\epsilon_5$ and $*_5$ are the volume form of $ds_5^2$ and the Hodge dual in $ds_5$ respectively, while $F^I = dA^I$ are the field strengths of the gauge fields given by Eq.~\eqref{eq:22gaugefield}. $\phi^I$ ($I = 1, 2, 3$) are three angles with the range $[0, 2\pi)$, which are not related to the scalar fields $\phi_{1, 2}$ appearing in the action \eqref{eq:5DGravL}.

The quantization condition is still given by \eqref{eq:Fquantization}:
\be
  \frac{1}{2 \kappa_{10}^2} \int_{\mathcal{M}_5} F_5 = N\, T_3\, .
\ee
We see that the only contribution to the integral comes from the term $\sim * \epsilon_5$. More precisely, the quantization condition in this case becomes
\be
  \frac{1}{2 \kappa_{10}^2} \int_{\mathcal{M}_5} 2 m \sum_{I=1}^3 X^I (X^I \mu_I^2 - \Delta) \left(* \epsilon_5\right) = N\, T_3\, .
\ee
For the special case with $\widetilde{c} = 1/2$, based on our analysis in Appendix~\ref{app:10Dmetric} there are the following results:
\be
  2 m \sum_{I=1}^3 X^I (X^I \mu_I^2 - \Delta) = - 2 m \left(e^{\hat{\varphi}} + e^{-\hat{\varphi}} \Delta \right)\, ,
\ee
\be
  * \epsilon_5 = \frac{1}{m^5 \Delta^2} \textrm{sin} \widetilde{\theta}\, \textrm{cos}^3 \widetilde{\theta}\, d\widetilde{\theta}  \wedge \omega_3' \wedge d\phi^2 \, ,
\ee
where
\be
  \omega_3' = \textrm{sin} \widetilde{\psi}\, \textrm{cos} \widetilde{\psi} \, d\widetilde{\psi} \wedge \left(d\phi^1 + \frac{1}{2} \textrm{cos}\theta\, d\phi \right) \wedge \left(d\phi^3 + \frac{1}{2} \textrm{cos}\theta\, d\phi \right)\, .
\ee
We see that in the limit $\tau \to \infty$ the term $\sim * \epsilon_5$ gives the exactly same contribution as the first term in Eq.~\eqref{eq:appF5} for the $\mathcal{N}=(4,4)$ case. Hence, the quantization condition for the $\mathcal{N}=(2,2)^*$ case with $\widetilde{c}=1/2$ imposes the same condition \eqref{eq:constm} on the constant $m$:
\be
  \frac{1}{m^2} = \sqrt{4 \pi g_s\, N} \alpha'\, .
\ee

\bibliographystyle{utphys}
\bibliography{IntegModel}

\providecommand{\href}[2]{#2}\begingroup\raggedright\begin{thebibliography}{10}

\bibitem{Maldacena}
J.~M. Maldacena, ``{The Large N limit of superconformal field theories and
  supergravity},'' \href{http://dx.doi.org/10.1023/A:1026654312961}{{\em Int.
  J. Theor. Phys.} {\bfseries 38} (1999) 1113--1133},
  \href{http://arxiv.org/abs/hep-th/9711200}{{\ttfamily arXiv:hep-th/9711200
  [hep-th]}}.
[Adv. Theor. Math. Phys.2,231(1998)].

\bibitem{AdSCFTinteg}
N.~Beisert {\em et~al.}, ``{Review of AdS/CFT Integrability: An Overview},''
  \href{http://dx.doi.org/10.1007/s11005-011-0529-2}{{\em Lett. Math. Phys.}
  {\bfseries 99} (2012) 3--32},
\href{http://arxiv.org/abs/1012.3982}{{\ttfamily arXiv:1012.3982 [hep-th]}}.

\bibitem{NS-1}
N.~A. Nekrasov and S.~L. Shatashvili, ``{Supersymmetric vacua and Bethe
  ansatz},'' \href{http://dx.doi.org/10.1016/j.nuclphysbps.2009.07.047}{{\em
  Nucl. Phys. Proc. Suppl.} {\bfseries 192-193} (2009) 91--112},
\href{http://arxiv.org/abs/0901.4744}{{\ttfamily arXiv:0901.4744 [hep-th]}}.

\bibitem{NS-2}
N.~A. Nekrasov and S.~L. Shatashvili, ``{Quantum integrability and
  supersymmetric vacua},'' \href{http://dx.doi.org/10.1143/PTPS.177.105}{{\em
  Prog. Theor. Phys. Suppl.} {\bfseries 177} (2009) 105--119},
\href{http://arxiv.org/abs/0901.4748}{{\ttfamily arXiv:0901.4748 [hep-th]}}.

\bibitem{NS-3}
N.~A. Nekrasov and S.~L. Shatashvili, ``{Quantization of Integrable Systems and
  Four Dimensional Gauge Theories},'' in {\em {Proceedings, 16th International
  Congress on Mathematical Physics (ICMP09): Prague, Czech Republic, August
  3-8, 2009}}, pp.~265--289.
\newblock 2009.
\newblock
\href{http://arxiv.org/abs/0908.4052}{{\ttfamily arXiv:0908.4052 [hep-th]}}.
\newblock

\bibitem{HOR-1}
S.~Hellerman, D.~Orlando, and S.~Reffert, ``{String theory of the Omega
  deformation},'' \href{http://dx.doi.org/10.1007/JHEP01(2012)148}{{\em JHEP}
  {\bfseries 01} (2012) 148},
\href{http://arxiv.org/abs/1106.0279}{{\ttfamily arXiv:1106.0279 [hep-th]}}.

\bibitem{HOR-2}
S.~Hellerman, D.~Orlando, and S.~Reffert, ``{The Omega Deformation From String
  and M-Theory},'' \href{http://dx.doi.org/10.1007/JHEP07(2012)061}{{\em JHEP}
  {\bfseries 07} (2012) 061},
\href{http://arxiv.org/abs/1204.4192}{{\ttfamily arXiv:1204.4192 [hep-th]}}.

\bibitem{OR-2}
D.~Orlando and S.~Reffert, ``{Deformed supersymmetric gauge theories from the
  fluxtrap background},''
  \href{http://dx.doi.org/10.1142/S0217751X13300445}{{\em Int. J. Mod. Phys.}
  {\bfseries A28} (2013) 1330044},
\href{http://arxiv.org/abs/1309.7350}{{\ttfamily arXiv:1309.7350 [hep-th]}}.

\bibitem{GS-1}
A.~A. Gerasimov and S.~L. Shatashvili, ``{Higgs Bundles, Gauge Theories and
  Quantum Groups},'' \href{http://dx.doi.org/10.1007/s00220-007-0369-1}{{\em
  Commun. Math. Phys.} {\bfseries 277} (2008) 323--367},
\href{http://arxiv.org/abs/hep-th/0609024}{{\ttfamily arXiv:hep-th/0609024
  [hep-th]}}.

\bibitem{GS-2}
A.~A. Gerasimov and S.~L. Shatashvili, ``{Two-dimensional gauge theories and
  quantum integrable systems},'' in {\em {Proceedings of Symposia in Pure
  Mathematics, May 25-29 2007, University of Augsburg, Germany}}.
\newblock 2007.
\newblock
\href{http://arxiv.org/abs/0711.1472}{{\ttfamily arXiv:0711.1472 [hep-th]}}.
\newblock

\bibitem{HiggsBundle}
G.~W. Moore, N.~Nekrasov, and S.~Shatashvili, ``{Integrating over Higgs
  branches},'' \href{http://dx.doi.org/10.1007/PL00005525}{{\em Commun. Math.
  Phys.} {\bfseries 209} (2000) 97--121},
\href{http://arxiv.org/abs/hep-th/9712241}{{\ttfamily arXiv:hep-th/9712241
  [hep-th]}}.

\bibitem{GravDual-1}
D.~Arean, P.~Merlatti, C.~Nunez, and A.~V. Ramallo, ``{String duals of
  two-dimensional (4,4) supersymmetric gauge theories},''
  \href{http://dx.doi.org/10.1088/1126-6708/2008/12/054}{{\em JHEP} {\bfseries
  12} (2008) 054},
\href{http://arxiv.org/abs/0810.1053}{{\ttfamily arXiv:0810.1053 [hep-th]}}.

\bibitem{IntegProceeding}
J.~Nian, ``{Nonlinear Schr\"odinger Equation, 2D $N = (2, 2)^*$ Topological
  Yang-Mills-Higgs Theory and Their Gravity Dual},''
\href{http://dx.doi.org/10.1088/1742-6596/804/1/012033}{{\em J. Phys. Conf.
  Ser.} {\bfseries 804} no.~1, (2017) 012033}.

\bibitem{Witten-2}
E.~Witten, ``{Two-dimensional gauge theories revisited},''
  \href{http://dx.doi.org/10.1016/0393-0440(92)90034-X}{{\em J. Geom. Phys.}
  {\bfseries 9} (1992) 303--368},
\href{http://arxiv.org/abs/hep-th/9204083}{{\ttfamily arXiv:hep-th/9204083
  [hep-th]}}.

\bibitem{Migdal}
A.~A. Migdal, ``{Recursion Equations in Gauge Theories},'' {\em Sov. Phys.
  JETP} {\bfseries 42} (1975) 413.
[Zh. Eksp. Teor. Fiz.69,810(1975)].

\bibitem{Rusakov}
B.~E. Rusakov, ``{Loop averages and partition functions in U(N) gauge theory on
  two-dimensional manifolds},''
\href{http://dx.doi.org/10.1142/S0217732390000780}{{\em Mod. Phys. Lett.}
  {\bfseries A5} (1990) 693--703}.

\bibitem{Fine}
D.~S. Fine, ``{Quantum Yang-Mills on the two-sphere},''
\href{http://dx.doi.org/10.1007/BF02097703}{{\em Commun. Math. Phys.}
  {\bfseries 134} (1990) 273--292}.

\bibitem{Witten-1}
E.~Witten, ``{On quantum gauge theories in two-dimensions},''
\href{http://dx.doi.org/10.1007/BF02100009}{{\em Commun. Math. Phys.}
  {\bfseries 141} (1991) 153--209}.

\bibitem{Blau}
M.~Blau and G.~Thompson, ``{Quantum Yang-Mills theory on arbitrary surfaces},''
  \href{http://dx.doi.org/10.1142/S0217751X9200168X}{{\em Int. J. Mod. Phys.}
  {\bfseries A7} (1992) 3781--3806}.
[,28(1991)].

\bibitem{Gross-1}
D.~J. Gross, ``{Two-dimensional QCD as a string theory},''
  \href{http://dx.doi.org/10.1016/0550-3213(93)90402-B}{{\em Nucl. Phys.}
  {\bfseries B400} (1993) 161--180},
\href{http://arxiv.org/abs/hep-th/9212149}{{\ttfamily arXiv:hep-th/9212149
  [hep-th]}}.

\bibitem{Gross-2}
D.~J. Gross and W.~Taylor, ``{Two-dimensional QCD is a string theory},''
  \href{http://dx.doi.org/10.1016/0550-3213(93)90403-C}{{\em Nucl. Phys.}
  {\bfseries B400} (1993) 181--208},
\href{http://arxiv.org/abs/hep-th/9301068}{{\ttfamily arXiv:hep-th/9301068
  [hep-th]}}.

\bibitem{Gross-3}
D.~J. Gross and W.~Taylor, ``{Twists and Wilson loops in the string theory of
  two-dimensional QCD},''
  \href{http://dx.doi.org/10.1016/0550-3213(93)90042-N}{{\em Nucl. Phys.}
  {\bfseries B403} (1993) 395--452},
\href{http://arxiv.org/abs/hep-th/9303046}{{\ttfamily arXiv:hep-th/9303046
  [hep-th]}}.

\bibitem{Moore-1}
G.~W. Moore, ``{2-D Yang-Mills theory and topological field theory},''
\newblock 1994.
\newblock
\href{http://arxiv.org/abs/hep-th/9409044}{{\ttfamily arXiv:hep-th/9409044
  [hep-th]}}.
\newblock

\bibitem{Moore-2}
S.~Cordes, G.~W. Moore, and S.~Ramgoolam, ``{Lectures on 2-d Yang-Mills theory,
  equivariant cohomology and topological field theories},''
  \href{http://dx.doi.org/10.1016/0920-5632(95)00434-B}{{\em Nucl. Phys. Proc.
  Suppl.} {\bfseries 41} (1995) 184--244},
\href{http://arxiv.org/abs/hep-th/9411210}{{\ttfamily arXiv:hep-th/9411210
  [hep-th]}}.

\bibitem{Minahan}
J.~A. Minahan and A.~P. Polychronakos, ``{Equivalence of two-dimensional QCD
  and the C = 1 matrix model},''
  \href{http://dx.doi.org/10.1016/0370-2693(93)90504-B}{{\em Phys. Lett.}
  {\bfseries B312} (1993) 155--165},
\href{http://arxiv.org/abs/hep-th/9303153}{{\ttfamily arXiv:hep-th/9303153
  [hep-th]}}.

\bibitem{Gorsky}
A.~Gorsky and N.~Nekrasov, ``{Hamiltonian systems of Calogero type and
  two-dimensional Yang-Mills theory},''
  \href{http://dx.doi.org/10.1016/0550-3213(94)90429-4}{{\em Nucl. Phys.}
  {\bfseries B414} (1994) 213--238},
\href{http://arxiv.org/abs/hep-th/9304047}{{\ttfamily arXiv:hep-th/9304047
  [hep-th]}}.

\bibitem{Douglas}
M.~R. Douglas, ``{Conformal field theory techniques in large N Yang-Mills
  theory},'' in {\em {NATO Advanced Research Workshop on New Developments in
  String Theory, Conformal Models and Topological Field Theory Cargese, France,
  May 12-21, 1993}}.
\newblock 1993.
\newblock
\href{http://arxiv.org/abs/hep-th/9311130}{{\ttfamily arXiv:hep-th/9311130
  [hep-th]}}.
\newblock

\bibitem{Rudd}
R.~E. Rudd, ``{The String partition function for QCD on the torus},''
\href{http://arxiv.org/abs/hep-th/9407176}{{\ttfamily arXiv:hep-th/9407176
  [hep-th]}}.

\bibitem{Vafa-1}
C.~Vafa, ``{Two dimensional Yang-Mills, black holes and topological strings},''
\href{http://arxiv.org/abs/hep-th/0406058}{{\ttfamily arXiv:hep-th/0406058
  [hep-th]}}.

\bibitem{Vafa-2}
M.~Aganagic, H.~Ooguri, N.~Saulina, and C.~Vafa, ``{Black holes, q-deformed 2d
  Yang-Mills, and non-perturbative topological strings},''
  \href{http://dx.doi.org/10.1016/j.nuclphysb.2005.02.035}{{\em Nucl. Phys.}
  {\bfseries B715} (2005) 304--348},
\href{http://arxiv.org/abs/hep-th/0411280}{{\ttfamily arXiv:hep-th/0411280
  [hep-th]}}.

\bibitem{PestunThesis}
V.~Pestun, {\em {Wilson loops in Supersymmetric Gauge Theories}}.
\newblock PhD thesis, Princeton U.,
2008.
\newblock

\bibitem{S2-1}
F.~Benini and S.~Cremonesi, ``{Partition Functions of ${\mathcal{N}=(2,2)}$
  Gauge Theories on S$^{2}$ and Vortices},''
  \href{http://dx.doi.org/10.1007/s00220-014-2112-z}{{\em Commun. Math. Phys.}
  {\bfseries 334} no.~3, (2015) 1483--1527},
\href{http://arxiv.org/abs/1206.2356}{{\ttfamily arXiv:1206.2356 [hep-th]}}.

\bibitem{S2-2}
N.~Doroud, J.~Gomis, B.~Le~Floch, and S.~Lee, ``{Exact Results in D=2
  Supersymmetric Gauge Theories},''
  \href{http://dx.doi.org/10.1007/JHEP05(2013)093}{{\em JHEP} {\bfseries 05}
  (2013) 093},
\href{http://arxiv.org/abs/1206.2606}{{\ttfamily arXiv:1206.2606 [hep-th]}}.

\bibitem{TopQFT}
E.~Witten, ``{Topological Quantum Field Theory},''
\href{http://dx.doi.org/10.1007/BF01223371}{{\em Commun. Math. Phys.}
  {\bfseries 117} (1988) 353}.

\bibitem{PestunReview}
V.~Pestun, \href{http://dx.doi.org/10.1007/978-3-319-18769-3_6}{``{Localization
  for $\mathcal{N}=2$ Supersymmetric Gauge Theories in Four Dimensions},''} in
  {\em New Dualities of Supersymmetric Gauge Theories}, {Teschner, J\"org},
  ed., pp.~159--194.
\newblock 2016.
\newblock
\href{http://arxiv.org/abs/1412.7134}{{\ttfamily arXiv:1412.7134 [hep-th]}}.
\newblock

\bibitem{TopSYM}
A.~Galperin and O.~Ogievetsky, ``{Holonomy Groups, Complex Structures and $D=4$
  Topological {Yang-Mills} Theory},''
\href{http://dx.doi.org/10.1007/BF02352500}{{\em Commun. Math. Phys.}
  {\bfseries 139} (1991) 377--394}.

\bibitem{PW-2}
D.~Z. Freedman, S.~S. Gubser, K.~Pilch, and N.~P. Warner, ``{Renormalization
  group flows from holography supersymmetry and a c theorem},'' {\em Adv.
  Theor. Math. Phys.} {\bfseries 3} (1999) 363--417,
\href{http://arxiv.org/abs/hep-th/9904017}{{\ttfamily arXiv:hep-th/9904017
  [hep-th]}}.

\bibitem{PW-1}
K.~Pilch and N.~P. Warner, ``{N=2 supersymmetric RG flows and the IIB
  dilaton},'' \href{http://dx.doi.org/10.1016/S0550-3213(00)00656-8}{{\em Nucl.
  Phys.} {\bfseries B594} (2001) 209--228},
\href{http://arxiv.org/abs/hep-th/0004063}{{\ttfamily arXiv:hep-th/0004063
  [hep-th]}}.

\bibitem{MN-1}
J.~M. Maldacena and C.~Nunez, ``{Supergravity description of field theories on
  curved manifolds and a no go theorem},''
  \href{http://dx.doi.org/10.1142/S0217751X01003935,
  10.1142/S0217751X01003937}{{\em Int. J. Mod. Phys.} {\bfseries A16} (2001)
  822--855}, \href{http://arxiv.org/abs/hep-th/0007018}{{\ttfamily
  arXiv:hep-th/0007018 [hep-th]}}.
[182(2000)].

\bibitem{Cvetic}
M.~Cvetic, M.~J. Duff, P.~Hoxha, J.~T. Liu, H.~Lu, J.~X. Lu,
  R.~Martinez-Acosta, C.~N. Pope, H.~Sati, and T.~A. Tran, ``{Embedding AdS
  black holes in ten-dimensions and eleven-dimensions},''
  \href{http://dx.doi.org/10.1016/S0550-3213(99)00419-8}{{\em Nucl. Phys.}
  {\bfseries B558} (1999) 96--126},
\href{http://arxiv.org/abs/hep-th/9903214}{{\ttfamily arXiv:hep-th/9903214
  [hep-th]}}.

\bibitem{Chamseddine}
A.~H. Chamseddine and W.~A. Sabra, ``{Magnetic and dyonic black holes in D = 4
  gauged supergravity},''
  \href{http://dx.doi.org/10.1016/S0370-2693(00)00652-3}{{\em Phys. Lett.}
  {\bfseries B485} (2000) 301--307},
\href{http://arxiv.org/abs/hep-th/0003213}{{\ttfamily arXiv:hep-th/0003213
  [hep-th]}}.

\bibitem{Eoin-1}
{\'O Colg\'ain, Eoin}, ``{Warped AdS$_3$, dS$_3$ and flows from $\mathcal{N} =
  (0,2)$ SCFTs},'' \href{http://dx.doi.org/10.1103/PhysRevD.91.105029}{{\em
  Phys. Rev.} {\bfseries D91} no.~10, (2015) 105029},
\href{http://arxiv.org/abs/1501.04355}{{\ttfamily arXiv:1501.04355 [hep-th]}}.

\bibitem{Eoin-2}
{\'O Colg\'ain, Eoin}, ``{All supersymmetric solutions of 3D U(1)$^3$ gauged
  supergravity},'' \href{http://dx.doi.org/10.1007/JHEP11(2015)116}{{\em JHEP}
  {\bfseries 11} (2015) 116},
\href{http://arxiv.org/abs/1502.04668}{{\ttfamily arXiv:1502.04668 [hep-th]}}.

\bibitem{Eoin-3}
{Karndumri, Parinya and \'O Colg\'ain, Eoin}, ``{3D supergravity from wrapped
  M5-branes},'' \href{http://dx.doi.org/10.1007/JHEP03(2016)188}{{\em JHEP}
  {\bfseries 03} (2016) 188},
\href{http://arxiv.org/abs/1508.00963}{{\ttfamily arXiv:1508.00963 [hep-th]}}.

\bibitem{GravDual-3}
D.~Arean, E.~Conde, A.~V. Ramallo, and D.~Zoakos, ``{Holographic duals of SQCD
  models in low dimensions},''
  \href{http://dx.doi.org/10.1007/JHEP06(2010)095}{{\em JHEP} {\bfseries 06}
  (2010) 095},
\href{http://arxiv.org/abs/1004.4212}{{\ttfamily arXiv:1004.4212 [hep-th]}}.

\bibitem{Bertolini}
M.~Bertolini, ``{Four lectures on the gauge / gravity correspondence},''
  \href{http://dx.doi.org/10.1142/S0217751X03016811}{{\em Int. J. Mod. Phys.}
  {\bfseries A18} (2003) 5647--5712},
\href{http://arxiv.org/abs/hep-th/0303160}{{\ttfamily arXiv:hep-th/0303160
  [hep-th]}}.

\bibitem{BB}
F.~Benini and N.~Bobev, ``{Two-dimensional SCFTs from wrapped branes and
  c-extremization},'' \href{http://dx.doi.org/10.1007/JHEP06(2013)005}{{\em
  JHEP} {\bfseries 06} (2013) 005},
\href{http://arxiv.org/abs/1302.4451}{{\ttfamily arXiv:1302.4451 [hep-th]}}.

\bibitem{Martin-1}
T.~Buscher, U.~Lindstrom, and M.~Rocek, ``{New Supersymmetric $\sigma$ Models
  With {Wess-Zumino} Terms},''
\href{http://dx.doi.org/10.1016/0370-2693(88)90859-3}{{\em Phys. Lett.}
  {\bfseries B202} (1988) 94--98}.

\bibitem{Martin-2}
U.~Lindstrom, M.~Rocek, R.~von Unge, and M.~Zabzine, ``{Generalized Kahler
  geometry and manifest N = (2,2) supersymmetric nonlinear sigma-models},''
  \href{http://dx.doi.org/10.1088/1126-6708/2005/07/067}{{\em JHEP} {\bfseries
  07} (2005) 067},
\href{http://arxiv.org/abs/hep-th/0411186}{{\ttfamily arXiv:hep-th/0411186
  [hep-th]}}.

\bibitem{Martin-3}
U.~Lindstrom, M.~Rocek, R.~von Unge, and M.~Zabzine, ``{Generalized Kahler
  manifolds and off-shell supersymmetry},''
  \href{http://dx.doi.org/10.1007/s00220-006-0149-3}{{\em Commun. Math. Phys.}
  {\bfseries 269} (2007) 833--849},
\href{http://arxiv.org/abs/hep-th/0512164}{{\ttfamily arXiv:hep-th/0512164
  [hep-th]}}.

\bibitem{Martin-4}
U.~Lindstrom, M.~Rocek, I.~Ryb, R.~von Unge, and M.~Zabzine, ``{New N = (2,2)
  vector multiplets},''
  \href{http://dx.doi.org/10.1088/1126-6708/2007/08/008}{{\em JHEP} {\bfseries
  08} (2007) 008},
\href{http://arxiv.org/abs/0705.3201}{{\ttfamily arXiv:0705.3201 [hep-th]}}.

\bibitem{Martin-5}
U.~Lindstrom, M.~Rocek, I.~Ryb, R.~von Unge, and M.~Zabzine, ``{Nonabelian
  Generalized Gauge Multiplets},''
  \href{http://dx.doi.org/10.1088/1126-6708/2009/02/020}{{\em JHEP} {\bfseries
  02} (2009) 020},
\href{http://arxiv.org/abs/0808.1535}{{\ttfamily arXiv:0808.1535 [hep-th]}}.

\bibitem{T2}
J.~Nian and X.~Zhang, ``{Dynamics of two-dimensional $
  \mathcal{N}=\left(2,\;2\right) $ theories with semichiral superfields I},''
  \href{http://dx.doi.org/10.1007/JHEP11(2015)047}{{\em JHEP} {\bfseries 11}
  (2015) 047},
\href{http://arxiv.org/abs/1411.4694}{{\ttfamily arXiv:1411.4694 [hep-th]}}.

\bibitem{S2}
F.~Benini, P.~M. Crichigno, D.~Jain, and J.~Nian, ``{Semichiral fields on
  S$^{2}$ and generalized K\"ahler geometry},''
  \href{http://dx.doi.org/10.1007/JHEP01(2016)060}{{\em JHEP} {\bfseries 01}
  (2016) 060},
\href{http://arxiv.org/abs/1505.06207}{{\ttfamily arXiv:1505.06207 [hep-th]}}.

\bibitem{HananyHori}
A.~Hanany and K.~Hori, ``{Branes and N=2 theories in two-dimensions},''
  \href{http://dx.doi.org/10.1016/S0550-3213(97)00754-2}{{\em Nucl. Phys.}
  {\bfseries B513} (1998) 119--174},
\href{http://arxiv.org/abs/hep-th/9707192}{{\ttfamily arXiv:hep-th/9707192
  [hep-th]}}.

\bibitem{Reffert}
S.~Reffert, ``{General Omega Deformations from Closed String Backgrounds},''
  \href{http://dx.doi.org/10.1007/JHEP04(2012)059}{{\em JHEP} {\bfseries 04}
  (2012) 059},
\href{http://arxiv.org/abs/1108.0644}{{\ttfamily arXiv:1108.0644 [hep-th]}}.

\bibitem{OR-1}
D.~Orlando and S.~Reffert, ``{Twisted Masses and Enhanced Symmetries: the A\&D
  Series},'' \href{http://dx.doi.org/10.1007/JHEP02(2012)060}{{\em JHEP}
  {\bfseries 02} (2012) 060},
\href{http://arxiv.org/abs/1111.4811}{{\ttfamily arXiv:1111.4811 [hep-th]}}.

\bibitem{2DYMcoupling}
P.~Di~Vecchia, H.~Enger, E.~Imeroni, and E.~Lozano-Tellechea, ``{Gauge theories
  from wrapped and fractional branes},''
  \href{http://dx.doi.org/10.1016/S0550-3213(02)00200-6}{{\em Nucl. Phys.}
  {\bfseries B631} (2002) 95--127},
\href{http://arxiv.org/abs/hep-th/0112126}{{\ttfamily arXiv:hep-th/0112126
  [hep-th]}}.

\bibitem{RussoZarembo-1}
J.~G. Russo and K.~Zarembo, ``{Large N Limit of N=2 SU(N) Gauge Theories from
  Localization},'' \href{http://dx.doi.org/10.1007/JHEP10(2012)082}{{\em JHEP}
  {\bfseries 10} (2012) 082},
\href{http://arxiv.org/abs/1207.3806}{{\ttfamily arXiv:1207.3806 [hep-th]}}.

\bibitem{RussoZarembo-2}
J.~G. Russo and K.~Zarembo, ``{Massive N=2 Gauge Theories at Large N},''
  \href{http://dx.doi.org/10.1007/JHEP11(2013)130}{{\em JHEP} {\bfseries 11}
  (2013) 130},
\href{http://arxiv.org/abs/1309.1004}{{\ttfamily arXiv:1309.1004 [hep-th]}}.

\bibitem{RT}
S.~Ryu and T.~Takayanagi, ``{Holographic derivation of entanglement entropy
  from AdS/CFT},'' \href{http://dx.doi.org/10.1103/PhysRevLett.96.181602}{{\em
  Phys. Rev. Lett.} {\bfseries 96} (2006) 181602},
\href{http://arxiv.org/abs/hep-th/0603001}{{\ttfamily arXiv:hep-th/0603001
  [hep-th]}}.

\bibitem{Klebanov}
I.~R. Klebanov, D.~Kutasov, and A.~Murugan, ``{Entanglement as a probe of
  confinement},'' \href{http://dx.doi.org/10.1016/j.nuclphysb.2007.12.017}{{\em
  Nucl. Phys.} {\bfseries B796} (2008) 274--293},
\href{http://arxiv.org/abs/0709.2140}{{\ttfamily arXiv:0709.2140 [hep-th]}}.

\bibitem{Korepin}
V.~E. Korepin, ``{Universality of Entropy Scaling in One Dimensional Gapless
  Models},''
\href{http://dx.doi.org/10.1103/PhysRevLett.92.096402}{{\em Phys. Rev. Lett.}
  {\bfseries 92} (2004) 096402}.

\bibitem{NLS-1}
Y.~Castin, ``Internal structure of a quantum soliton and classical excitations
  due to trap opening,'' {\em The European Physical Journal B} {\bfseries 68}
  no.~3, (2009) 317--328.

\bibitem{NLS-2}
F.~Calogero and A.~Degasperis, ``Comparison between the exact and hartree
  solutions of a one-dimensional many-body problem,'' {\em Physical Review A}
  {\bfseries 11} no.~1, (1975) 265.

\bibitem{NLS-thesis}
D.~I.~H. Holdaway, {\em Many body effects in one-dimensional attractive Bose
  gases}.
\newblock PhD thesis, Durham University, 2013.

\bibitem{GravDual-2}
D.~Arean, E.~Conde, and A.~V. Ramallo, ``{Gravity duals of 2d supersymmetric
  gauge theories},''
  \href{http://dx.doi.org/10.1088/1126-6708/2009/12/006}{{\em JHEP} {\bfseries
  12} (2009) 006},
\href{http://arxiv.org/abs/0909.3106}{{\ttfamily arXiv:0909.3106 [hep-th]}}.

\bibitem{CYAdS-1}
O.~A.~P. Mac~Conamhna, ``{Inverting geometric transitions: Explicit Calabi-Yau
  metrics for the Maldacena-Nunez solutions},''
  \href{http://dx.doi.org/10.1103/PhysRevD.76.106010}{{\em Phys. Rev.}
  {\bfseries D76} (2007) 106010},
\href{http://arxiv.org/abs/0706.1795}{{\ttfamily arXiv:0706.1795 [hep-th]}}.

\bibitem{CYAdS-2}
J.~P. Gauntlett and O.~A.~P. Mac~Conamhna, ``{AdS spacetimes from wrapped
  D3-branes},'' \href{http://dx.doi.org/10.1088/0264-9381/24/24/009}{{\em
  Class. Quant. Grav.} {\bfseries 24} (2007) 6267--6286},
\href{http://arxiv.org/abs/0707.3105}{{\ttfamily arXiv:0707.3105 [hep-th]}}.

\bibitem{Hitchin}
N.~J. Hitchin, ``{Stable bundles and integrable systems},''
\href{http://dx.doi.org/10.1215/S0012-7094-87-05408-1}{{\em Duke Math. J.}
  {\bfseries 54} (1987) 91--114}.

\bibitem{Mason}
L.~J. Mason and N.~M.~J. Woodhouse, {\em {Integrability, selfduality, and
  twistor theory}}.
\newblock
1991.
\newblock

\bibitem{Zee}
A.~Zee, ``{Vortex strings and the antisymmetric gauge potential},''
\href{http://dx.doi.org/10.1016/0550-3213(94)90226-7}{{\em Nucl.Phys.}
  {\bfseries B421} (1994) 111--124}.

\bibitem{Gubser}
S.~S. Gubser, R.~Nayar, and S.~Parikh, ``{Strings, vortex rings, and modes of
  instability},'' \href{http://dx.doi.org/10.1016/j.nuclphysb.2015.01.005}{{\em
  Nucl. Phys.} {\bfseries B892} (2015) 156--180},
\href{http://arxiv.org/abs/1408.2246}{{\ttfamily arXiv:1408.2246 [hep-th]}}.

\bibitem{BEC}
{A. Mu\~{n}oz Mateo}, X.~Yu, and J.~Nian, ``{Vortex lines attached to dark
  solitons in Bose-Einstein condensates and Boson-Vortex Duality in 3+1
  Dimensions},'' \href{http://dx.doi.org/10.1103/PhysRevA.94.063623}{{\em Phys.
  Rev.} {\bfseries A94} no.~6, (2016) 063623},
\href{http://arxiv.org/abs/1606.02776}{{\ttfamily arXiv:1606.02776
  [cond-mat.quant-gas]}}.

\bibitem{BECstring}
S.~Giaccari and J.~Nian, ``{Dark Solitons, D-branes and Noncommutative Tachyon
  Field Theory},'' \href{http://dx.doi.org/10.1142/S0217751X17502013}{{\em Int.
  J. Mod. Phys.} {\bfseries A32} no.~33, (2017) 1750201},
\href{http://arxiv.org/abs/1608.07262}{{\ttfamily arXiv:1608.07262 [hep-th]}}.

\bibitem{KdV}
J.~Nian, ``{Note on Nonlinear Schr\"odinger Equation, KdV Equation and 2D
  Topological Yang-Mills-Higgs Theory},''
\href{http://arxiv.org/abs/1611.04562}{{\ttfamily arXiv:1611.04562 [hep-th]}}.

\bibitem{OSV}
H.~Ooguri, A.~Strominger, and C.~Vafa, ``{Black hole attractors and the
  topological string},''
  \href{http://dx.doi.org/10.1103/PhysRevD.70.106007}{{\em Phys. Rev.}
  {\bfseries D70} (2004) 106007},
\href{http://arxiv.org/abs/hep-th/0405146}{{\ttfamily arXiv:hep-th/0405146
  [hep-th]}}.

\bibitem{5DSUGRA-1}
M.~Gunaydin, L.~J. Romans, and N.~P. Warner, ``{Gauged N=8 Supergravity in
  Five-Dimensions},''
\href{http://dx.doi.org/10.1016/0370-2693(85)90361-2}{{\em Phys. Lett.}
  {\bfseries B154} (1985) 268--274}.

\bibitem{5DSUGRA-2}
M.~Pernici, K.~Pilch, and P.~van Nieuwenhuizen, ``{Gauged N=8 D=5
  Supergravity},''
\href{http://dx.doi.org/10.1016/0550-3213(85)90645-5}{{\em Nucl. Phys.}
  {\bfseries B259} (1985) 460}.

\bibitem{5DSUGRA-3}
M.~Gunaydin, L.~J. Romans, and N.~P. Warner, ``{Compact and Noncompact Gauged
  Supergravity Theories in Five-Dimensions},''
\href{http://dx.doi.org/10.1016/0550-3213(86)90237-3}{{\em Nucl. Phys.}
  {\bfseries B272} (1986) 598--646}.

\bibitem{5DSUGRA-4}
K.~Behrndt, A.~H. Chamseddine, and W.~A. Sabra, ``{BPS black holes in N=2
  five-dimensional AdS supergravity},''
  \href{http://dx.doi.org/10.1016/S0370-2693(98)01208-8}{{\em Phys. Lett.}
  {\bfseries B442} (1998) 97--101},
\href{http://arxiv.org/abs/hep-th/9807187}{{\ttfamily arXiv:hep-th/9807187
  [hep-th]}}.

\bibitem{BershadskyVafa}
M.~Bershadsky, A.~Johansen, V.~Sadov, and C.~Vafa, ``{Topological reduction of
  4-d SYM to 2-d sigma models},''
  \href{http://dx.doi.org/10.1016/0550-3213(95)00242-K}{{\em Nucl. Phys.}
  {\bfseries B448} (1995) 166--186},
\href{http://arxiv.org/abs/hep-th/9501096}{{\ttfamily arXiv:hep-th/9501096
  [hep-th]}}.

\end{thebibliography}\endgroup

\end{document}